\documentclass[aps]{revtex4}
\usepackage[english]{babel}
\usepackage{amsfonts}
\usepackage{amsmath}
\usepackage{amssymb}
\usepackage{latexsym}
\usepackage{hyperref}
\usepackage{grffile}
\usepackage{graphicx}
\usepackage{epstopdf}
\newcommand{\I}{{\rm i}}
\renewcommand{\le}{\leqslant}
\renewcommand{\ge}{\geqslant}
\newcommand{\kk}{\mathbf{k}}
\newcommand{\qq}{\mathbf{Q}}
\newcommand{\zu}{z_{\uparrow}}
\newcommand{\zd}{z_{\downarrow}}
\newcommand{\es}{e_+(\kk)}
\newcommand{\ea}{e_-(\kk)}
\newcommand{\ek}{t_{\kk}}
\newcommand{\eap}[1]{e^{#1}_-(\kk)}
\newcommand{\efsq}{{{\tilde{E}}^{\rm sq}_{\rm F}}}
\newcommand{\lint}{\int\limits}
\newcommand{\ebt}{{\varepsilon_1(\tau)}}
\newcommand{\sumk}{\sum_{\mathbf{k}}}
\newcommand{\efsc}{{E^{\rm sc}_{\rm F}}}

\begin{document}

\title{Metal-insulator transition and antiferromagnetism  in the generalized Hubbard model: 
Treatment of correlation effects}

\author{P.A. Igoshev$^{1,2}$, V.Yu. Irkhin$^{1}$}
\email{igoshev\_pa@imp.uran.ru}
\affiliation{$^{1}$Institute of Metal Physics, Kovalevskaya str. 18, 620108, Ekaterinburg, Russia\\
$^{2}$Ural Federal University, 620002, Ekaterinburg, Russia}

\begin{abstract}
The ground state for the half-filled $t-t'$ Hubbard model is treated within the Hartree-Fock approximation and the slave boson approach including 
correlations. The criterium for the metal-insulator transition in the Slater scenario is formulated using an analytical free-energy expansion in the next-nearest-neighbor transfer integral $t'$ and in direct antiferromagnetic gap $\Delta$. 
The correlation effects are generally demonstrated to favor the first-order transition.
For a square lattice with a strong van Hove singularity, accidental close 
degeneracy of antiferromagnetic and paramagnetic phases is analytically found in a wide 
parameter region. As a result, there exists an
interval of $t'$ values for which the metal-insulator transition is of the first order due to the existence of the van Hove singularity. 
This interval is very sensitive to model parameters (direct exchange integral) or external parameters.  
For the simple and body-centered cubic lattices, the transition from the insulator antiferromagnetic state with  increasing $t'$  occurs to the phase of an antiferromagnetic metal and is a second-order transition which is followed by a transition to a paramagnetic metal. These results are quantitatively modified when taking into 
account the intersite Heisenberg interaction, which can induce first-order 
transitions. A comparison with the Monte Carlo results is performed.
\end{abstract}

\maketitle

\section{Introduction}
The nature of metal-insulator transitions (MITs) is a long-standing problem in condensed matter physics. An important unclear aspect of this problem is theoretical description of MIT and its order. Important challenges in this field are the role of lattice geometry, relevant physical interactions, etc.~\cite{1998:Imada}. 
An exhaustive review of mechanisms underlying MIT is presented in Ref.~\onlinecite{book:Gebhard}. 
%direct exchange interaction and a charge density wave instability effects

%The physical nature of first-order transition is important issue, compounds??.
Experimentally, MIT is usually a first-order transition~\cite{book:Mott}, 
but the role of the electron-lattice coupling, and dominating on-site Coulomb interaction or long-range interactions~\cite{2018:Schuler} in this phenomenon  should be clarified. 
Looking aside of the Peierls mechanism, we focus on the mechanisms of MIT 
originating purely within the electron subsystem:  Slater~\cite{1951:Slater} and Mott~\cite{1949:Mott} scenarios. The first scenario corresponds to antiferromagnetic (AFM) insulating state, and the second to paramagnetic (PM) state.
The   Slater scenarios are realized in NaOsO$_3$~\cite{2009:Shi,2012:Calder,2013:Jung}, pyrochlore oxides $Ln_2$Ir$_2$O$_7$, $Ln = $~Nd, Sm, Eu, Gd, Tb, Dy, and Ho~\cite{2011:Matsuhira}, Cd$_2$Os$_2$O$_7$~\cite{2001:Mandrus}, Pb$_2$CaOsO$_6$~\cite{2020:Pb2CaOsO6},  V$_{2-x}$O$_3$~\cite{1991:Carter:V2O3,2020:Trastoy}, and NiS$_{2-x}$Se$_x$~\cite{1992:Sudo:NiSSex}, in  two latter compounds, metallic AFM phases being found. 
A remarkable transition from Mott to Slater electronic structure was recently observed in the AFM layer-ordered compound Sr$_2$Ir$_{1-x}$Rh$_x$O$_4$\cite{2020:Xu}, which is possibly related to interplay of the Hubbard and Hund interactions and orbital selective physics. 
Another important issue is the role of intersite interaction effects~---~the exchange and charge interactions. 
Whereas a considerable part of MIT physics  is manifested itself in an interorbital interaction being probably relevant for the MIT problem in real compounds, we suppose that some important aspects of MIT problem within the single-orbital picture remain still unsolved. 
The solution of this problem looks useful in two lines: the application to some single orbital compounds (e.g.~copper-oxide systems) on bipartite lattice and solving the classical theoretical MIT problem in the nondegenerate half-filled Hubbard model~\cite{Hubbard-I,Hubbard-III}.   
%Indeed, the  Hubbard model in the case of one electron per site (a \textit{half-filling}) which takes into account on-site Coulomb interaction yields a proper tool for a~route into~the~problem. 
Thus we deal with the properties of a~\textit{generalized} nondegenerate 
Hubbard model including on-site Coulomb interaction  and direct intersite 
exchange interaction which turns out to be qualitatively important. 

Whereas the electron density in the model is fixed, the role of other parameters (Coulomb and direct exchange interaction strengths as well as hopping integrals configuration) should be discussed in detail.  
It is well known that in the weak-coupling limit MIT typically follows  the Slater scenario  originating from the AFM gap formation~\cite{1984:Katsnelson,Spalek}. 
In terms of the renormalization-group  loop expansion, the electron interaction can be generally decomposed in one-loop level as the sum of three channels: particle-particle (Cooper), direct, and crossed (magnetic) particle-hole contributions~\cite{fRG_review:Salmhofer}. 
%The parquet equations~\cite{Parquet} arguments show that for bipartite lattices
Direct analysis of the momentum dependence of (bubble) one-loop susceptibilities implies that for bipartite lattices the nesting property $t_{\mathbf{k}+\mathbf{Q}} = -\ek$ of the electron spectrum $\ek$ in the nearest-neighbor approximation with integral $t$ with respect to nesting vector 
$\mathbf{Q}$ results in dominating of crossed particle-hole channel provided that the Coulomb interaction parameter $U$ is sufficiently small~\cite{fRG_T_flow}. 
This justifies the application of the mean-field (Hartree-Fock) approximation for bipartite lattices at small $U$ in a general way:  
The AFM gap in the electron spectrum (and, hence, the insulator state) appears at infinitely small values of the Coulomb interaction
parameter $U$. At this level of approximation, both the Hubbard on-site and Heisenberg intersite interactions behave similarly. 
However, in the presence of hopping between the next-nearest neighbors (integral $t'$, $\tau = t'/t$ being dimensionless parameter) with characteristic
energy $D'$, the instability of the paramagnetic metal state (and, consequently, the MIT formation in the Slater scenario) occurs at a finite value of $U$. This circumstance poses three issues: (i) a destruction of nesting property of the Fermi surface, which possibly results in the transition into incommensurate state~\cite{1990:Schulz,2015:Igoshev};  (ii) the increase of the roles of alternative to crossed particle-hole channel, particle-particle and direct particle-hole channels resulting in worse applicability of Hartree-Fock approximation (HFA); and~(iii) the increasing difference in 
the roles of on-site (Hubbard) and intersite (Heisenberg) exchange interactions manifesting on many-electron level.

The dramatic role of lattice geometry manifests itself in the occurrence of the van Hove singularities (vHS), 
%of the electronic spectrum in two ways\textbf{[???not clear]}: in the case of AFM insulator (AFM-I) case when the next-nearest-neighbor hopping integral value does not affect the free energy and the gap, and paramagnetic and AFM metal phases where the spectrum enters within the next-nearest-neighbor approximation.   
which can be very different and change significantly  physical properties 
of the system (in particular, thermodynamics of the phase
transition)~\cite{2014:Markiewicz,2007:Igoshev,2010:Igoshev,2011:Igoshev,2015:Igoshev}: The coefficients of the Landau-like expansion in powers of 
 order parameter (e.g.,~gap) acquires additional logarithmic factors.   
This is also true for MIT, so that inverse critical interaction on the MIT line within the AFM phase acquires an additional logarithmic correction~\cite{1998:Hofstetter} for the square lattice due to the presence of vHS 
in electron density of state (DOS) $\rho(E, \tau)$. %being supported by~topological reasons~\cite{1953:vanHove} in the two-dimensional case\textbf{[Wrong???]}. 
Generally, in the main logarithmic
approximation, we have the following estimations for the critical value:
\begin{equation}
	1/U_\textrm{MIT} = \begin{cases}
	\rho(0)\ln(D/D'),& \rho(E) \sim \rho(0) \\
	(a/2)\ln^2(D/D'),& \rho(E) \sim a\ln(D/|E|)\\
	(a'/3)\ln^3(D/D'),& \rho(E) \sim a'\ln^2(D/|E|)	
	\end{cases},
	\label{eq:Stoner_criterion}
\end{equation}
where $\rho(E) = \rho(E, \tau = 0)$, for three bipartite lattices: simple cubic (sc, nonsingular DOS),
square (logarithmic singularity of DOS), and body-centered cubic (bcc, $\ln^2$ singularity of DOS) lattices,
\begin{figure}[h]
\includegraphics[angle=-90,width=0.45\textwidth]{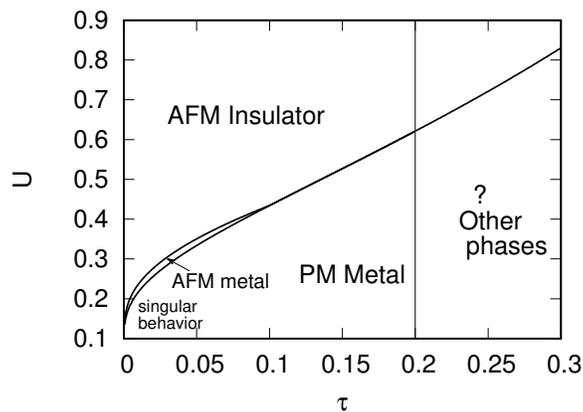}
\caption{
(Color online) Schematic phase diagram of MIT in the ground state in $\tau - U$ terms within the Slater scenario. In the vicinity of $U = 0$, $\tau = 0$, the AFM order is stable and its treatment within HFA is well justified. The region of AFM metal between AFM insulator and paramagnetic regions is present for some lattices only. Away from close vicinity of~$U = 
0$, $\tau = 0$, other phases may present. 
}
\label{fig:PD_scheme}
\end{figure}
respectively~\cite{2019:Igoshev_JETP_MIT}.  Since in the phenomenological 
approach of~Ref.~\onlinecite{2007:Misawa} logarithmic dependence of zero-temperature coefficients in the free-energy expansion was missed this can 
result in some subtle errors connected with the loss of a universality property (pure quadratic contribution to the free energy),  especially for the problems where small energy scales are actual.  
Here, $D$ is the band half-width, $\rho(E)$ is the bare density of states 
near the Fermi level for the electron spectrum in the nearest-neighbor approximation, and $a$ and $a'$ are positive coefficients at singular contributions to $\rho(E)$. Therefore,  presence of these contributions changes significantly the dependence of $U_{\rm MIT}$  on $D'$, or actually on $t'$. However, we will demonstrate that the result (\ref{eq:Stoner_criterion}) is valid only at extremely small $D'$ and is not quantitatively applicable when the condition $D'/D \ll 1$ is not valid.

%In the antiferromagnetic state of the Hubbard model with the half-filled  band,  in the nearest-neighbor approximation, when the electron spectrum $\ek$ satisfies the nesting condition  $t_{\mathbf{k}+\mathbf{Q}} = -\ek$ ($\mathbf{Q}$  is the vector of the AFM structure), the AFM gap in the electron spectrum (and, hence, the insulator state) appears at infinitely small values of the Coulomb interaction parameter $U$. However, in the presence of hopping between the next-nearest neighbors with characteristic energy $D'$, the instability of the paramagnetic metal state occurs at a finite value of $U$. 

Basing on theoretical investigations, it is commonly believed that MIT in 
the nondegenerate Hubbard model for bipartite lattices without vHS in DOS (e.g.,~simple cubic lattice) occurs as continuous phase transition AFM insulator --- AFM metal in a wide region of spectrum parameters, namely next-nearest-neighbor (nnn) hopping integral, 
%the ??frustration parameter $\tau = t'/t$ 
and accordingly to Slater scenario~\cite{1998:Hofstetter,1999:Chitra,2016:Timirgazin}. 
As a consequence of simple quadratic dependence of free energy in all phases,  universal behavior occurs, resulting in impossibility of MIT order change, at least at small gap value.  
However, the investigation of this problem on the square lattice poses a challenge: In early investigations within the simplest HFA approximation it was found that the order of the transition changes from second to the first as $\tau$ increases up to rather small value $\tau\sim 0.1$~\cite{1996:Kondo,1997:Duffy,2010:Yu} and change back from first to second order was found in latter investigation. 
The remarkable feature of the square lattice  is nearly degenerate energies of AFM insulator and PM metal phases at small $\tau$ at MIT transition 
line within AFM phase. 
A schematic phase diagram for MIT within the Slater scenario drawn by hand is shown in Fig.~\ref{fig:PD_scheme}. 
%Completely different situation is  for the square lattice. Here the energies of AFM-I and PM phases are strongly degenerate and accurate calculations of these is an important theoretical problem.
Physically, the difference of two- and three-dimensional cases is crucial: While in three-dimensional case strong van Hove singularity in the density of states (DOS), generally speaking, does not exist (or, for body-centered cubic lattice, occurs only in nearest-neighbor hopping approximation, apart from the  giant van Hove singularity line in the peculiar case $t' = t$~\cite{2019:Igoshev_FMM,2019:Igoshev_JETP}), for the square lattice the logarithmic van Hove singularity is always present~\cite{1953:vanHove}, which results in nonanalytic dependence on $t'$ and, therefore, strong lowering the energy of PM metal phase at finite $t'$. At the same time, the energy of AFM insulator phase also acquires  nonanalytical contributions from van Hove singularity of DOS in nearest-neighbor approximation at the center of the band. 
Qualitative validity of these results were supported by simulations on finite lattices within quantum Monte-Carlo approximation~\cite{1987:Hirsch,1997:Duffy,2007:Misawa} where critical interaction for paramagnetic metal~
---~antiferromagnetic insulator transition were found to be larger in about 25\% than within HFA. 
More heavy tools, e.g.,~variational cluster approximation~\cite{2008:Nevidomskyy,2013:Yamada}, variation Monte Carlo approximation~\cite{2006:Yokoyama,2008:Tocchio,2009:Becca}, path integral renormalization group approach~\cite{2001:Kashima,2001:Morita,2006:Mizusaki} were later used for the solution of this problem. 
Within the Kotliar-Ruckenstein slave boson approximation (SBA)~\cite{1986:Kotliar} applied to nondegenerate Hubbard model (without exchange interaction included) it was found numerically that at small coupling the Slater scenario for the square lattice holds but the point $\tau$ of order change substantially decreases (critical $\tau\sim0.07$): The system demonstrates anomalous sensitivity to the parameter change even in small coupling regime~\cite{2000:Yang}!  It is notable that the improvement of HFA attainable within the SBA allows %to  To take into account the corrections beyond the HFA we apply the well-established slave--boson approximation (SBA) method which allows not only to take the correlation correction to HFA, but also 
to distinguish contributions into the energy from singly- and doubly-occupied local (on-site) many-electron states. 

The consideration within the single-site dynamical mean-field theory (DMFT)~\cite{1999:Chitra,2003:Pruschke,2004:Zitzler,2009:Peters} and cellular 
DMFT $2\times2$-cluster approximations~\cite{2017:Frantino} yields the picture of MIT in local self-energy approximation and allows to trace the intersite exchange interaction impact. 
An increase of $U$ within DMFT approximation results in continuous change 
of physical picture of AFM state from Slater one at small $U$: A Kondo resonance peak at Fermi level and lower and upper Hubband band precursors well away from Fermi level,  well-defined AFM Hubbard bands and a gap between them  are formed, which is a manifestation of the insulator AFM phase. 
The difference between Gutzwiller-like picture  quasiparticle residue and 
DMFT including quasiparticle renormalization is pure quantitative. 
It is commonly believed that local quantum fluctuations result in incoherent picture of the spectrum weight instead of accounting of many-electron 
effects via static slave boson amplitudes (SBA). 
One can state that numerical complexity of  DMFT calculations  for the Hubbard model with nonzero $t'$ does not allow to solve some problems (AFM 
metal phase, hysteresis, transition order, phase separation~\cite{2007:Eckstein}, spiral magnetic states, coexistence of different phases). In this context, the MIT problem revives the interest in more simple techniques. 

During recent years, an issue of existence of the AFM metal phase between 
AFM insulator and PM metal has been repeatedly discussed. It was found that within the DMFT for the Hubbard model for the Bethe lattice at arbitrary degree of frustration metal the existence of AFM metal phase is possibly an artifact of numerical solution of the effective single-site Anderson model (see \cite{2004:Zitzler,2009:Peters} and references therein). 
However, the verification of this conclusion, as well as its validation for another lattices is still needed. 

We will also show that, for the lattices with vHS in electron spectrum,  nonanalytic (logarithmic) corrections to expansion coefficients occur for all quantities, which should generally change the order of MIT transition provided that some degeneracy is present. 
Thus, it is very instructive to construct an analytical theory of the MIT, where the van Hove singularities of DOS play an important role. 
We will  demonstrate that  
the correlation effects modify considerably the phase diagram  to favor the first-order transition. 
We also investigate the role of combined action of correlation effects and the intersite (``direct'') exchange interaction $J$,   which turns out to be  important for the phase diagram. At the same time, the simplest Hartree-Fock approach fully misses the considerable influence of  exchange interactions. 
%which however can be taken into account as a kind of correlation effect. 

In Sec. 2 we present the equations of the Hartree-Fock approximation and 
the slave boson approach in the half-filled generalized $t-t'$ Hubbard model with inclusion of the intersite exchange (Heisenberg) interaction. 
In Sec. 3 we derive the equation for anticipated MIT transition within AFM phase in terms of $t'$ and AFM gap $\Delta_\ast$ valid in both Hartree-Fock and slave boson approximations. 
We treat  free energies of AFM insulator and PM phases in HFA and SBA approximations for square and three-dimensional lattices on the MIT line within AFM phase.  
We develop an analytic expansion in $t'$ for the PM phase and analytic expansion of AFM insulator state free energy with respect to AFM gap $\Delta$. 
This enables us to investigate in detail the order of MIT and the analytical origins of its nature. 
We investigate analytically and numerically the impact of correlation effects and direct intersite exchange interaction beyond the Hartree-Fock approximation on the MIT and found its relation to the MIT order and a sign 
of exchange interaction. 
In the Appendix~\ref{appendix:G_expansion}, we present an~general derivation of useful expansion of~the~lattice sum $G(\Delta)$ with 
respect to $\Delta$ and the connection of singularity of $G(\Delta)$ at $\Delta = 0$ and the singularity of DOS $\rho(\epsilon) $ at $\epsilon = 
0$. 
Also, we consider the asymptotics for the density of states for the square, simple cubic and body-centered cubic lattices and use them for analytic investigation of $G(\Delta)$ for these lattices. 
%In the Appendix analytical results for some lattice sums and free energy for paramagnetic and antiferromagnetic insulator phases are derived.
In the Appendix~\ref{appendix:PM_expansion} an analytical expansion of free energy of the paramagnetic phase for the square and simple cubic lattice with respect to $t'$ is presented.

\section{Theoretical Setup}
\subsection{Model}
\label{sec:General}
We start from the generalized  Hubbard model Hamiltonian on a bipartite lattice with direct intersite exchange
\begin{equation}
\mathcal{H}=\sum_{ij\sigma}t_{ij}c^\dag_{i\sigma}c_{j\sigma}
	+U\sum_in_{i\uparrow}n_{i\downarrow} + \frac12\sum_{ij}J_{ij}\mathbf{s}_i\mathbf{s}_j,
\end{equation}
where $c_{i\sigma}/c^\dag_{i\sigma}$ is annihilation/creation Fermi operators, hopping integrals $t_{ij} = -t(t')$ for nearest~(next-nearest)-neighbor site (Wannier) states $i$, $j$, and zero otherwise, $\mathbf{s}_i = 
(1/2)\sum_{\sigma\sigma'}c^\dag_{i\sigma}\vec\sigma_{\sigma\sigma'} c_{i\sigma'}$ is site spin operator, $\sigma,\sigma'$ label spin projections. Whereas the first term describes the kinetic energy of electron states moving in the lattice environment, there are two types of many-body interactions: Coulomb (the second term) and exchange (the third term) interactions. 
%with very different nature being
They are proportional to the numbers of doubly occupied sites and of exchange links of singly occupied state, respectively. 
We still assume nothing concerning a concrete form for site dependence of 
exchange interaction integral $J_{ij}$.
%The notations for the exchange integrals $J_{ij}, J, J'$ are analogous to $t_{ij},t,t'$.

A lot of variants of the ground state magnetic ordering, including spiral, within this model was considered in earlier papers~\cite{1992:Fresard,2010:Igoshev,2015:Igoshev} where the HFA and SBA treatment of the many-body interaction was applied. %spiral magnetic order of the nondegenerate Hubbard model were applied. 
Being motivated by earlier problems within 
the context of MIT context in the small coupling limit, here we focus our 
attention on more concrete case, the Neel AFM ordering in the ground state at half-filling 
\begin{equation}\label{eq:magnetic_order_form}
	\mathbf{m}_i = \hat{\mathbf{z}}m\exp(\I\mathbf{QR}_i),
\end{equation}
where $\mathbf{m}_i = 2\langle \mathbf{s}_i\rangle$, 
$\hat{\mathbf{a}}$ is $a$ axis unit vector, $a = x,z$ labels axes here and below, 
$\vec\sigma = (\sigma^x,\sigma^y,\sigma^z)$ is the vector of the Pauli matrices, 
$\mathbf{Q}$ being AFM wave vector, $m$  the (staggered) magnetization amplitude,  so that $\exp(\I\mathbf{QR}_i) = \pm 1$ when $\mathbf{R}_i$ belongs to the  first (second) sublattice. 
%To treat spiral ordering  
%with wave vector $\mathbf{Q}$ and amplitude $m$
We perform the SU(2) rotation around the $x$ axis on an angle $\mathbf{QR}_i$. 
Therefore the $c$ operators acquire the following transformation
\begin{equation}
	c_{i\sigma}\rightarrow \sum_{\sigma'}U_{\sigma\sigma'}({\bf R}_i)c_{i\sigma'},
\end{equation}
with the spin matrix $U({\bf R}_i)=\exp[\I(\qq{\bf R}_i)({\bf n}\vec\sigma/2)]=\sigma^0\cos(\qq{\bf R}_i/2)+\I(\mathbf{n}\vec\sigma)\sin(\qq{\bf R}_i/2)$, we choose ${\bf n}=\hat{\bf x}$, $\sigma^0$ is unity spin matrix. In terms of transformed $c$ operators, the AFM state looks like usual ferromagnetic order with magnetization directed along $z$ axis. 
Spin operators transform accordingly (Rodrigues's formula) $\mathbf{s}_i \rightarrow \tilde{\mathbf{s}}_i$, where
$\tilde{\mathbf{s}}_i \equiv \mathbf{s}_i \cos \qq{\bf R}_i + \mathbf{n} (\mathbf{n}\mathbf{s}_i)(1 - \cos \qq{\bf R}_i)$. 
After such a transformation the Hamiltonian takes the form
\begin{equation}\label{eq:H_rotated}
\mathcal{H}'=\sum_{ij\sigma\sigma'}t_{ij}^{\sigma\sigma'}c^\dag_{i\sigma}c_{j\sigma'}
	+U\sum_in_{i\uparrow}n_{i\downarrow} + \frac12\sum_{ij}J_{ij}\tilde{\mathbf{s}}_i\tilde{\mathbf{s}}_j,
\end{equation}
with $t_{ij}^{\sigma\sigma'}=e^+_{ij}\sigma^0_{\sigma\sigma'}+e^-_{ij}\sigma^x_{\sigma\sigma'}$, 
where $e_{ij}^\pm$ is Fourier transform of $e_{\pm}(\kk)=\frac12(t_{\kk+\qq/2}\pm t_{\kk-\qq/2})$, 
and $t_{\mathbf{k}}(\tau) = (1/N)\sum_{ij}t_{ij}\exp[\I\mathbf{k}(\mathbf{R}_i-\mathbf{R}_j)]$
being the bare spectrum, $N$ being the site number. %with account of nearest and next-nearest approximation (the primed (doubly primed) summation corresponds to nearest (next-nearest) neighbors sites $i$ and $j$). 
Rewriting the first term in Eq.~(\ref{eq:H_rotated}) in  the Bloch basis $c_\mathbf{k} = N^{-1/2}\sum_i c_i\exp(\mathrm{i}\mathbf{kR}_i)$, we obtain
%The Hamiltonian in the reciprocal space reads
\begin{equation}\label{Hk}	 \mathcal{H}'=\sum_{\kk\sigma\sigma'}t_{\kk}^{\sigma\sigma'}c^\dag_{\kk\sigma}c_{\kk\sigma'}
	+U\sum_in_{i\uparrow}n_{i\downarrow}
	+ \frac12\sum_{ij}J_{ij}\tilde{\mathbf{s}}_i\tilde{\mathbf{s}}_j.
\end{equation}

There are two strategies of treatment of the Hamiltonian~(\ref{eq:H_rotated}): robust application of the mean-field approximation (Hartree-Fock approximation, HFA (Sec.~\ref{sec:HFA})), and more accurate Kotliar-Ruckenstein slave boson approximation (SBA, Sec.~\ref{sec:SBA})~\cite{1986:Kotliar,1992:Fresard}.
\subsection{Hartree-Fock approximation}\label{sec:HFA}
It is widely believed that the Overhauser-type mean-field treatment of the Hamiltonian~(\ref{eq:H_rotated}) yields quantitatively correct results in the case of small coupling $U\ll D$ due to nesting feature of the Fermi surface. 
Namely, we can apply a parquet-like argument: nesting peculiarity of the Fermi surface makes the crossed particle-hole channel dominating over the Cooper (superconducting) and direct particle-hole (screening) channels~\cite{fRG_review:Salmhofer}, which justifies the mean-field (Hartree-Fock) ansatz of the interaction term in~Eq.~(\ref{eq:H_rotated}):
%\begin{equation}
%	n_{i\uparrow} n_{i\downarrow} = n^2_i/4 - \tilde{\mathbf{s}}^2_i \rightarrow n\cdot n_i/2 -  \mathbf{m}_i \tilde{\mathbf{s}}_i - n^2/4 + \mathbf{m}^2_i/4,
%\end{equation}
\begin{equation}
	n_{i\uparrow} n_{i\downarrow} = n^2_i/4 - \tilde{\mathbf{s}}^2_i \rightarrow n\cdot n_i/2 -  2\langle\tilde{\mathbf{s}}_i\rangle \tilde{\mathbf{s}}_i - n^2/4 + \langle\tilde{\mathbf{s}}_i\rangle^2,
\end{equation}
\begin{equation}
	\tilde{\mathbf{s}}_i\tilde{\mathbf{s}}_j \rightarrow \langle\tilde{\mathbf{s}}_i\rangle\tilde{\mathbf{s}}_j + \tilde{\mathbf{s}}_i\langle\tilde{\mathbf{s}}_j\rangle - \langle\tilde{\mathbf{s}}_i\rangle\langle\tilde{\mathbf{s}}_j\rangle,
\end{equation}
with the operator $n_i = n_{i\uparrow} + n_{i\downarrow}$, and electron filling $n = \langle n_i \rangle$, being assumed site independent, is~introduced and $\langle\tilde{\mathbf{s}}_i\rangle = m\hat{\mathbf{z}}/2$ according to Eq.~(\ref{eq:magnetic_order_form}). 
The resulting Hamiltonian reads
\begin{equation}\label{eq:Hk_MF}
\mathcal{H}_{\rm HFA} =\sum_{\kk\sigma\sigma'}((\es + Un/2)\sigma^0_{\sigma\sigma'} - U_{\rm eff}^{\rm HFA}m\sigma^z_{\sigma\sigma'}/2 + \ea\sigma^x_{\sigma\sigma'} )c^\dag_{\kk\sigma}c_{\kk\sigma'}
- \frac{N}4(Un^2 - U_{\rm eff}^{\rm HFA}m^2),
\end{equation}
where
\begin{equation}\label{eq:U_eff_HFA_def}
U_{\rm eff}^{\rm HFA} = U - J_{\qq}/2,
\end{equation}
$J_{\qq} = (1/N)\sum_{ij}J_{ij}\exp[\I\mathbf{k}(\mathbf{R}_i-\mathbf{R}_j)]$. 
%where $\bar\sigma = -\sigma$. 
It is clear that despite  different nature of local in-site and exchange interaction, HFA treats them both  in the same way, through the introducing the effective interaction (\ref{eq:U_eff_HFA_def}). 
 
Diagonalizing the Hamiltonian (\ref{eq:Hk_MF}) yields the spectrum branches of AFM subbands:    
\begin{equation}
	E^{\rm HFA}_{\nu}(\kk)=\es + Un/2 + (-1)^\nu\sqrt{\Delta^2+\eap{2}},\; 
\nu = 1,2,
\end{equation}
where AFM gap is $\Delta = U_{\rm eff}^{\rm HFA}m/2$.

Below in this section we consider the insulator state for which the problem acquires a pretty form. The upper subband ($\nu = 2$) is empty and the diagonalization of the Hamiltonian (\ref{eq:Hk_MF}) results in the equation of self-consistency:
\begin{equation}\label{eq:m_vs_Delta}
	m = \Delta\Phi_1(\Delta),
\end{equation}
and the following equation for the free energy:
%The ground state (and free energy at $T = 0$) reads
\begin{equation}
F^{\rm HFA}_{\rm AFM}(\Delta) = U/4 -\Delta m/2  - \Phi_2(\Delta),
\end{equation}
where the lattice sums 
\begin{eqnarray}
\label{eq:Phi1_def}
	\Phi_1(\Delta) &=& \frac{1}{N}\sum_{\kk}	 \frac{1}{\sqrt{\Delta^2+\eap{2}}}, \\
\label{eq:Phi2_def}	
	\Phi_2(\Delta) &=& \frac{1}{N}\sum_{\kk}	 \frac{\eap{2}}{\sqrt{\Delta^2+\eap{2}}}
\end{eqnarray}
are introduced. 
%This results in $z^2_{\rm A} = 1, \Delta = \Delta_\ast$. 
While Eq.~(\ref{eq:m_vs_Delta}) is equation on $m$ only, 
it is convenient to choose $\Delta$ as a natural control parameter since 
%It is clear that all for the insulator case 
all quantities are expressed via this. 
Since we are interested in precise information about the behaviour of the 
free energy as a function of system parameters 
it is convenient to count the free energy from its zero $\Delta$ value $F_0 = F^{\rm HFA}_{\rm AFM}(0) = U/4 - \Phi_2(0)$: $\delta F^{\rm HFA}_{\rm AFM}(\Delta) = F^{\rm HFA}_{\rm AFM}(\Delta) - F_0$, so that we get
\begin{equation}	
\label{eq:F_AFM_HFA}
	\delta F^{\rm HFA}_{\rm AFM}(\Delta) = -\Delta m/2  - \delta\Phi_2(\Delta),
\end{equation}
where $\delta\Phi_2(\Delta) = \Phi_2(\Delta) - \Phi_2(0)$. 
Both lattice sums (\ref{eq:Phi1_def}) and (\ref{eq:Phi2_def})
can be expressed in terms of the auxiliary lattice sum 
\begin{equation}\label{eq:G_def}
G(\Delta) = \frac{1}{2N}\sum_{\kk} \frac{1}{\sqrt{\Delta^2+\eap{2}} + |\ea|},
\end{equation}
through 
\begin{eqnarray}
\label{eq:Phi1_in_terms_of_G}
\Phi_1(\Delta) &=& 4G(\Delta) + 2\Delta\cdot G'(\Delta),\\
\label{eq:Phi2_in_terms_of_G}
\delta\Phi_2(\Delta) &=& -2\Delta^2\left(G(\Delta) + \Delta\cdot G'(\Delta)\right),
\end{eqnarray}
where primes stand for the derivative with respect to $\Delta$. 
The dependence of $G$ on $\Delta$ is fully determined by 
the spectrum $\ea$ only, coinciding with full electron spectrum at~$\tau = 0$.  
Hence, the investigation of density of states for $\ea$
allows to investigate in a convenient way analytical properties of $G(\Delta)$ and, through it, all other quantities. In the Appendix~\ref{appendix:G_expansion} we derive a~general expansion of the~lattice sum $G(\Delta)$ at small $\Delta$. 
%Using the lattice sum (\ref{eq:G_def}) we can rewrite the latter Equation
For this purpose we recast Eq.~(\ref{eq:F_AFM_HFA}) as
\begin{equation}\label{eq:F_AFM_HFA_final}
 \delta F^{\rm HFA}_{\rm AFM}(\Delta) = \Delta^3 G'(\Delta),
\end{equation}
and from~Eq.~(\ref{eq:m_vs_Delta}) we derive
\begin{equation}\label{eq:U_eff_HFA}
U^{\rm HFA}_{\rm eff} = 2/\Phi_1(\Delta).
\end{equation}

\subsection{Slave boson approximation setup}\label{sec:SBA}
A simple way of taking into account the local correlation
effects on a qualitative level is to introduce the auxiliary slave boson states~\cite{1986:Kotliar,1992:Fresard}.
%We denote the Kotliar-Ruckenstein slave boson amplitudes as $e$~(empty state), $p_\sigma$~(singly occupied state with spin projection~$\sigma$), $d$~(double occupied state).
This extends the configuration space of the Hamiltonian~(\ref{eq:H_rotated}) to a bosonic sector by introducing the {\it slave boson}  annihilation (creation) operators $e_i(e_i^\dag)$, $p_{i\sigma}(p_{i\sigma}^\dag),
d_i(d_i^\dag)$ for empty, singly and doubly occupied states, respectively. 
The transitions between the site states originating from intersite electron transfer are now accompanied by corresponding transitions in bosonic sector.
%The transitions between the site states originating from intersite electron transfer can be formulated in alternative terms, but {\it simultaneously} (**) with the conventional Slater determinant formalism  (related to 
%the electron
%creation/annihilation operators).
%Conceptually this is close to the Hubbard's $X$-operator formalism\cite{Hubbard-II:1964,Zarubin:2004}, where, however, the site transition $X$-operators are introduced {\it instead of} the original one-electron operators.
The equivalence of the original and new description is achieved by
the replacement $c_{i\sigma}\rightarrow \mathsf{z}_{i\sigma}c_{i\sigma}$,
through a boson transfer operator $\mathsf{z}_{i\sigma} = (1 - d^\dag_i 
d^{}_i - p^\dag_{i\sigma}p^{}_{i\sigma})^{-1/2}\left(e^\dag_ip^{}_{i\sigma} + p^\dag_{i\bar\sigma}d^{}_{i} \right)(1 - e^\dag_i e^{}_i - p^\dag_{i\bar\sigma}p^{}_{i\bar\sigma} )^{-1/2}$, %(and $\mathsf{z}^\dag_{i\sigma}$)
which complements the action of $c_{i\sigma}$ on the bosonic subspace. 
The % in conjunction with %the introducing of
constraints
\begin{eqnarray}
      \label{eq:eta-constraint}
      e^\dag_ie^{}_i+\sum_\sigma p^\dag_{i\sigma}p^{}_{i\sigma}+d^\dag_id^{}_i&=&1,\\
      \label{eq:lmb-constraint}
      2d^\dag_id^{}_i+\sum_\sigma p^\dag_{i\sigma}p^{}_{i\sigma}&=&\sum_\sigma c^\dag_{i\sigma}c^{}_{i\sigma}, \\
      \label{eq:delta-constraint}
	  p^\dag_{i\uparrow}p^{}_{i\uparrow} - p^\dag_{i\downarrow}p^{}_{i\downarrow} &=&  c^\dag_{i\uparrow}c^{}_{i\uparrow} - c^\dag_{i\downarrow}c^{}_{i\downarrow}
\end{eqnarray}
guarantee formal equivalence of SBA action to that of the original model. 

This allows to  recast exactly the interaction terms in the Hamiltonian (\ref{eq:H_rotated}) in the bosonic language
\begin{equation}
Un_{i\uparrow}n_{i\downarrow} \rightarrow Ud^\dag_id_i,
\end{equation}
\begin{equation}
J_{ij}\tilde{\bf s}_{i}\cdot\tilde{\bf s}_{j} \rightarrow J_{ij}\tilde{\bf s}^p_{i}\cdot\tilde{\bf s}^p_{j},
\end{equation}
where $\tilde{\bf s}^p_{i} = (1/2)\sum_{\sigma\sigma'}p^\dag_{i\sigma}\vec\sigma_{\sigma\sigma'} p_{i\sigma'}$. 
Presence of the constraints can be taken into account within the functional integral formalism via the Lagrange
multipliers [$\eta_i$ for Eq.~(\ref{eq:eta-constraint}), %the on-site
%electron-boson ``force''
$\lambda_i$ for Eq.~(\ref{eq:lmb-constraint}) and $\Delta_i$ for Eq.~(\ref{eq:delta-constraint})] which are introduced into the action. 
Since we assume no any inhomogeneity, within the saddle-point approximation the operators $e_i(e^\dag_i),p_{i\sigma}(p^\dag_{i\sigma}),d_i(d^\dag_i)$ become $i$-independent slave boson amplitudes $e,p_{\sigma},d$, and~$\mathsf{z}_{i\sigma}$ by $z_\sigma =(d^2+p_\sigma^2)^{-1/2}(ep_\sigma+p_{\bar\sigma}d)(e^2+p_{\bar\sigma}^2)^{-1/2}\le 1$. The same argument holds for $\lambda_i$ and $\Delta_i$. 
Smallness of local electron spin-dependent quasiparticle residue $z_\sigma$ reflects the average incoherence of single and double states on a pair 
of sites between which a transfer occurs~\cite{2018:Igoshev}.  %narrowed $\uparrow$ and $\downarrow$ spin subbands are mixed due to AFM order formation. 
The difference of $z_\uparrow$ and $z_\downarrow$ residues can be induced 
by magnetic ordering away of half-filling. 
The grand potential $\Omega = -T\ln Z$, $T$ being the temperature, $Z$ being the partition function, %where $Z = \int D[c,c^\dag]\exp(-\mathcal{S}_{\rm sp})$
can be presented as a sum of two contribution from fermion and boson subsystems,
%\begin{equation}\label{eq:Omega_two_parts}
	$\Omega^{\rm SBA}_{\rm AFM} = \Omega_c + \Omega_b$, where
%\end{equation}
%where fermion contribution
the fermion contribution to the grand potential  
\begin{equation}
\Omega_c = -\frac{T}{N}\sum_{\nu\kk} \ln(1 + \exp(-(E_\nu(\kk)-\mu)/T)),
\end{equation}
%\begin{equation}
where $\mu$ is the chemical potential, is that of 
%\end{equation}
%\begin{equation}
%	\hat{H}= \sum_{\kk\sigma\sigma'} H^c_{\sigma\sigma'}(\kk)c^\dag_{\kk\sigma}c_{\kk\sigma'},
%\end{equation}
effective free-fermion Hamiltonian
\begin{equation}\label{eq:H_eff}
	\mathcal{H}_{\rm SBA} = \sum_{\kk\sigma\sigma'}(\lambda\sigma^0_{\sigma\sigma'} - \Delta\sigma^z_{\sigma\sigma'} + z_\sigma z_{\sigma'}(\es\sigma^0_{\sigma\sigma'} + \ea\sigma^x_{\sigma\sigma'}))c^\dag_{\kk\sigma}c^{}_{\kk\sigma'},
\end{equation}
which has AFM spectrum branches
\begin{equation}\label{eq:Ek_def}
	E^{\rm SBA}_{\nu}(\kk)=\frac{(z^2_\uparrow+z^2_\downarrow)\es}2+\lambda + (-1)^{\nu}\sqrt{\Delta^2(\kk)+\left(z_\uparrow z_\downarrow\ea\right)^2},
\end{equation}
where $\Delta(\kk) = \Delta - (\zu^2 - \zd^2)\es/2$. 

The boson contribution to grand potential reads
\begin{equation}
\Omega_b =  - 2\lambda d^2 + (1/4)J(\qq)(p^2_{\uparrow} - p^2_\downarrow)^2 - \lambda (p^2_{\uparrow} + p^2_\downarrow) + \Delta (p^2_{\uparrow} 
- p^2_\downarrow).
\end{equation}
%Coulomb interaction can be rewritten as a bilinear of $d$, see details in Ref.~\cite{1986:Kotliar}. 

From Eqs.~(\ref{eq:lmb-constraint}) and (\ref{eq:delta-constraint})  we get 
\begin{eqnarray}
	n & = & p^2_\uparrow + p^2_\downarrow + 2d^2, \\
	m & = & p^2_\uparrow - p^2_\downarrow,
\end{eqnarray}
which relates the electron filling~$n$ and magnetization amplitude~$m$ to 
boson amplitudes. 
Explicit equation determining slave boson parameters is given by general slave boson equations for the mean-field ansatz of slave boson amplitude 
and Lagrange multipliers (see,  e.g.,~Ref.~\onlinecite{2015:Igoshev}). 
%\begin{eqnarray}
%	\label{eq:constraint_general}
%	1 &=& e^2+p_\uparrow^2+p_\downarrow^2+d^2,\\
%	\label{eq:general_U}
%	U &=& -\frac{\zeta/2}{edp_\uparrow p_\downarrow}\sum_\sigma (ep_{\bar\sigma} + p_\sigma d)\Phi_\sigma, \\
%	\label{eq:general_n}
%	n &=& \frac1{N}\sum_{\kk\nu}f[E_{\nu}(\kk)],\\
%	\label{eq:general_m}
%	m &=& \frac1{N}\sum_{\kk\nu}(-1)^{\nu + 1}\frac{\Delta(\kk)}{\sqrt{\Delta^2(\kk)+ \zu^2\zd^2\eap{2}}}f[E_{\nu}(\kk)],\\
%	\label{eq:general_lmb}
%	\lambda &=&\frac{U}2+\frac{\zeta/2}{2ed}\sum_\sigma \Phi_\sigma\left(
%\frac{ ep_{\sigma}}{p_{\sigma}^2+d^2}-\frac{ p_{\bar\sigma}d}{e^2+p^2_{\bar\sigma}}\right),\\
%	\label{eq:general_Delta}
%	\Delta
%	&=& \Delta_J -\frac{\zeta}{4p_\uparrow p_\downarrow}\sum_{\sigma}\sigma\Phi_{\sigma}
%	\left(\frac{ep_\sigma}{e^2+p_{\bar\sigma}^2} - \frac{p_{\bar\sigma}d}{p_{\sigma}^2+d^2}\right),
%\end{eqnarray}
%\begin{equation}\label{eq:Phi_def}
%\Phi_{\sigma}=\frac1{2N}\frac{ep_\sigma+p_{\bar\sigma}d}{(e^2+p_{\bar\sigma}^2)(p_{\sigma}^2+d^2)}\sum_{\kk\nu}\left(\es %+\frac{(-1)^\nu\left(-\sigma\es\Delta(\kk)+\eap{2} z_{\bar\sigma}^2\right)}{\sqrt{\Delta^2(\kk)+ \zu^2\zd^2\eap{2}}}\right)f[E_\nu(\kk)],
%\end{equation}
%Generally, non-trivial effects can occur as a result of particle-hole asymmetry which manifests itself as a difference of $\zu^2$ and $\zd^2$. 
In the following we focus our attention on the case of insulator state at 
half-filling. 
%%%%%%%%%
%%%%%%%%%
%An alternative form of the Equation (\ref{eq:general_lmb}) reads
%\begin{equation}
%	\lambda = -\frac{\zeta}{2ep_\uparrow p_\downarrow}\sum_{\sigma}\left(\frac{p_\sigma }{e^2 + p^2_{\bar\sigma}} + d\left(\frac{ep_{\bar\sigma}}{p^2_{\sigma}+d^2}-\frac{p_\sigma d}{e^2 + p^2_{\bar\sigma}}\right) \right)\Phi_\sigma.
%\end{equation}

%We have to add also the equations 

%We can rewrite  Eq.~(\ref{eq:Omega_two_parts}) as
%\begin{equation}\label{eq:Omega_general}
%	\Omega = Ud^2 + \frac{J_\qq}8m^2  +\Delta\cdot m + \sum_{\kk \nu} (E_{\nu}(\kk) -\lambda -\mu)f[E_\nu(\kk)].
%\end{equation}

\subsection{Slave boson equations: Half-filled insulator case}
%\section{Half-filling case equations}
\label{sec:AFI}
At half-filling ($n = 1$), particle-hole symmetry results in the relations $e = d$, $\zu^2 = \zd^2 = z^2$, where
\begin{equation}\label{eq:bosons_z2}
	z^2 = \frac{1 - m^2 - \zeta^2 }{1 - m^2}.
\end{equation}
As a consequence,  $\Delta(\kk) = \text{const}$,  so that the subband spectrum acquires $\kk$-independent narrowing. 
Here a slave boson amplitude correlation parameter $\zeta = 2(p_\uparrow p_\downarrow - e^2)$ indicates the difference of single and double electron state motion, which is neglected in the Hartree-Fock approximation. 
All the boson variables can be expressed through $m$ and $\zeta$, i.e. %using $(p_\uparrow + p_\downarrow)^2 = 1 + \zeta$, $(p_\uparrow - p_\downarrow)^2 = m^2/(1 + \zeta)$,
\begin{eqnarray}
    \label{eq:bosons_pp}		
	p_\uparrow p_\downarrow &=& \frac14\frac{(1 + \zeta)^2 - m^2}{1 + \zeta},\\
	\label{eq:bosons_e2}
	e^2 &=& \frac{1 -m^2 - \zeta^2}{4(1 + \zeta)},
\end{eqnarray}
thereby we fully exclude boson variables in the consideration below. 
Equation~(\ref{eq:bosons_z2}) yields the natural bound for $\zeta$: $\zeta < \zeta_{\rm max} = \sqrt{1 - m^2}$. If $\zeta\lesssim\zeta_{\rm max}$, we 
have a strongly correlated regime with small quasiparticle weight, otherwise we are in the regime of a usual Fermi liquid.
For AFM insulator state, we assume that only lower subband ($\nu = 1$) is filled, whereas upper one is empty. For this case, we call $z$ as $z_{\rm A}$ and $\zeta$ as $\zeta_{\rm A}$.  %Then %Equation (\ref{eq:lmb_eq_hf}) is not needed and the other
Equations of Ref.~\onlinecite{2015:Igoshev} for the case of AFM insulator 
state can be strongly simplified:
\begin{eqnarray}
%	\label{eq:n_func}
%	n &=& 2d^2+p_\uparrow^2+p_\downarrow^2,\\
%	\label{eq:m_func}
%	m &=& p_\uparrow^2-p_\downarrow^2,\\
	\label{eq:constraint}
	1 &=& 2e^2+p_\uparrow^2+p_\downarrow^2,\\
	\label{eq:m_eq}
	m &=& \Delta_\ast\Phi_1(\Delta_\ast),\\
%	\label{eq:Delta_eq}
%	\Delta_\ast - \Delta_{J\ast}	&=&  \frac{Um}2\frac{1 + \zeta_{\rm A}/2}{(1 + \zeta_{\rm A})^2},\\
	%\label{eq:main_p}
	%U &=& +\frac{\zeta_{\rm A} z^2_{\rm A}}{2e^2 p_\uparrow p_\downarrow}\Phi_2(\Delta_\ast),
	\label{eq:Delta_eq}
    \Delta_\ast &=&  -(1/4)J_{\qq}m/z^2_{\rm A} + \frac{4m\zeta_{\rm A}(1 + \zeta_{\rm A}/2)}{(1 - m^2)\left((1 + \zeta_{\rm A})^2 - m^2\right)}\Phi_2(\Delta_\ast),\\
	\label{eq:U_in_AFM}
	U &=& \frac{8\zeta_{\rm A} (1 + \zeta_{\rm A})^2}{(1 - m^2)\left((1 + \zeta_{\rm A})^2 - m^2\right)}\Phi_2(\Delta_\ast),	
\end{eqnarray}
with rescaled parameter $\Delta_\ast = \Delta/z^2_{\rm A}$. We see from Eq.~(\ref{eq:Delta_eq}) that the parameter $\Delta$ contains two terms: The first originates from intersite exchange and the second originates from local quantities~(boson amplitudes).

From  Eq.~(\ref{eq:m_eq}), it is clear that $m$ is a function of $\Delta_\ast$ only. %Note that for ``antiferromagnetic'' sign of exchange integrals ??configuration is $J_{\qq} < 0$.
We pass from $\Omega_{\rm AFM}^{\rm SBA}$ to the free energy 
%From the equation (\ref{eq:Omega_two_parts}) we get the free energy 
$F_{\rm AFM}^{\rm SBA} = \Omega_{\rm AFM}^{\rm SBA} + \mu\cdot n$ at half-filling
\begin{equation}	
\label{eq:F_AFM2}
	F_{\rm AFM}^{\rm SBA} = \frac{J_\qq}8m^2 + z^2_{\rm A}\left(\frac{U}4\frac{1 - m^2}{1 + \zeta_{\rm A}} - \Phi_2(\Delta_\ast)\right).
\end{equation}
%\begin{equation}	
%\label{eq:F_AFM_another}
%	F_{\rm AFM} = \frac{J_\qq}8m^2 - \Phi_2(\Delta_\ast) + \frac{(1 - \zeta_{\rm A}/2)U}4 -  \frac{Um^2}{4(1 + \zeta_{\rm A})}\left(1 + \frac{\zeta_{\rm A}/2}{1 + \zeta_{\rm A}}\right).
%\end{equation}
%$$
%U = -\frac{16\zeta (1 - 2\zeta)^2}{(1 - 2\zeta)^2 - m^2}\frac{1}{N}\sum_{\kk}	 \frac{\left(\ea\right)^2}{\sqrt{\Delta_\ast^2+(\ea )^2}}
%$$
%From Eqs.~(\ref{eq:bosons}) we can state that
%\begin{equation}
%	\frac12\left(-1 + m\right)< \zeta_{\rm A} < \frac12\sqrt{1 - m^2}.
%\end{equation}
The Eq.~(\ref{eq:U_in_AFM}) imply $\zeta > 0$, which imply positive definiteness of effective field $\Pi(\Delta_\ast) = \Phi^{-1}_1(\Delta_\ast) 
+ J_{\qq}/(4z^2_{\rm A}) > 0$
%Also, since $\Delta > 0$ Eq.~(\ref{eq:m_eq}) implies $m > 0$ and we find 
from Eq.~(\ref{eq:Delta_eq}) $\Delta_\ast > \Delta_{J\ast}$, which gives with the use of~Eq.~(\ref{eq:m_eq})
%\begin{equation}
%	\Pi(\Delta_\ast) > 0.
%\end{equation}
%Excluding boson variables with the use of Eqs.~(\ref{eq:bosons_pp}),~(\ref{eq:bosons_z2}),~(\ref{eq:bosons_e2}) we rewrite Eq.~(\ref{eq:main_p}) as
%Excluding  $U$ by using Eq.~(\ref{eq:Delta_eq}) yields
Introducing $\xi = \zeta_{\rm A}(2 + \zeta_{\rm A})$, 
%inverse transformation is 
%\begin{equation}\label{eq:zeta_through_xi}
%\zeta_{\rm A} = \sqrt{1 + \xi} - 1,
%\end{equation}
%\begin{equation}
we solve Eq.~(\ref{eq:Delta_eq}) with respect to $\xi$ to obtain
%	\xi = \frac{(\Delta_\ast - \Delta_{J\ast})(1 - m^2)^2}{2m\Phi_2(\Delta_\ast) - (\Delta_\ast - \Delta_{J\ast})(1 - m^2)}.
%\end{equation}
\begin{equation}\label{eq:xi_general}
	\xi = \frac{\Pi(\Delta_\ast)(1 - m^2)^2}{2\Phi_2(\Delta_\ast) - \Pi(\Delta_\ast)(1 - m^2)}.
\end{equation}
Expanding $\Phi_1(\Delta_\ast)$, $\Phi_2(\Delta_\ast)$ and $\Pi(\Delta_\ast)$ and, in turn, the~Eq.~(\ref{eq:xi_general}) in powers of $\Delta_\ast$ allows to find expansions of all other quantities. 

The Hartree-Fock approximation neglecting the difference between singly and doubly occupied states (see Sec.~\ref{sec:HFA}) can be simply obtained in this way by replacing Eq.~(\ref{eq:Delta_eq}) by $\zeta = 0$. Thereby $\zeta$ can be considered as a small parameter of the expansion and HFA is zero-order approximation of it. 
%Using the Eq.~(\ref{eq:U_in_AFM}) we rewrite the~Eq.~(\ref{eq:F_AFM2}) explicitly separating the Hartree-Fock contribution
%\begin{equation}
%\label{eq:F_AFM3}
%	F_{\rm AFM}(\Delta_\ast) = \frac{J_\mathbf{\qq}}8m^2 - \frac{U\zeta_{\rm A}}{4}\left(1 - \frac{m^2(1 + 4\zeta_{\rm A})}{(1 + 2\zeta_{\rm A})^2}  \right) + F^{\rm HFA}_{\rm AFM}(\Delta_\ast).
%\end{equation}
Here and below it is convenient to consider all quantities as functions of the rescaled gap $\Delta_\ast$.
Analogously, introducing the definition $\delta F_{\rm AFM}^{\rm SBA} = 
F_{\rm AFM}^{\rm SBA} - F_0$ and %and using  Eq.~(\ref{eq:Delta_eq}), we get
%\begin{equation}
%	\delta F_{\rm AFM} =\frac{J_\mathbf{\qq}}8m^2 - \delta\Phi_2(\Delta_\ast) - \frac{\zeta U}4 -  \frac{Um^2}{4(1 + 2\zeta)}\left(1 + \frac{\zeta}{1 + 2\zeta}\right).
%\end{equation}
%\begin{equation}
%	\delta F_{\rm AFM} =\frac{J_\mathbf{\qq}}8m^2 - \delta\Phi_2(\Delta_\ast) - \frac{\zeta U}4 -  \frac{Um^2}4\left(1  - \frac{\zeta(1 + 4\zeta)}{(1 + 2\zeta)^2}\right).
%\end{equation}
%\begin{equation}\label{eq:F_AFM_SBA}
%	\delta F_{\rm AFM} =\frac{J_\qq}8m^2 - \delta\Phi_2(\Delta_\ast) - \frac{\zeta_{\rm A} U}8 -  \frac{m(1 + 3\zeta_{\rm A}/2)}{2 + \zeta_{\rm A}}(\Delta_\ast - \Delta_{J\ast}).
%\end{equation}
%It is convenient to 
picking up explicitly the HFA contribution given by~Eq.~(\ref{eq:F_AFM_HFA}) we get %in the expression (\ref{eq:F_AFM_SBA}). 
%Expressing $\Delta_\ast$ through $\Phi_{1}$, see Eq.~(\ref{eq:m_eq}) we obtain
%\begin{equation}
%	\delta F_{\rm AFM} =-\frac{J_\mathbf{\qq}m^2}4\frac{\zeta}{1 - m^2 - 4\zeta^2}\left(\frac{1 - m^2}{1 + \zeta} + 2\zeta\right) - \delta\Phi_2(\Delta_\ast) - \frac{\zeta U}4 -  \frac{m(1 + 3\zeta)}{2(1 + \zeta)}\Delta_\ast.
%\end{equation}
%\begin{equation}
%	\delta F_{\rm AFM} = \delta F^{\rm HFA}_{\rm AFM}(\Delta_\ast) -\frac{J_\mathbf{\qq}m^2}4\frac{\zeta}{1 + \zeta}\left(1 + \frac{2\zeta(1 + 3\zeta)}{1 - m^2 - 4\zeta^2}\right)  - \frac{\zeta U}4 -  \frac{m\zeta}{1 + \zeta}\Delta_\ast,
%\end{equation}
%\begin{equation}
%	\delta F_{\rm AFM} = \delta F^{\rm HFA}_{\rm AFM}(\Delta_\ast) + \Phi^{-1}_{1J}m^2\zeta\left(\frac{1}{1 + \zeta} + \frac{2\zeta}{1 - m^2}\right)  - \frac{\zeta U}4 -  \frac{m\zeta}{1 + \zeta}\Delta_\ast,
%\end{equation}
%\begin{equation}
%	\delta F_{\rm AFM} = \delta F^{\rm HFA}_{\rm AFM}(\Delta_\ast) - \frac{\zeta(\Phi^{-1}_{1}-\Phi^{-1}_{1J})m^2}{1 + \zeta} + \frac{2\zeta^2\Phi^{-1}_{1J}m^2}{1 - m^2}  - \frac{\zeta U}4.
%\end{equation}
%where $m$ is definitely determined by~the~Eq.~(\ref{eq:m_eq}).
\begin{equation}\label{eq:delta_F_AFM}
	\delta F_{\rm AFM}^{\rm SBA}(\Delta_\ast) = \delta F^{\rm HFA}_{\rm AFM}(\Delta_\ast) - \frac18\frac{\zeta^2_{\rm A}J_{\qq}m^2}{1 - m^2 - \zeta_{\rm A}^2}  - \frac{\zeta_{\rm A} U}8\left(
1 + \frac{2m^2}{(1 + \zeta_{\rm A})^2}	
	\right),
\end{equation}

Thus an improvement of HFA by correlation effects results not 
only in rescaling of AFM gap $\Delta\rightarrow\Delta_{\ast}$, but also in the occurrence of exchange (second) and correlation (third) terms, which both yield a manifestation of delicate many-electron effects.  

%Below in this section we focus on the behavior of $\delta F_{\rm AFM}(\Delta_\ast)$ at small $\Delta$. We will see that an additional logarithmic correction that occurs is governed by singular contributions to the density of states of the spectrum at $\tau = 0$, see Appendix \ref{Appendix:G_expansion}. 

\subsection{Treatment of paramagnetic case}\label{sec:PM_treatment}
In this section we consider free energy of the paramagnetic phase as a function of $\tau$ strictly at half-filling for all the considered bipartite (square, sc, and bcc) lattices. % and further focus special attention on the two-dimensional case (square lattice). 
The peculiarity of the square lattice is the presence of persistent van Hove singularity in the electronic spectrum due to topological reason. 

\begin{figure}[h]
\noindent
\includegraphics[angle=-90,width=0.32\textwidth]{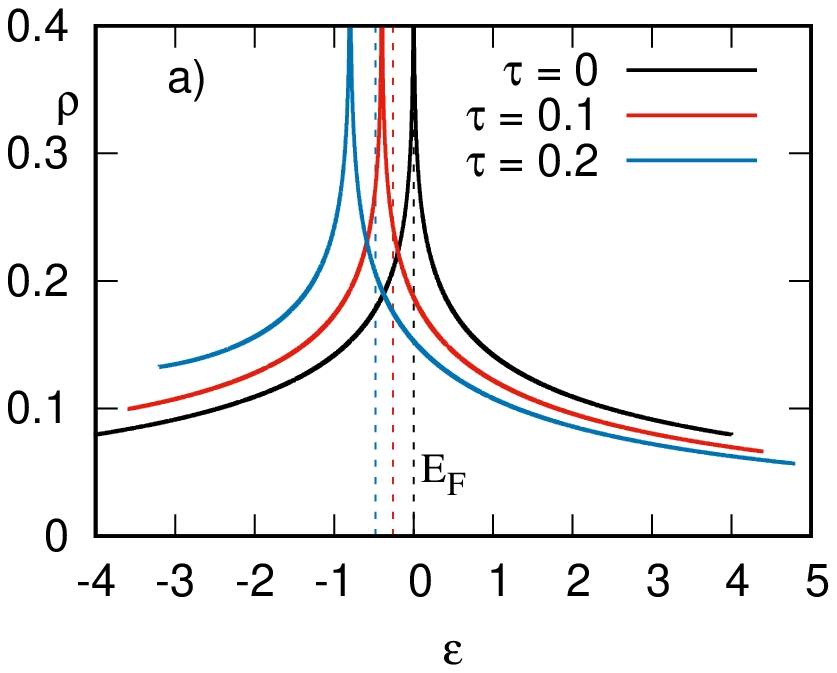}
\includegraphics[angle=-90,width=0.32\textwidth]{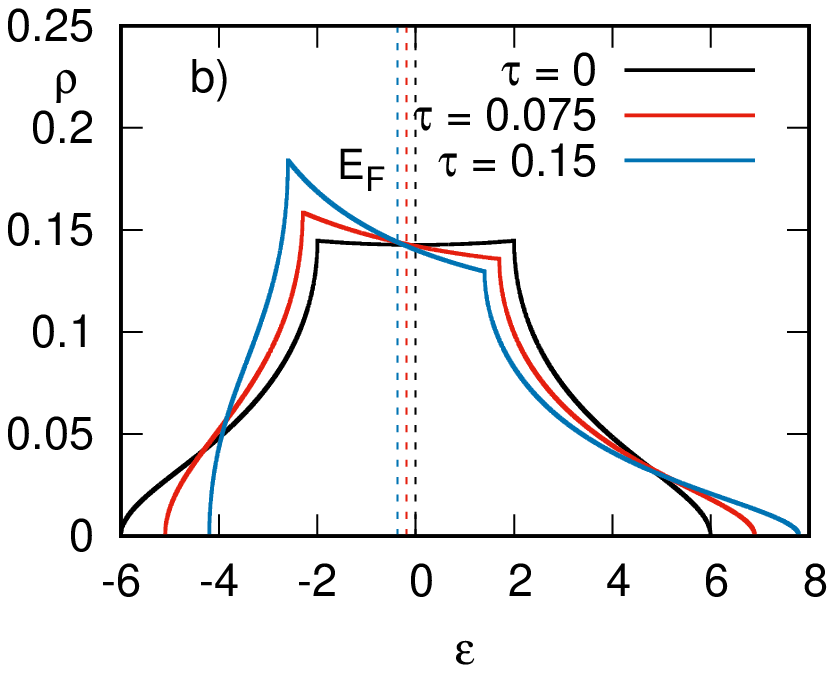}
\includegraphics[angle=-90,width=0.32\textwidth]{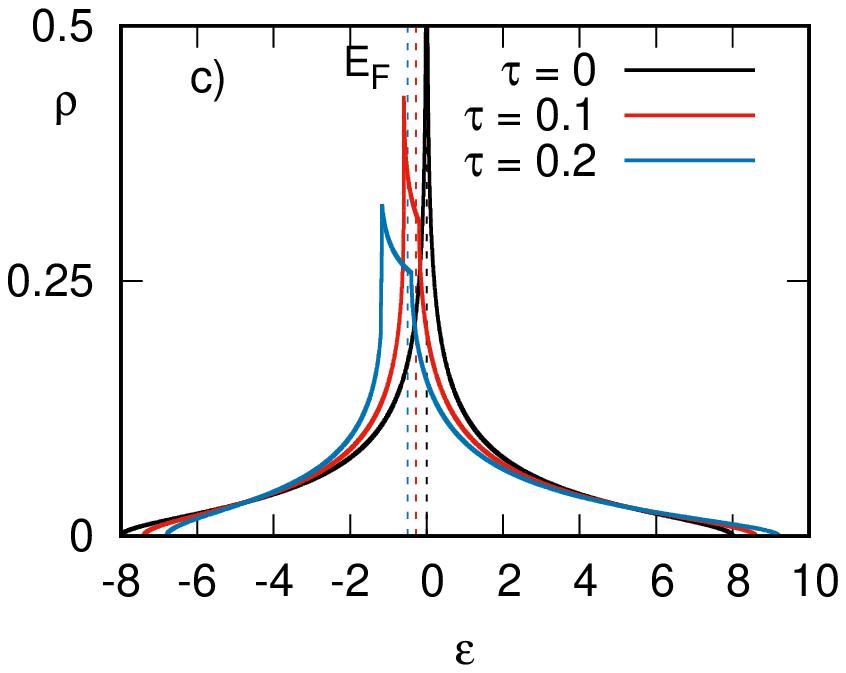}
\caption{
(Color online)
Density of states for a) square at $\tau = 0$, 0.1, and~0.2; b) sc at $\tau = 0, 0.075$, and~$0.15$; c) bcc lattice at $\tau = 0, 0.1$, and~0.2. The positions of the Fermi level corresponding to half-filling are shown by 
vertical dashed lines with corresponding colors. %For the square lattice, 
Electron energy is counted from the van Hove singularity level. 
The Fermi level $E_{\rm F}$ corresponding to half-filling is shown by dashed lines.
}
\label{fig:DOS_3d}
\end{figure}

For bipartite lattices, at zero $\tau$ the Fermi level coincides with the 
van Hove singularity at the center of the band. The deviation of $\tau$ from zero shifts the van Hove singularity from the Fermi level, see~Fig.~\ref{fig:DOS_3d}. 

For the paramagnetic phase we get within SBA (see Ref.~\onlinecite{2015:Igoshev}, $\zeta = \zeta_{\rm P}$, $z^2 = z^2_{\rm P}$)
\begin{eqnarray}
\label{eq:PM_n}
	\frac12 &=& \frac1{N}\sum_{\kk}f[z^2_{\rm P}t_{\kk}(\tau)],\\
\label{eq:PM_U}	
	U &=& 8\zeta_{\rm P}\Phi_{2\text{P}}(\tau),\\
\label{eq:PM_z^2}
	z^2_{\rm P} &=& 1 - \zeta^2_{\rm P},
\end{eqnarray}
where $f[E] = (\exp[(E - \mu)/T] + 1)^{-1}$ is the Fermi function and an PM phase analog of $\Phi_2$
\begin{equation}
		\Phi_{2\text{P}}(\tau) = -\frac2{N}\sum_{\kk}t_{\kk}(\tau)f[z^2_{\rm P}t_{\kk}(\tau)].
\end{equation}
The Fermi level $E_{\rm F}$ differs from $\mu$ by nonvaluable in a current context constant and is determined by Eq.~(\ref{eq:PM_n}) and
%\begin{equation}	
%\label{eq:F_PM1}
%	F_{\rm PM} = Ue^2 - z^2_{\rm P}\Phi_{2\text{P}}(\tau).
%\end{equation}
%\begin{equation}	
%\label{eq:F_PM2}
%	F_{\rm PM} = \frac{U}4(1 - 2\zeta_{\rm P}) + (1 - 4\zeta^2_{\rm P})T(\tau),
%\end{equation}
%or, using Eq.~(\ref{eq:PM_U}),
the free energy of PM phase reads 
\begin{equation}
\label{eq:F_PM2}
	F_{\rm PM}^{\rm SBA} = \frac{U}4(1 - \zeta_{\rm P}/2) - \Phi_{2\mathrm{P}}(\tau).
\end{equation}
As above, introducing the difference $\delta F_{\rm PM}^{\rm SBA} = F_{\rm PM}^{\rm SBA} - F_0$ we write down
\begin{equation}
\label{eq:delta_F_PM}
	\delta F_{\rm PM}^{\rm SBA} =  - \delta\Phi_{2\mathrm{P}}(\tau) -\frac{U\zeta_{\rm P}}8,
\end{equation}
where $\delta\Phi_{2\mathrm{P}}(\tau) = \Phi_{2\mathrm{P}}(\tau) - \Phi_2(0)$. 
We also directly obtain the analogous HFA expression by setting in the latter equation $\zeta_{\rm P} = 0$, and $z_{\rm P} = 1$ in Eq.~(\ref{eq:PM_n}),
\begin{equation}
\label{eq:delta_F_PM_HFA}
	\delta F^{\rm HFA}_{\rm PM}(\tau) = - \delta\Phi_{2\mathrm{P}}(\tau).
\end{equation}
It is clear that accounting the difference of single and double states within SBA,  cf. Eqs.~(\ref{eq:delta_F_PM}) and~(\ref{eq:delta_F_PM_HFA}), allows to lower the energy of the paramagnetic state.

Vanishing of $e^2$ in Eq.~(\ref{eq:bosons_e2}) at $m = 0$ means 
the condition of zero mobility of carriers, which yields 
 in the paramagnetic phase critical $\zeta_{\rm P} = \zeta_{\rm BR} = 
1$ (the Brinkman-Rice realization of Mott scenario of metal-insulator transition), $\zeta_{\rm P} = \zeta$ in PM phase. Thus the transition to the metal state occurs at $U < U_{\rm BR}$ where
%From Eq.~(\ref{eq:U_in_AFM}) at $m = 0$ we get the condition of full suppressing  quasiparticle weight $z = 0$ which implies $\zeta_{\rm A} = 1/2$: $U = U_{\rm BR}$, where
\begin{equation}\label{eq:U_BR_def}
	U_{\rm BR} = 8\Phi_2(0).
\end{equation}
Direct calculations give $U^{\rm sq}_{\rm BR} = 128/\pi^2 = 12.97$, $U^{\rm sc}_{\rm BR} = 16.04$, $U^{\rm bcc}_{\rm BR} = 16.51$.

\begin{figure}[h]
\includegraphics[angle=-90,width=0.49\textwidth]{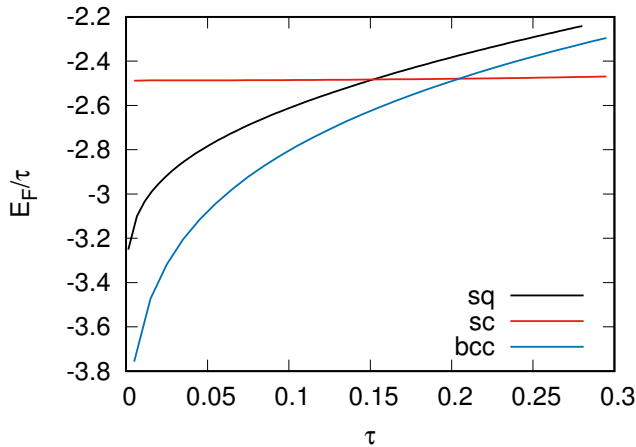}
\caption{
(Color online)
$E_{\rm F}/\tau$ as a function of $\tau$ for the square, sc and bcc lattices.  
}
\label{fig:Ef}
\end{figure}

To calculate the ground-state free energy of paramagnetic phase, see~Eqs.~(\ref{eq:delta_F_PM}) and~(\ref{eq:delta_F_PM_HFA}), we use the explicit expressions for exact density of states at finite $\tau$ for the square, sc, and bcc lattices derived in Refs.~\onlinecite{2019:Igoshev_JETP,2019:Igoshev_FMM}, %(see the results in Appendix~\ref{Appendix:DOS}??), 
where the presence of van Hove singularity lines at finite $\tau$ was found (at $\tau = 0.25$ for sc lattice and $\tau = 1.0$ for bcc lattice), which allows to solve Eqs.~(\ref{eq:PM_n}) and (\ref{eq:PM_U}) numerically with extremely high precision.

For the square lattice, a finite $\tau$ value results in the shift of position of the van Hove singularity from $\varepsilon = 0$: in Fig.~\ref{fig:DOS_3d}a) the plots of density of states for the square lattice at different $\tau$ are shown. This dramatically distinguishes this case from three-dimensional cases [sc and bcc lattices, see Figs.~\ref{fig:DOS_3d}b and~\ref{fig:DOS_3d}c], where the deviation of $\tau$ from special values $\tau_\ast$ corresponding to topological transitions (for which the van Hove singularity line 
is present) results in destroying van Hove singularity lines~(see details 
in~Refs.~\onlinecite{2019:Igoshev_FMM,2019:Igoshev_JETP}). 
For the square lattice at $\tau = 0$ the Fermi level corresponding to half-filling coincides with the van Hove singularity position; at finite $\tau$ this is not the case, but the van Hove singularity holds its impact. 
For sc lattice the van Hove singularity is well away from zero, which results in an analytic dependence $E^{\rm sc}_{\rm F}$ on $\tau$; as a consequence one can see equidistant positions of the Fermi level at different $\tau$ with the same difference, see Fig.~\ref{fig:DOS_3d}b). 
For bcc lattice, despite  that van Hove singularity is absent at finite $\tau$, the peak below $\epsilon = E^{\rm bcc}_{\rm F}$ originates from heavy mass at the (saddle) van Hove point $\Lambda^\ast$ with large ($\propto \tau^{-1}$) three masses at the diagonal of the Brillouin zone, split off from P point as $\tau$ becomes nonzero. 
Nonequidistant positions of Fermi level can be seen even by eye for both 
square and bcc lattices, see Figs.~\ref{fig:DOS_3d}a and \ref{fig:DOS_3d}c. We can therefore state that heavy mass of diagonal (saddle) van Hove singularity point enhancement holds its impact on thermodynamic quantities, e.g.,~the~free energy, which indicates the similarity of the bcc and square lattices. 
In Fig.~\ref{fig:Ef} the dependence of $E_{\rm F}/\tau$ in the PM phase as a function of $\tau$ is shown for the square, sc and bcc lattices. For the square and bcc lattices the dependence of $E_{\rm F}$  on $\tau$ is nonanalytical in the vicinity of $\tau = 0$, and the van Hove DOS singularity  for bcc lattice present at $\tau = 0$  retains to great extent its  impact  on the Fermi energy due to strong  mass enhancement at $\Lambda^\ast$ point, see~Refs.~\onlinecite{2019:Igoshev_FMM,2019:Igoshev_JETP}.

\section{Investigation of the phase competition at the MIT line}\label{sec:MIT_treatment}

\begin{figure}[h!]%\center
\noindent
\includegraphics[width=0.32\textwidth]{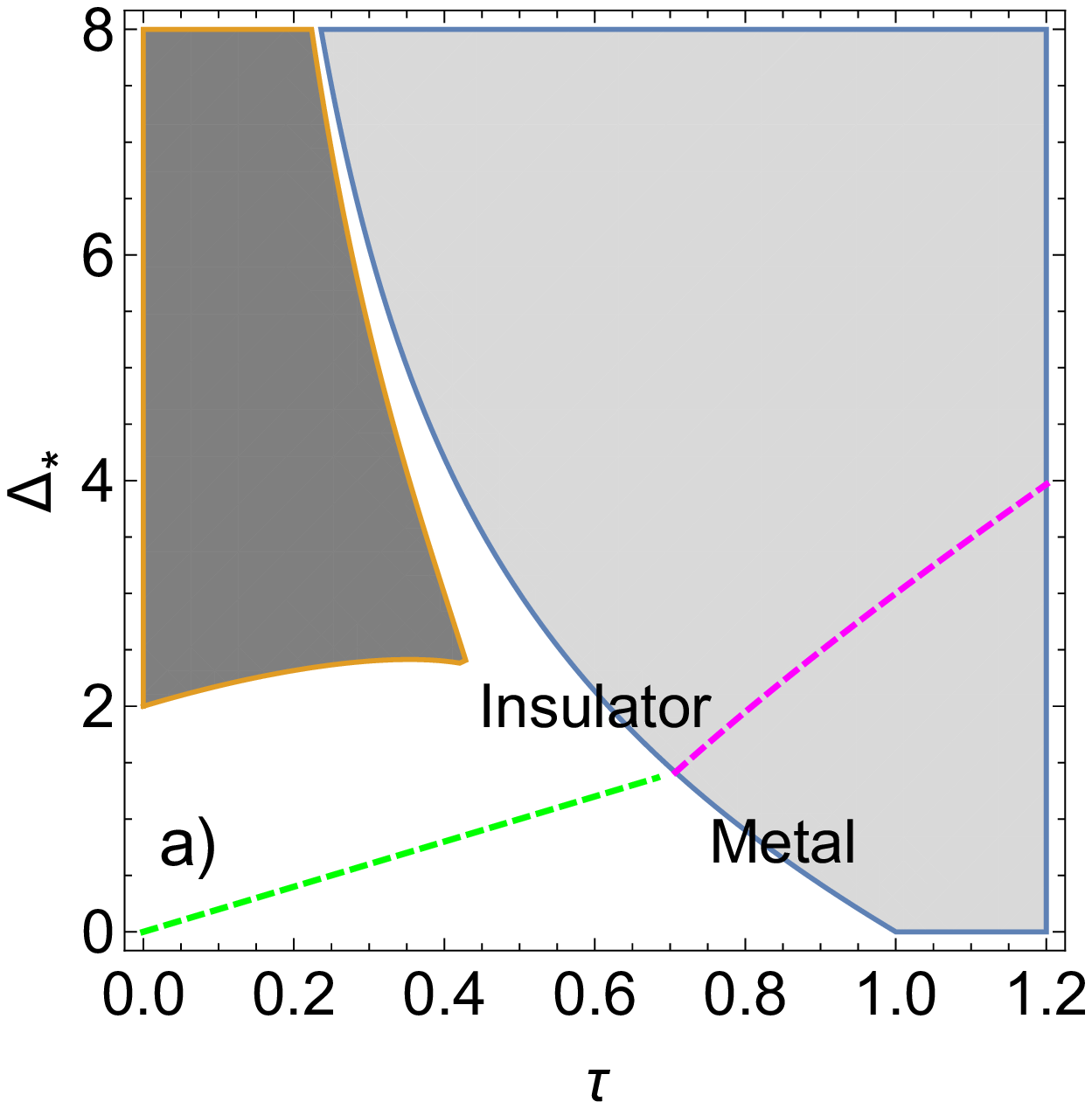}
\includegraphics[width=0.32\textwidth]{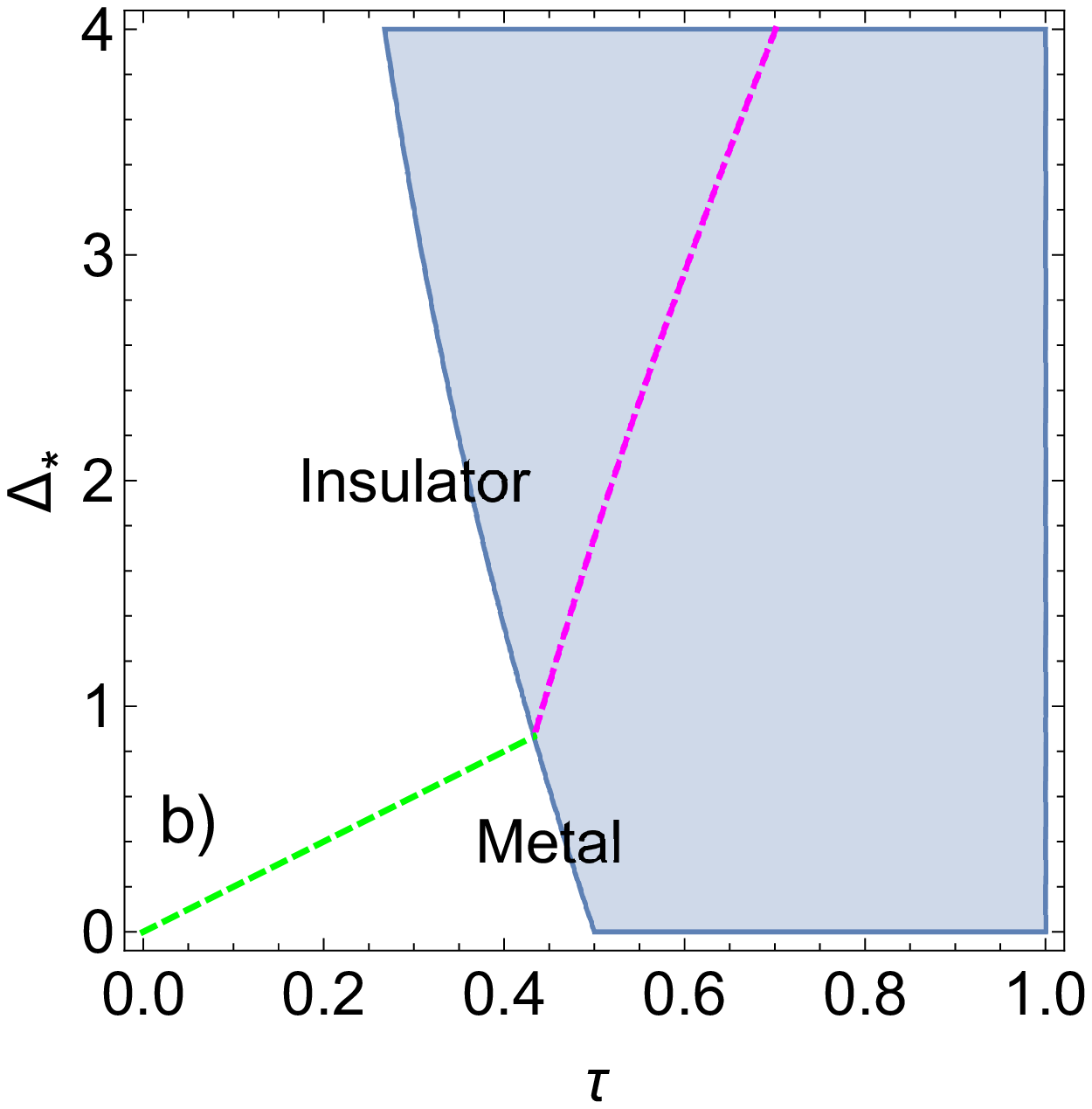}
\includegraphics[width=0.32\textwidth]{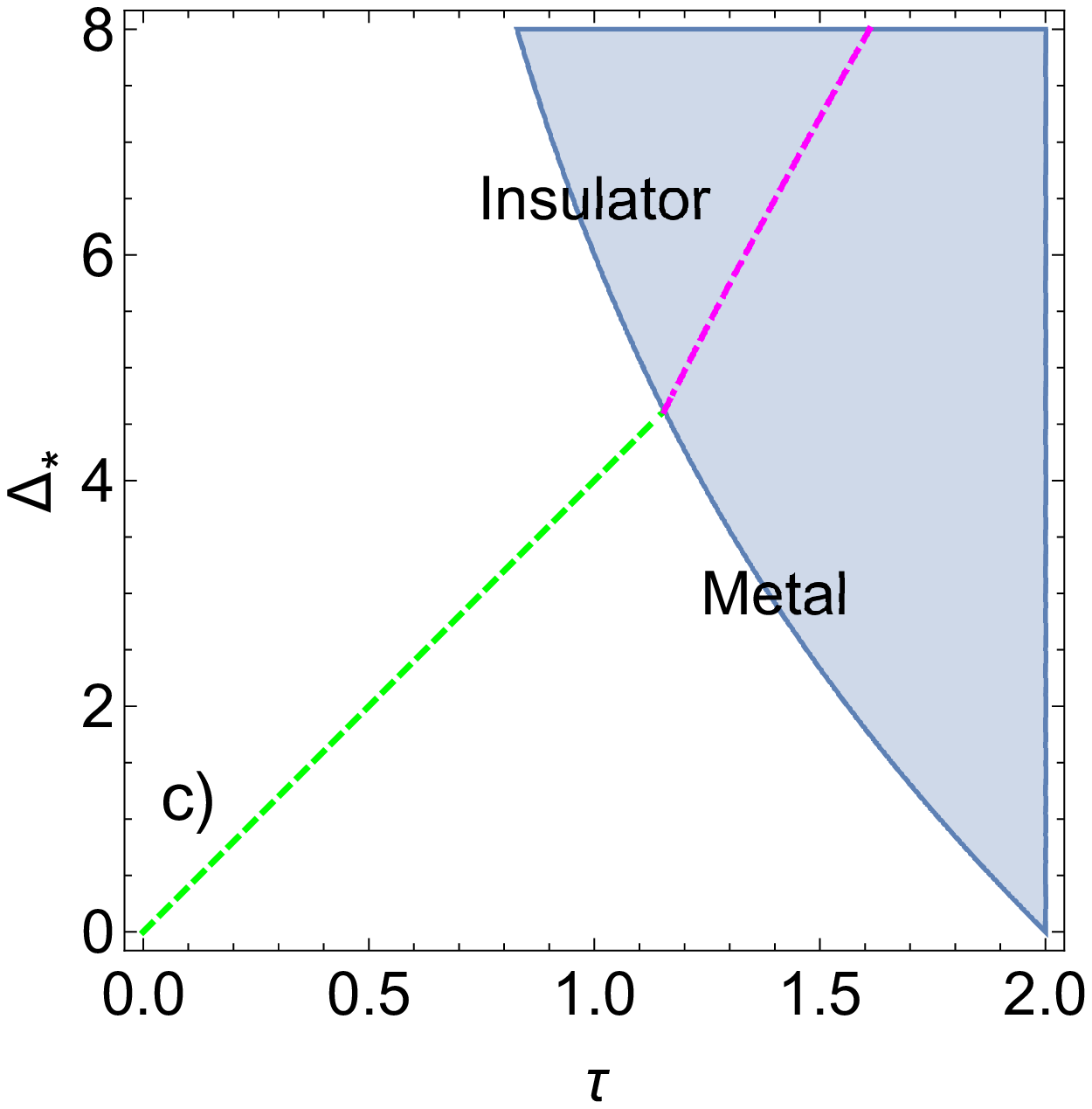}
\caption{\label{fig:MIT_phase_tau-Delta}
(Color online) Ground state phase diagram of  $\tau-\Delta_\ast$  within AFM phase for the square~(a), sc (b) and bcc (c) lattices. (a) In the white region~$E^{\rm sq}_{1, \text{max}}/z^2 = -\Delta_\ast$, in the dark-gray region~$E^{\rm sq}_{1, \text{max}}/z^2 = -4\tau - \Delta_\ast\sqrt{\frac{1 - 2\tau}{1 + 2\tau}}$, in the light-gray region~$E^{\rm sq}_{1, \text{max}}/z^2 = 4\tau - \sqrt{16 + \Delta^2_\ast}$. $E^{\rm sq}_{2, \text{min}}/z^2 = -4\tau + \Delta_\ast$ The breakpoint of the MIT line is $\tau = 1/\sqrt{2}, \Delta_\ast = \sqrt{2}$. (b) White region $E^{\rm sc}_{1,\rm max}/z^2 = -\Delta_\ast$, at $\Delta_\ast < 3\tau^{-1}/2 - 6\tau$, blue region $E^{\rm sc}_{1,\rm max}/z^2 = 12\tau -\sqrt{36 + \Delta^2_\ast}$ otherwise; everywhere $E^{\rm sc}_{2,\rm min}/z^2 = -4\tau + \Delta_\ast$. (c) White region $E^{\rm bcc}_{1,\rm max}/z^2 = 8\tau - \Delta_\ast$ at $\Delta_\ast < 8\tau^{-1} - 2\tau$, blue region $E^{\rm bcc}_{1,\rm max}/z^2 =  12\tau - \sqrt{64 + \Delta^2_\ast}$; everywhere $E^{\rm bcc}_{2,\rm min}/z^2 = \Delta_\ast$
}
\end{figure}

In this section we write down an explicit MIT line equation within AFM state for different lattices and directly compare free energies of PM metal 
and AFM insulator states in both HFA and SBA, which yields the order of MIT transition.  
Whereas at $\tau = 0$ the magnetic subbands in AFM phase at fixed $\Delta_\ast$ are separated by~a~gap, an increase of $\tau$ results in non-coincidence of $\mathbf{k}$-point locations of the maximum of a lower and minimum of an upper AFM subband [see Eq.~(\ref{eq:Ek_def})]. This, in turn, 
results in  
a decrease of indirect gap between the subbands and eventually in its closing  at some critical $\tau$. 
The line of metal-insulator transition within the antiferromagnetic phase is given by the equation
\begin{equation}\label{eq:MIT_line}
	\Delta_\ast = \Delta_{\rm MIT}(\tau),
\end{equation}
and can be directly obtained from $\max_{\mathbf{k}}E_1(\mathbf{k}) = \min_{\mathbf{k}}E_2(\mathbf{k})$. % (overlapping of AFM spectrum branches, see Eq.~(\ref{eq:Ek_def})). 
This condition is valid within  HFA approximation ($\Delta_\ast = \Delta$).

Direct analysis of the electronic spectrum in the AFM phase yields for the square lattice 
\begin{equation}\label{eq:Delta_sq}
\Delta^{\rm sq}_{\rm MIT}(\tau) = 
\begin{cases}
2\tau,& \tau < 1/\sqrt{2},\\
4\tau - \tau^{-1},& \tau \ge 1/\sqrt{2},
\end{cases}
\end{equation}
for sc lattice
\begin{equation}\label{eq:Delta_sc}
\Delta^{\rm sc}_{\rm MIT}(\tau) = 
\begin{cases}
2\tau,& \tau < \sqrt{3}/4,\\
8\tau - 9\tau^{-1}/8,& \tau \ge \sqrt{3}/4,
\end{cases}
\end{equation}
and for bcc lattice
\begin{equation}\label{eq:Delta_bcc}
\Delta^{\rm bcc}_{\rm MIT}(\tau) = 
\begin{cases}
4\tau,& \tau < 2/\sqrt{3},\\
6\tau - 8\tau^{-1}/3,& \tau \ge 2/\sqrt{3}.
\end{cases}
\end{equation}
These MIT lines in variables $\tau - \Delta_\ast$ are shown in Fig.~\ref{fig:MIT_phase_tau-Delta}. The breakpoints  originate from the change in the position of the maximum of lower branch of AFM spectrum, see details in a caption of the figure and Ref.~\onlinecite{2019:Igoshev_JETP_MIT}. It 
is clear that at small $\tau$ we have linear relation between $\tau$ and $\Delta_\ast$:
\begin{equation}\label{eq:MIT_line_at_small_tau}
\Delta_\ast = \kappa_{\rm MIT}\tau,
\end{equation} 
where $\kappa_{\rm MIT}$ differs for different lattices. 
%??where uniform narrowing of the spectrum in AFM phase is taken into account. - povtor??

%Since at half-filling the electron spectrum acquires the $\kk$ space uniform narrowing due to particle-hole symmetry we can formulate the criterium.
%Equation of metal-insulator transition within the antiferromagnetic phase
%\begin{equation}
%\tau = \tau_{\rm MIT}(\Delta_\ast).
%\end{equation}
The parameter $\Delta_\ast$, or, alternatively, the parameter $\tau$ via the relation (\ref{eq:MIT_line}), fully determines all properties of both 
PM and AFM phase on the MIT line. 
To determine the order of MIT phase transition %within the formulation of the problem 
it is sufficient to consider the difference of free energies of paramagnetic and antiferromagnetic phases
%We consider the free energy difference
\begin{equation}\label{eq:F_MIT_line_difference}
	\Delta F_{\rm MIT}(\tau) = F_{\rm AFM}\left(\Delta_{\rm MIT}(\tau)\right) - F_{\rm PM}(\tau)
\end{equation}
on a MIT line as a function of $\tau$.
If $\Delta F_{\rm MIT}(\tau)$ is negative, an additional second-order transition from AFM insulator to AFM metal phase occurs when $U$ decreases, so that MIT appears to be a second-order transition. If $\Delta F_{\rm MIT}$ is positive, the first-order transition from  AFM insulator into PM metal phase occurs when $U$ decreases.

\subsection{Numerical results}
%%%%%% HFA treatment %%%%%
\begin{figure}[!h]
\includegraphics[angle=-90,width=0.49\textwidth]{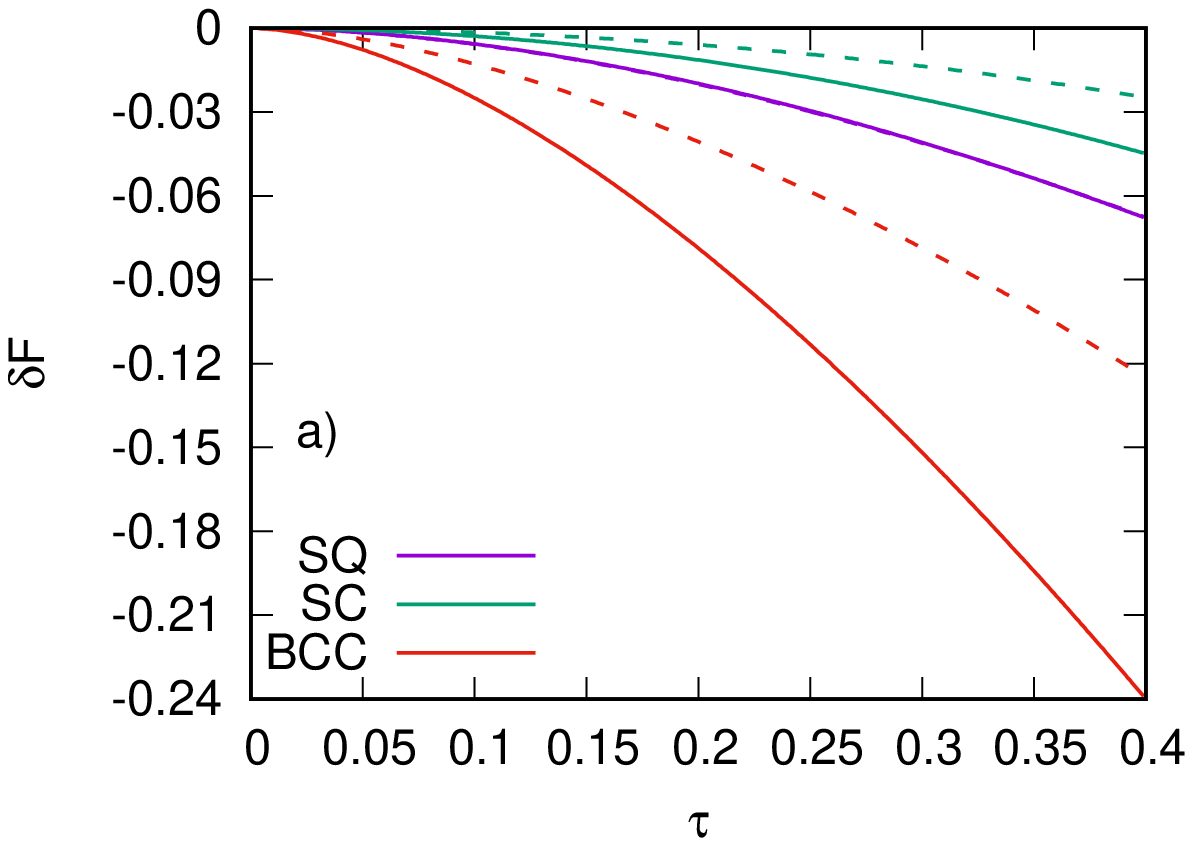}
\includegraphics[angle=-90,width=0.49\textwidth]{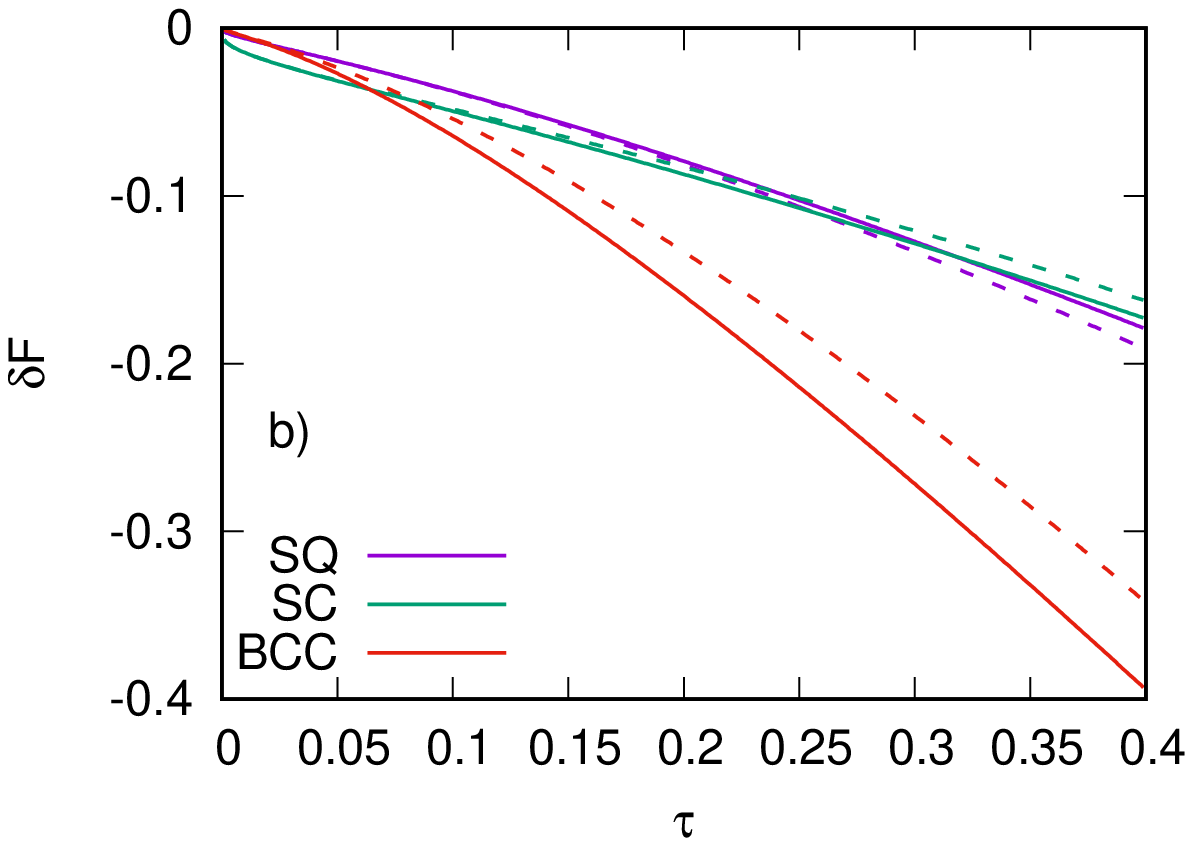}
\caption{
(Color online)
(a) $\delta F^{\rm HFA}_{\rm AFM}\left(\Delta_{\rm MIT}(\tau)\right)$ (solid line) and $\delta F^{\rm HFA}_{\rm PM}(\tau)$~(dashed line) within HFA; (b) $\delta F_{\rm AFM}\left(\Delta_{\rm MIT}(\tau)\right)$~(solid line) and $\delta F_{\rm PM}(\tau)$~(dashed line) within SBA.  The square, sc, and bcc lattices are 
considered.
}
\label{fig:F_MIT_all_lattices}
\end{figure}

%At first we consider the energy difference within the HFA approximation. 

In this subsection we present the results of numerical analysis of~the sign of expression~(\ref{eq:F_MIT_line_difference}) for different lattices and approximation used. 
In Fig.~\ref{fig:F_MIT_all_lattices} the free energies of AFM insulator and PM metal phases on MIT line, Eq.~(\ref{eq:MIT_line}), within both HFA (see the Sec.~\ref{sec:HFA}) and SBA (see~Sec.~\ref{sec:AFI}) 
are shown.  
Whereas for the square lattice the free energies of AFM and PM phases are 
found to be very close (especially within the HFA), for sc and especially 
for bcc (due to lowering of the free energy of AFM phase caused by vHS of 
DOS at $\tau = 0$) lattices, the energy of AFM insulator phase is considerably lower than the energy of PM metal phase in both the approximations used. 
We find that correlation effects considerably reduce the energy of both phases $\delta F_{\rm AFM}$ and $\delta F_{\rm PM}$ at MIT line. However, this reducing is substantially stronger for paramagnetic phase. 
Another correlation effect is the occurrence of nonanalytic contributions~$\sim \zeta^2$ (as a function of $\tau$) to the free energy, see~Eq.~(\ref{eq:U_in_AFM}), which yields considerable contribution at small $\tau$ 
[see Fig.~\ref{fig:F_MIT_all_lattices}b]. %The influence of this correction can be seen in Fig.\ref{fig:F_MIT_all_lattices}b at small $\tau$(povtor???). From Eq.~(\ref{eq:xi_general}) it is clear that at $J = 0$ in the leading order 
From  Eqs.~(\ref{eq:U_in_AFM}) and~(\ref{eq:PM_U}) we see  that at small~$\tau$ $\zeta_{\rm A}\sim\zeta_{\rm P}\sim 2/(\Phi_1(\Delta_{\rm MIT}(\tau))U_{\rm BR})$, therefore the nonanalytic behaviour of $\zeta_{\rm P,A}$ is determined by that of $\Phi_1$, which is very different for different lattices, see derivation in~the~Appendix~\ref{appendix:G_expansion}. 
 
\begin{figure}[h]
\includegraphics[angle=-90,width=0.44\textwidth]{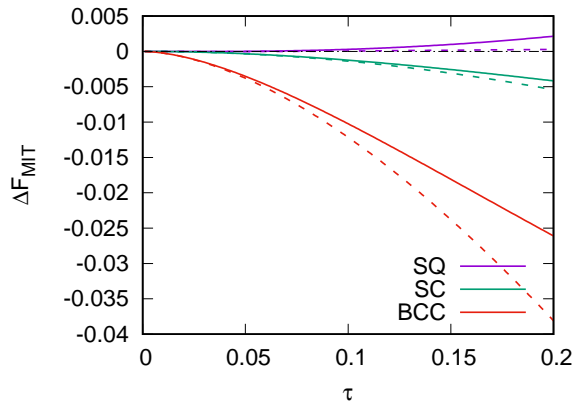}
\caption{
(Color online)
The difference of $\Delta F_{\rm MIT}(\tau)$ within SBA (solid lines) and 
HFA approximations (dashed lines). 
}
\label{fig:dF_MIT_all_lattices}
\end{figure}

In all the cases the correlation effects increase $\Delta F_{\rm MIT}$, which enhances the tendency towards first-order MIT. 
In~Fig.~\ref{fig:dF_MIT_all_lattices} the~free energy difference $\Delta F_{\rm MIT}(\tau)$ on the MIT line (\ref{eq:MIT_line}), see~Eq.~(\ref{eq:F_MIT_line_difference}), is shown for the square, sc and bcc lattices. We~find that whereas for considered three-dimensional lattices $\Delta F_{\rm MIT}(\tau) < 0$, for the square lattice the sign change of $\Delta F_{\rm MIT}(\tau)$ occurs at small $\tau$ since the free energies of AFM and 
PM phases are still very close, so that an accurate consideration is needed in both HFA and SBA approximation. 

In the following subsections we analyze the impact of correlations and exchange interaction effects on  the PM--AFM insulator free energy difference on the hypothetical MIT line within AFM phase~(see Sec.~\ref{sec:analysis_of_correlation_and_exchange}) and derive an expansion of the free-energy of AFM insulator phase for the square lattice to reveal the origin for smallness of free energy difference found above, see also Refs.~\onlinecite{2000:Yang,2010:Yu}~(also see Sec.~\ref{sec:analytic_expansion}).  
%However, for the square lattice the energies of AFM and PM phases are still very close. The understanding  of this fact requires an additional consideration.

\subsection{Analysis of correlation and exchange effects}\label{sec:analysis_of_correlation_and_exchange}
Here we analyze the contributions of correlations and exchange intersite interactions into $\Delta F_{\rm MIT}$ on the MIT line within AFM phase. 
Since both these contributions are pure correlation effects beyond HFA and are of fourth order with respect to $\tau$, it can be expected that they start to play role with increasing $\tau$.    

\begin{figure}[!h]
\includegraphics[angle=-90,width=0.49\textwidth]{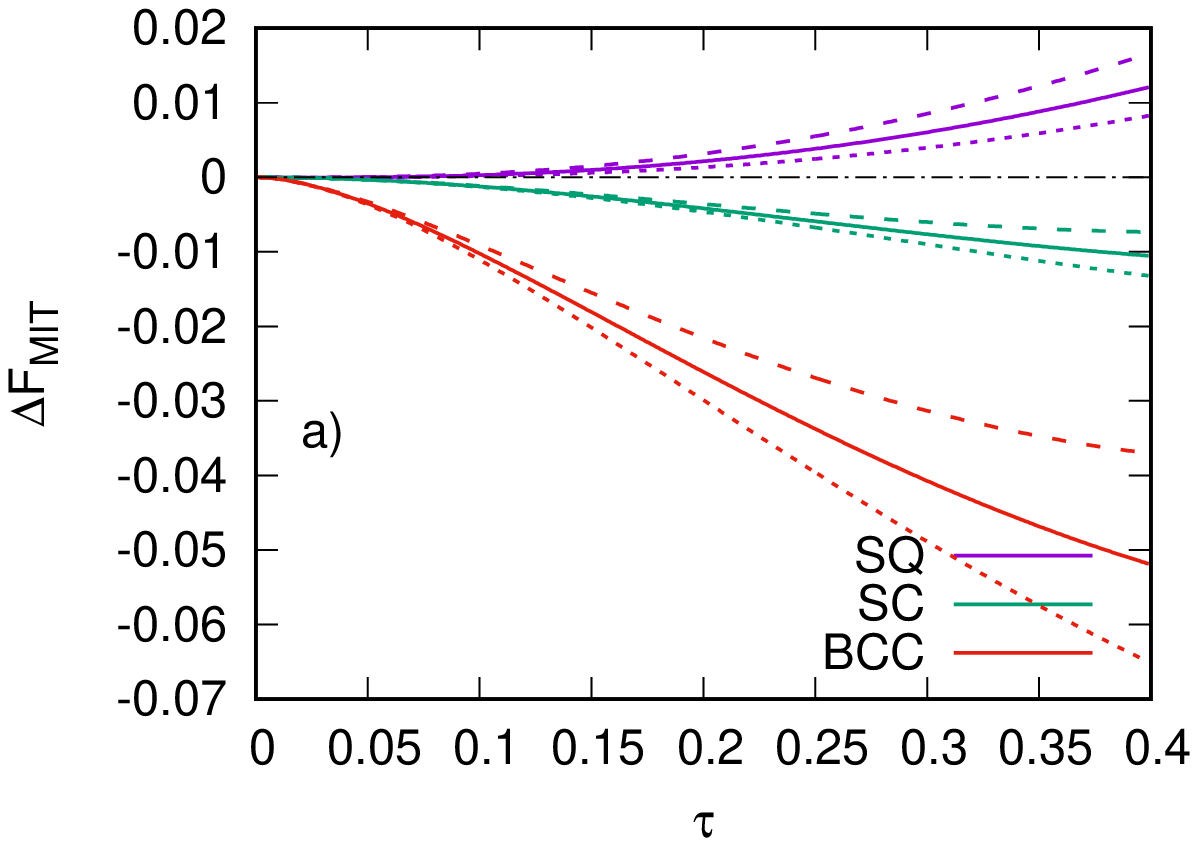}
\includegraphics[angle=-90, width=0.45\textwidth]{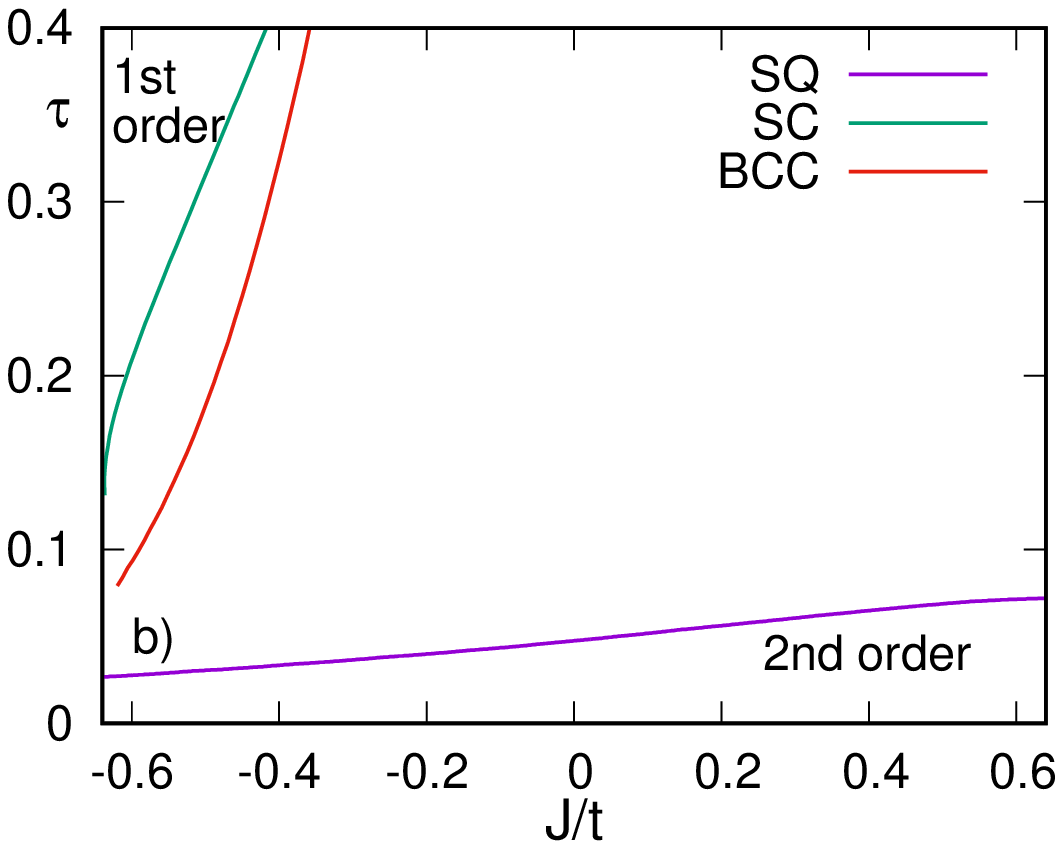}
\caption{
(Color online)
(a) The free-energy difference $\Delta F_{\rm MIT}(\tau)$ on the MIT line 
for square (SQ), simple cubic (sc), and body centered cubic (bcc) lattices 
within SBA in the presence of $J_{\qq} = 0$~(solid lines), $J_{\qq} = 
+z^2_{\rm A}$~(dashed lines), $J_{\qq} = -z^2_{\rm A}$~(dotted lines). 
(b) The phase diagram demonstrating  the MIT type in the $ J-\tau $ variables for square (SQ), simple cubic (sc), and body centered cubic (bcc) lattices, $ J $ being the exchange integral between the nearest neighbors. % 
The negative sign of $ J $ corresponds to~AFM, positive --- to~FM ordering.
To the left of the curves, a first-order transition takes place, to the right --- the second order transition.
}
\label{fig:delta_F_MIT_J_effect}
\end{figure}

Since the exchange interaction enters all equation through its Fourier transform, for simplicity we use the approximation for exchange integrals: $J_{\qq} = -Z_{\rm nn}J$, $J(Z_{\rm nn})$ being  the nearest-neighbor exchange integral (number). 
In Fig. \ref{fig:delta_F_MIT_J_effect}a we show $\Delta F_{\rm MIT}$ on the MIT line in the presence of  exchange interaction of different sign. It is worthwhile to note that the main part of the interaction is absorbed 
into the gap and does not affect  $\Delta F_{\rm MIT}$, but only its subleading part  plays a role. This subleading contribution is inaccessible within HFA and has a delicate many-electron nature.   
%%%%%% SBA treatment %%%%%
Applying the method considered above to the case of finite exchange interaction we find that $\Delta F_{\rm MIT}$ is rather sensitive to the value 
of $J$: positive (negative) $J$ results in decreasing (increasing) of $\Delta F_{\rm MIT}$.  
Thus, we may expect an intersite exchange interaction can change the MIT order transition,    
%the exchange interaction can change the order of MIT 
which is purely many-electron effect. 
Figure~\ref{fig:delta_F_MIT_J_effect}b provides the phase diagram in terms of $J-\tau_{\rm c}$, where $\tau_{\rm c}$ is determined by the equation $\Delta F_{\rm MIT}\left(\Delta_{\rm MIT}(\tau), J\right) = 0$ and is actually the critical $\tau$ separating the first- and second-order transition regions. 
One can see that the exchange interaction of ``ferromagnetic'' sign $J < 0$ can transform the second-order transition into the first-order one reducing the stripe of AFM metal in the phase diagrams~(see Fig.~\ref{fig:PD_scheme}): For the square lattice the change of exchange integral shifts $\tau$ point of MIT order change only weakly, whereas for sc and bcc lattices substantially large values of exchange integral are needed for the change. 
Also we note that the sensitivity of $\Delta F_{\rm MIT}$ with respect to 
$J$ value is very different for different lattices: the size of the effect is determined by DOS van Hove singularity at $\tau = 0$. 
%Note that the results for MIT line in Fig.~\ref{fig:delta_F_MIT_J_effect}b for sc and bcc lattices at large $\tau\sim 0.5$ should be considered with a caution, since other phases may occur destroying the bipartite character of the magnetic order of the lattice~\cite{2016:Timirgazin}, see discussion below. 
Note that the results for MIT line in~Fig.~7b for sc and bcc lattices should be considered with a caution, since at large $\tau \sim 0.5$ other (spiral) phases with lower free energy may occur destroying the two-sublattice antiferromagnetic order \cite{2016:Timirgazin}. 
At the same time, the results for small $\tau$ seem to~be~reliable. 

Now we analyze different contributions in 
$\Delta F_{\rm MIT}(\Delta_\ast)$ to trace explicitly the influence of many-electron effects and exchange interaction on the order of MIT. 
From Eqs.~(\ref{eq:delta_F_AFM}) and~(\ref{eq:delta_F_PM}) we have
\begin{equation}\label{eq:delta_F_tmp}
\Delta F_{\rm MIT}(\Delta_\ast) = \Delta F^{\rm HFA}_{\rm MIT}(\Delta_\ast) - \frac18\frac{\zeta^2_{\rm A}J_{\qq}m^2}{1 - m^2 - \zeta^2_{\rm A}} 
- \frac{U}8\left(
\zeta_{\rm A}\left(1 + \frac{2m^2}{(1 + 2\zeta)^2}\right) - \zeta_{\rm P}
\right).
\end{equation}
We exclude $\zeta_{\rm A, P}$ using  Eqs.~(\ref{eq:U_in_AFM}) and (\ref{eq:PM_U}),
\begin{eqnarray}
\label{eq:alpha_A}
\zeta_{\rm A} &=& u\left(1 - \frac{m^2}{(1 + \zeta_{\rm A})^2}\right)\frac{1 - m^2}{1 + \alpha_{\rm A}},\\
\label{eq:alpha_P}
\zeta_{\rm P} &=& \frac{u}{1 + \alpha_{\rm P}},
\end{eqnarray}
where $\alpha_{\rm A} = \delta\Phi_{2}(\Delta_\ast)/\Phi_2(0)$, $\alpha_{\rm P} = \delta\Phi_{2\rm P}\left(\tau_{\rm MIT}(\Delta_\ast)\right)/\Phi_2(0)$ and the dimensionless interaction parameter 
\begin{equation}\label{eq:u_def}
	u = U/U_{\rm BR}
\end{equation}
is introduced.

\begin{figure}[h!]
\includegraphics[angle=-90,width=0.5\textwidth]{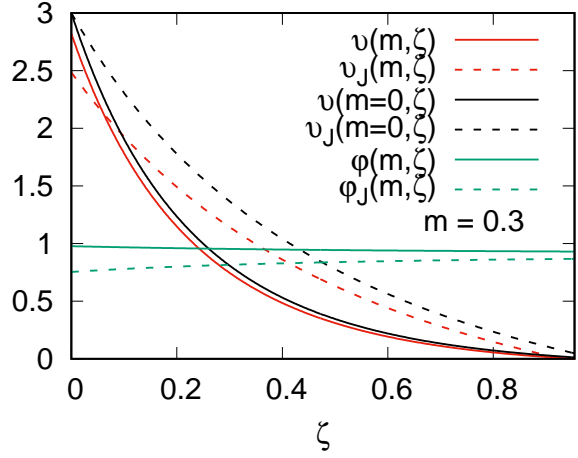}
\caption{
	(Color online)
	$\zeta$ plot of $\varphi(m,\zeta)$, $\varphi_J(m,\zeta)$, $v(m,\zeta)$ and $v_J(m,\zeta)$ at $m = 0.3$ and $m = 0$.
}
\label{fig:phi_and_v}
\end{figure}

We split the expression (\ref{eq:delta_F_tmp})
\begin{equation}\label{eq:final_delta_F_MIT}
\Delta F_{\rm MIT}(\Delta_\ast) = \Delta F^0_{\rm MIT}(\Delta_\ast) + \Delta F^{\rm c}_{\rm MIT}(\Delta_\ast)  + \Delta F^J_{\rm MIT}(\Delta_\ast),
\end{equation}
where
\begin{eqnarray}\label{eq:final_delta_F0_MIT}
\Delta F^0_{\rm MIT}(\Delta_\ast) &=& \Delta F^{\rm HFA}_{\rm MIT}(\Delta_\ast)\left(1 - \frac{u^2}{(1 + \alpha_{\rm P})(1 + \alpha_{\rm A})}\left(\varphi(m,\zeta_{\rm A}) + \frac{J_{\mathbf{Q}}m^2\varphi_J(m,\zeta_{\rm A})}{8z^2_{\rm A}\Phi_2(\Delta_\ast)} \right)
\right), \\
\label{eq:final_delta_Fc_MIT}
\Delta F^{\rm c}_{\rm MIT}(\Delta_\ast) &=& \frac{Uum^4v(m,\zeta_{\rm A})}{8(1 + \alpha_{\rm P})}, \\
\label{eq:final_delta_FJ_MIT}
\Delta F^J_{\rm MIT}(\Delta_\ast) &=& \frac{u^2J_{\mathbf{Q}}m^4}{8z^2_{\rm A}(1 + \alpha_{\rm P})(1 + \alpha_{\rm A})} \left(v_J(m, \zeta_{\rm A}) + \frac{J_{\mathbf{Q}}\varphi_J(m,\zeta_{\rm A})}{8z^2_{\rm A}\Phi_2(\Delta_\ast)} \right),
\end{eqnarray}
where 
$$\varphi (m,\zeta) = \left(1 - \frac{m^2}{(1 + \zeta)^2}\right)(1 - m^2)\left(1 + \frac{2m^2}{(1 + \zeta)^2}\right), 
\varphi_{J}(m, \zeta) = (1-m^2)\left(1 - \frac{m^2}{(1 + \zeta)^2}\right)^2,$$ 
$$v(m,\zeta)= \frac{(1 - \zeta)(3 + \zeta) - 2m^2}{(1 +  \zeta)^4}, 	v_J(m,\zeta)  =\frac{(1 - \zeta)(3 + \zeta)}{(1 + \zeta)^2} - \frac{2(3 + 2\zeta + \zeta^2)}{(1 + \zeta)^4}m^2 + \frac{3m^4}{(1 + \zeta)^4}.$$
There are three terms in~Eq.~(\ref{eq:final_delta_F_MIT}): The first one is renormalization of $\Delta F^{\rm HFA}_{\rm MIT}$, the second one has pure many-electron nature originating from the
difference of singly and doubly occupied states, and third one yields the 
exchange interaction contribution, which also has many-electron nature. 
A typical behavior of the functions $\varphi(m,\zeta)$, $\varphi_J(m,\zeta)$, $v(m,\zeta)$, $v_J(m,\zeta)$ is shown in Fig.~\ref{fig:phi_and_v}. 
%It is clear that typically $v, v_J > 0$, possibly except for  $\zeta \lesssim \sqrt{1 - m^2}$.\textbf{[??? is it really needed]}. 

%From an exact splitting of $\Delta F_{\rm MIT}$ into different contributions~(\ref{eq:final_delta_F_MIT}) we obtain 
We expand the expression (\ref{eq:final_delta_F_MIT}) taking into account 
the terms of $m^4$ order in  spirit of above analysis of the square lattice case. 
Since the expansion of $\alpha_{\rm A}, \alpha_{\rm P}$ starts from $m^2$ terms (with possible logarithmic prefactors), 
%\textbf{[??? how can we write that  starts from $m^2$ terms?]} 
$\varphi(m, \zeta_{\rm A}), \varphi_J(m, \zeta_{\rm A})\sim 1$, $v(m, \zeta_{\rm A})\sim v(0, u), v_J(m, \zeta_{\rm A})\sim v_J(0, u)$ and $\Phi^4_1(\Delta_\ast)\sim \bar{\Phi}^4_1(\Delta_\ast)$, where $\bar{\Phi}_1(\Delta)$ is leading contribution to $\Delta$-expansion series of $\Phi_1(\Delta)$
($\Delta$ dependence is realized through logarithms only).
Then we can rewrite Eq.~(\ref{eq:final_delta_F_MIT}) separating leading term in the $m$ expansion, thereby formulating the estimation
\begin{equation}\label{eq:delta_F_MIT_est}
\Delta F^{\rm est}_{\rm MIT}(\Delta_\ast) = \Delta F^0_{\rm MIT,est}(\Delta_\ast) 
+ \Delta F^{\rm c}_{\rm MIT,est}(\Delta_\ast)  + \Delta F^J_{\rm MIT,est}(\Delta_\ast)  + o(\Delta^5_\ast),
\end{equation}
where the contributions read
\begin{eqnarray}
\Delta F^0_{\rm MIT,est}(\Delta_\ast) &=& \Delta F^{\rm HFA}_{\rm MIT}(\Delta_\ast)\left(1 - u^2\right),  \\	
\label{eq:Fc_est}
\Delta F^{\rm c}_{\rm MIT,est}(\Delta_\ast) &=& U_{\rm BR}\frac{u^2\bar{\Phi}^4_1(\Delta_\ast)\Delta^4_\ast (1 - u)(3 + u)}{8(1 +  u)^4}, \\	
\Delta F^{J}_{\rm MIT,est}(\Delta_\ast) &=& U_{\rm BR}\frac{u^2\bar{\Phi}^4_1(\Delta_\ast)\Delta^4_\ast j}{4(1 - u^2)}
	\left(\frac{(1 - u)(3 + u)}{(1 +  u)^2} + \frac{j}{1 - u^2}\right),
\end{eqnarray}
at small $\Delta_{\ast}$ [we retain the terms of leading (second) and subleading (fourth) orders], where the dimensionless interaction exchange interaction parameter %charactering the strength of correlation effects 
$j = J_{\qq}/U_{\rm BR}$
is introduced. Physically $u, |j| \ll 1$, since we are far away from Mott 
transition. %Note that typically\textbf{[???]} 

%We see that $\Delta F_{\rm MIT}$ contains three terms: slightly ``renormalized'' HFA contribution (the first term),  many-electron term originating from the difference of singly  and doubly occupied site states (the second term), exchange term which originates from the latter difference (the third term). 

At $|j|\ll 1$ the third term in~Eq.~(\ref{eq:delta_F_MIT_est}) (its sign is determined by the sign of $j$) is by absolute value is much smaller than the second one (which is positively defined). 
We conclude therefore that the influence of exchange effects [which, being taken alone, tends to change the MIT order, $J > 0$~$(J < 0)$ to second 
(first) order] is compensated by correlation contribution. 
However, this statement can be violated in the limit of moderate and large $\Delta$ when the use of the expression (\ref{eq:delta_F_MIT_est}) is not valid. In this case one should use the exact expression~(\ref{eq:final_delta_F_MIT}): From Fig.~\ref{fig:delta_F_MIT_J_effect}b it is clear that for sc and bcc lattices the MIT order changes at large ``ferromagnetic'' $|J|\gtrsim 0.4$. 
%which is beyond weak-coupling limit\textbf{[???more detaily!]}. 

\begin{figure}[h]
\includegraphics[angle=-90,width=0.44\textwidth]{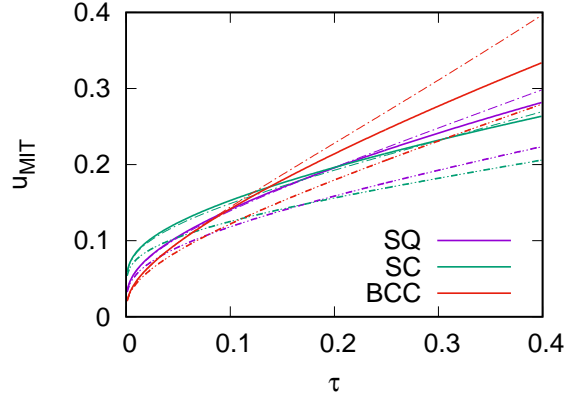}
\caption{
(Color online)
Critical $U$ at MIT line within the AFM phase in units of $U_{\rm BR}$ for square, sc and bcc lattices from numerical solution of SBA equations (solid lines), numerical solution of HFA equations (dashed lines) and asymptotic solution (\ref{eq:U_asymp_leading}) appropriate in HFA (dotted lines). Two latter lines practically coincide. Analytical solution (\ref{eq:U_asymp_subleading}) is also shown by dot-dashed lines.}
\label{fig:U_MIT_all_lattices}
\end{figure}

Consider now the second-order transition from AFM insulator into AFM metal phase [AFM MIT line $\Delta_\ast = \Delta_{\rm MIT}(\tau)$, Eq.~(\ref{eq:MIT_line}], see also Fig.~\ref{fig:PD_scheme}) to obtain some analytical results concerning the critical $U_{\rm MIT}$. 
The required expansion of $U_{\rm MIT}$ in powers of $\zeta_{\rm A}$ and $m$ can be derived directly from~Eq.~(\ref{eq:U_in_AFM}), with the use of expansion of $\delta\Phi_2(\Delta_{\rm MIT})$ (see Eq.~(\ref{eq:Phi2_through_G}) 
%this was Eq.~(5) of SM
in~the~Appendix~\ref{appendix:G_expansion}). We can obtain this expanding $\zeta_{\rm A}$  as a function of $\xi$ in~powers $\left(\Pi(\Delta_{\rm MIT}) U_{\rm BR}\right)^{-1}$ and $m$ using the~Eq.~(\ref{eq:xi_general}). $m$ should be excluded using the~Eq.~(\ref{eq:m_vs_Delta}). 
We restrict ourself by the leading (zero-order) with respect to $m$ contribution: this is realized by setting $m = 0$ and replacing $\Phi_2(\Delta_{\rm MIT})$ by $\Phi_2(0)$ in~Eqs.~(\ref{eq:U_in_AFM}) and (\ref{eq:xi_general}) and replacing in the latter $\Phi_1(\Delta_{\rm MIT})$ by $\bar{\Phi}_1(\Delta_{\rm MIT}) + (1/4)J_{\qq}/(1 - \zeta^2_{\rm A})$. 
Within this approximation we obtain 
%Consider the dependence of $U_{\rm MIT}$ on $\tau$. Since $\Phi_1(\Delta)$ depends on $\Delta$ as a polynomial of logarithms~(see the~subsection 1 of~Supplemental Material\cite{Supplemental}), whereas $m$  depends on $\Delta$ linearly (up to logarithmic factors), we obtain with account of  Eqs.~(\ref{eq:m_vs_Delta}) that in zeroth order in $\Delta_\ast$ one has $\zeta_{\rm A} = u$. 
%Retaining in Eq.~(\ref{eq:xi_general}) only zeroth order with respect to 
$\Delta_\ast$~(or $m$) we get
 \begin{equation}\label{eq:Uc_expansion}
	u_{\rm MIT} = \zeta_{\rm A} = u_{\rm eff} + 3u_{\rm eff}^2/2 + 5u_{\rm eff}^3/2 + \ldots,
\end{equation}
where $u_{\rm eff} = u_{\rm HFA} + \frac12j/(1-u^2)$, $u_{\rm HFA} = U_{\rm eff}^{\rm HFA}/U_{\rm BR}$, see Eq.~(\ref{eq:U_eff_HFA}). 
Equation~(\ref{eq:Uc_expansion}) yields the equation on $u$. The solution by the method of successive iterations yields
\begin{equation}\label{eq:Uc_expansion_solved}
	u_{\rm MIT} = u_{\rm HFA} + j/2  + \frac32(u_{\rm HFA} + j/2)^2 + \frac12ju^2_{\rm HFA} + \ldots,
\end{equation}   
so that exchange effect on $U_{\rm MIT}$ is not reduced to typical for HFA absorption $u_{\rm HFA} \rightarrow u_{\rm HFA} + j/2$.
The expression (\ref{eq:Uc_expansion}) yields direct correlation corrections to HFA's $U_{\rm c}$.
For different lattices the behaviour considerably differs due to vHS in the center of the band of different types at $\tau = 0$, see~Eq.~(\ref{eq:Stoner_criterion}).

As one can see above, the asymptotic behavior of lattice $\Phi_1(\Delta)$ 
is main feature determining all characteristics of the system in the zeroth order in $\tau$. 
In the~Appendix~\ref{appendix:G_expansion} a general way of treatment of the asymptotics of this quantity is developed, depending on the van Hove singularity type.
Below we present the analytical expansion for $u_{\rm HFA}(\tau)$ at $J = 
0$ up to leading with respect to $\tau$ terms for different lattices. 
For the square lattice we get %from the expansion~(\ref{eq:Phi1_sq_expansion}) %(see Supplemental Material~\ref{appendix:G_expansion})
\begin{equation}\label{eq:U_HFA_square}
u^{\rm sq}_{\rm HFA}(\tau) = \frac{4\pi^2}{U^{\rm sq}_{\rm BR}}\left(\ln^2\frac{16}{\tau} + \frac{\pi^2}6 + \frac12\ln2 - 4\ln^22- \frac18  + 2\pi^2\delta g^{\rm sq}_0\right)^{-1},
\end{equation}
where $\delta g^{\rm sq}_0 = 2.8\cdot10^{-3}$. 
For sc lattice we analogously get %the Eq.~(\ref{eq:Phi1_sc_expansion}) yields
\begin{equation}\label{eq:U_HFA_sc}%d = tau/3
u^{\rm sc}_{\rm HFA}(\tau) = \frac{12/U^{\rm sc}_{\rm BR}}{2a^{\rm sc}_0\ln\frac{2}{\tau} + a^{\rm sc}_2/9 + 4\delta g^{\rm sc}_0},
\end{equation}
see values $a^{\rm sc}_0 = 0.86, a^{\rm sc}_2 = 0.10$, $\delta g^{\rm 
sc}_0 = 0.21$, the derivation details are presented in~the~Appendix~\ref{appendix:G_expansion}.
An analogous procedure for the bcc lattice yields%the Eq.~(\ref{eq:Phi1_bcc_expansion}) yields
%\begin{multline}\label{eq:U_HFA_bcc}
%u^{\rm bcc}_{\rm HFA}(\tau) = \frac{12\pi^3}{U^{\rm bcc}_{\rm BR}}\\
%\left(\ln^3\frac{32}{\tau} - \frac{\pi^2}4\ln\frac{32}{\tau} + \frac{9\pi^2}4\ln2 - 27\ln^32 + \frac32\zeta(3) - \frac{3}{16} - \frac{3\pi^2}{64} + \frac{27}{16}\ln^22 - \frac98\ln2 + 3\pi^3\delta g^{\rm bcc}_0\right)^{-1}.
%\end{multline}
\begin{equation}\label{eq:U_HFA_bcc}
u^{\rm bcc}_{\rm HFA}(\tau) = \frac{12\pi^3}{U^{\rm bcc}_{\rm BR}}
\left(\ln^3\frac{32}{\tau} - \frac{\pi^2}4\ln\frac{32}{\tau} + C_{\rm bcc}\right)^{-1},
\end{equation}
where $C_{\rm bcc} = \frac{9\pi^2}4\ln2 - 27\ln^32 + \frac32\zeta(3) - \frac{3}{16} - \frac{3\pi^2}{64} + \frac{27}{16}\ln^22 - \frac98\ln2 + 3\pi^3\delta g^{\rm bcc}_0 = 7.2$, $\zeta(s)$ being Riemann zeta function, $\delta g^{\rm bcc}_0 = -4\cdot10^{-3}$. 
%see also Eqs.~(\ref{eq:bcc_phi_0,3}-\ref{eq:bcc_phi_0,0}). 
These expressions substantially improve the simple Eq.~(\ref{eq:Stoner_criterion}) obtained  within leading logarithmic approximation.

The plot of critical $U_{\rm MIT}$ $\tau$ dependence at $J = 0$ for MIT 
line within both HFA and SBA together with the asymptotic expressions obtained from (\ref{eq:Uc_expansion_solved}) in leading (corresponding to HFA):   
\begin{equation}
\label{eq:U_asymp_leading}
U^{\rm lead}_{\rm MIT} = 2/\bar{\Phi}_1\left(\Delta_{\rm MIT}(\tau)\right),
\end{equation}
and subleading with respect to $\zeta_{\rm A}$ approximation (corresponding to SBA corrections)
\begin{equation}
\label{eq:U_asymp_subleading}
U^{\rm sublead}_{\rm MIT} = 2/\bar{\Phi}_1\left(\Delta_{\rm MIT}(\tau)\right) + 6/\left(U_{\rm BR}\bar{\Phi}^2_1\left(\Delta_{\rm MIT}(\tau)\right)\right), 
\end{equation}
is shown in~Fig.~\ref{fig:U_MIT_all_lattices}, see Eq.~(\ref{eq:U_eff_HFA}). 
Using the leading logarithm contribution to $\bar{\Phi}_1$(see the~Appendix~\ref{appendix:G_expansion}),  we obtain~Eq.~(\ref{eq:Stoner_criterion}), thereby 
$U_{\rm MIT}$ is mainly formed by inverse logarithmic contributions. For  
HFA, the agreement of leading contribution asympotics and numerical result at $J = 0$ is very good up to $\tau \sim 0.4$. At the same time, for SBA the agreement between numerical result and approximation~(\ref{eq:U_asymp_subleading}) is good for sc lattice,  worse for the square lattice and  bad for bcc lattice (in all the cases an~overestimation is present). This issue is closely related to the question of  applicability of zero-order  approximation with respect to $\Delta$ for $\Phi_1(\Delta)$, $\delta\Phi_2(\Delta)$, and solutions of Eqs.~(\ref{eq:U_in_AFM}) and $m$~corrections to $\zeta$ for the system with van Hove singularity at $\tau = 0$.

The result  $U_{\rm MIT}=2.16$ in AFM phase for the square lattice at $\tau = 0.2$ can be compared with the Monte Carlo calculations at $N=8\times8$ and $T = 1/6$~\cite{1997:Duffy}, where the transition from PM to AFM metal phase was found at $U_{\rm MIT} = 2.5\pm0.5$; however, the 
transition from AFM metal to AFM insulator occurs at $U_{\rm MIT} > 4$. 
We see that there is some discrepancy with our results. However, we believe that the Monte Carlo calculations do not allow to to treat precisely  the ground-state properties due to rather high temperature involved. 

\subsection{Analytical expansion}\label{sec:analytic_expansion}
%%%%%%%%%%%%%%%%%%%%%%%%%%%%%%%%
%%%%%%% Square lattice:
%%%%%%%%%%%%%%%%%%%%%%%%%%%%%%%%
In this section we focus  attention on the case of the square lattice, 
where the energies of AFM and PM phases were numerically found to be nearly degenerate, which causes first-order MIT  from AFM insulator  into PM metal phase~\cite{1997:Duffy,2010:Yu,2000:Yang,2016:Timirgazin}. 
Some results will be also obtained for sc and bcc lattices.   
The presence of the van Hove singularity in the density of states is the origin of possible numerical errors and results in the absence of an universal energy scale (which is determined by constant quadratic coefficient 
of expansion in powers of $\tau$ for the case of regular DOS). % and in non-universal behaviour energy of the  parameters $\Delta$ and $\tau$~(neponyatno???). 
%We get the following asymptitics for the paramagetic energy deviation
We develop expansions for both PM and AFM insulator phases in powers of $\tau$.

At first, we consider the expansion for the free energy of PM phase for the square lattice within HFA using the~Eqs.~(\ref{eq:PM_n}) and (\ref{eq:delta_F_PM_HFA}). 
The solution of~Eq.~(\ref{eq:PM_n}) within SBA reduces to the solution of this equation within HFA by rescaling $E_{\rm F}^{\rm SBA}/z^2_{\rm P} = E_{\rm F}^{\rm HFA}$. 
In~the Appendix~\ref{appendix:PM_expansion}, we derive an asymptotic solution of Eq.~(\ref{eq:PM_n}) within HFA in the limit $\efsq\rightarrow 0$ for the square lattice [the Fermi level $\efsq = E^{\rm sq}_{\rm F} + 4\tau$ counted from the position of van Hove singularity ($-4\tau$) is chosen as a small parameter]: 
%Using the result of Appendix??? we get
\begin{equation}\label{eq:ef_vs_tau}
\tau(\efsq) \simeq \left[w_0(\efsq) + \gamma\left(w_0(\efsq)\right)(\efsq)^2\right]\efsq,
\end{equation}
where 
\begin{equation}\label{eq:w0_def}
	w_0(E) = \frac18\left(1 + \ln\frac{16}{E}\right)
\end{equation}
and 
\begin{equation}\label{eq:gamma_def}
\gamma(w) = \frac18\left(B_{\rm sq}w^3 - 4w^2 + 5w/12 - 5/576\right),
\end{equation}
where $B_{\rm sq} = 7.11$, see the~Appendix~\ref{appendix:PM_expansion}.  We state that even in the small $\tau$ regime there 
is no linear relation between $\tau$ and $\efsq$ due to logarithmic factors originating from van Hove singularity of the square-lattice DOS. 
The simplest estimate derived from Eq.(\ref{eq:ef_vs_tau}) and valid in the case of small $\tau$ and $\efsq$ is
\begin{equation}
\label{eq:ef_vs_tau_simple}
\tau(\efsq) \simeq w_0(\efsq)\efsq.
\end{equation}

\begin{figure}[h]
\includegraphics[angle=-90,width=0.49\textwidth]{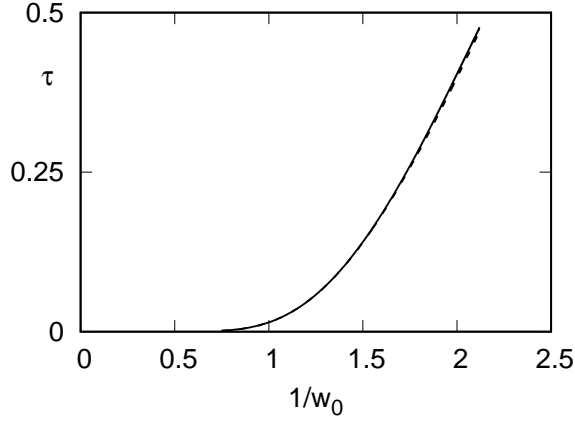}
\caption{
(Color online)
The dependence of $\tau$ as a function of $1/w_0(\efsq)$,  Eq.~(\ref{eq:ef_vs_tau}), is shown by solid line, a simple approximation (\ref{eq:ef_vs_tau_simple})  by dashed line.
}
\label{fig:w0}
\end{figure}
The relation of $w_0$ and $\tau$ is shown in~Fig.~\ref{fig:w0}. It is clear that the most relevant case corresponds to the interval $0.8 \lesssim 1/w_0 < 1.8$: below this interval $\tau$ tends to zero exponentially, above it $\tau$ is too large being beyond the case of small $\tau$ range under the scope in this work. The approximation (\ref{eq:ef_vs_tau_simple}) works very well in the relevant inverval $\tau\in (0,0.5)$. 
%Since $\delta\Phi_{2\rm P}(\tau) = -2\mathcal{E}$ 
We get from~the~Appendix~\ref{appendix:PM_expansion} %for the PM phase free energy
\begin{equation}\label{eq:PM_energy_final}
\delta\Phi^{\rm sq}_{2\rm P}(\tau) = -2\sum_{k = 2,4}\left(\efsq\right)^k\sum_{n = 0}^k a^{\rm sq}_{kn}w^n_0(\efsq),
\end{equation}
where
$a^{\rm sq}_{20} = -0.013$, 
$a^{\rm sq}_{21} = +0.20$,
$a^{\rm sq}_{22} = -0.54$,
$a^{\rm sq}_{40} = -0.00035$,
$a^{\rm sq}_{41} = +0.017$,
$a^{\rm sq}_{42} = -0.20$,
$a^{\rm sq}_{43} = +0.95$,
$a^{\rm sq}_{44} = -1.36$,
where an explicit calculation of these values is presented in~the~Appendix~\ref{appendix:PM_expansion}.

We state that $\delta\Phi^{\rm sq}_{2\rm P}(\tau)$ has a nonquadratic dependence on $\tau$: logarithmic corrections occur due to the presence of van Hove singularity in the vicinity of the Fermi level. 
To analyze the effect of van Hove singularity we express $\efsq$ through $\tau$ and get
$\efsq \approx \frac{\tau}{w_0(\efsq)}\left(1 - \frac{\gamma(w_0(\efsq))}{w^3_0(\efsq)}\tau^2\right)$. 
Substituting this in Eq.~(\ref{eq:PM_energy_final}) and using Eq.~(\ref{eq:delta_F_PM_HFA}) we obtain
%\begin{equation}\label{eq:PM_energy_final_another}
%\delta \mathcal{F}^{\rm sq}_{\rm PM}(\tau) = 2\sum_{k = 2,4}\tau^k\left(1 - k\frac{\gamma(\efsq)}{w^3_0(\efsq)}\tau^2\right)\sum_{n = 0}^k a_{kn}w^{n-k}_0(\efsq).
%\end{equation}
%\begin{equation}\label{eq:PM_energy_final_another}
%\delta \mathcal{F}^{\rm sq}_{\rm PM}(\tau) = 2\sum_{k = 2,4}\tau^k\sum_{n = 0}^k a_{kn}w^{n-k}_0(\efsq) - 4\tau^4\frac{\gamma(\efsq)}{w^3_0(\efsq)}\sum_{n = 0}^2 a_{2n}w^{n-2}_0(\efsq).
%\end{equation}
\begin{equation}\label{eq:delta_F_PM_coefs_intro}
	\delta F^{\rm HFA}_{\rm PM, sq, analytic}\left(\tau(\efsq)\right) = -f^{\rm eff, sq}_{2,\rm PM}\left(w_0(\efsq)\right)\tau^2(\efsq),
\end{equation}
where
\begin{equation}\label{eq:f2_PM_eff_def}
	f^{\rm eff, sq}_{2,\rm PM}\left(w_0(\efsq)\right) = f^{\rm sq}_{2,\rm PM}\left(w_0(\efsq)\right) + f^{\rm sq}_{4,\rm PM}\left(w_0(\efsq)\right)\tau^2(\efsq),
\end{equation}
\begin{eqnarray}\label{eq:f_PM_coefs.2}
f^{\rm sq}_{2, \rm PM}(w) &=& -2(a^{\rm sq}_{20}w^{-2} + a^{\rm sq}_{21}w^{-1} + a^{\rm sq}_{22}), \\
\label{eq:f_PM_coefs.4}
f^{\rm sq}_{4, \rm PM}(w) &=& -2\gamma(w)w^{-3}f^{\rm sq}_{2, \rm PM}(w) - 2(a^{\rm sq}_{40}w^{-4} + a^{\rm sq}_{41}w^{-3}  + a^{\rm sq}_{42}w^{-2} + a^{\rm sq}_{43}w^{-1} + a^{\rm sq}_{44})
\end{eqnarray}
are polynomials in  $w^{-1}$. The expansion of paramagnetic free energy in inverse logarithmic factor is a general characteristics for systems with van Hove singularity  (two-dimensional lattices). 

To designate the difference we make the expression of free energy expansion for sc case. 
In~the~Appendix~\ref{appendix:PM_expansion} an expansion of paramagnetic-phase free energy is derived, and it is shown that the leading contribution is
\begin{equation}\label{eq:delta_F_HFA_PM_sc}
\delta F^{\rm HFA}_{\rm PM, sc, analytic}(\tau) = - a^{\rm sc}_{2,\rm PM}\tau^2 + \mathcal{O}(\tau^4).
\end{equation}
Since the electron spectrum of sc lattice in paramagnetic phase has no  strong  van Hove singularities,  $\delta F^{\rm HFA}_{\rm PM, sc}(\tau)$ is an analytical function of $\tau$. 

The case of bcc lattice we expect some nonanalytic dependence of $\delta F^{\rm HFA}_{\rm PM, bcc}(\tau)$ caused by the closeness of heavy mass points to the Fermi surface, see discussion in~Sec.~\ref{sec:PM_treatment}. 
%An application of this expansion to analysis of energetical favourability of PM phase with respect to AFM insulator is presented in~Sec.~\ref{sec:analytic_expansion}.

Non we consider the expansion for the case of AFM insulator case based on 
Eq.~(\ref{eq:F_AFM_HFA_final}) and MIT line equation (\ref{eq:Delta_sq}). 

For the square lattice we derive an expansion of $\delta F^{\rm sq}_{\rm AFM}$ in powers of $\tau$ up to fourth order at $J = 0$ at MIT line, see 
Eq.~(\ref{eq:Delta_sq}), using the derived expansion in~Eq.~(\ref{eq:delta_F_HFA_sq_expansion}) 
%this was Eq. (43) in SM
within~the~Appendix~\ref{appendix:G_expansion},
\begin{multline}\label{eq:delta_F_AFM_asymp_square}
	\delta F^{\rm sq, HFA, analytic}_{\rm AFM}\left(\Delta^{\rm sq}_{\rm MIT}(\tau)\right) = \\
	-\frac{\tau^2}{2\pi^2}\left(2\ln\frac{16}{\tau} + 1\right) 
	 - \frac{\tau^4}{64\pi^2}\left(
		\ln^2\frac{16}{\tau} - \frac72\ln\frac{16}{\tau}  - 4\ln^22- 4\ln2 + \frac{25}8 + \frac{\pi^2}6
	  - 32\pi^2\delta g^{\rm sq}_2 
	\right),
\end{multline}
where $\delta g^{\rm sq}_2 = -1.7\cdot10^{-3}$.

Expressing here the logarithms through $w_0$ analogously to Eq.~(\ref{eq:delta_F_PM_coefs_intro}), we write down
\begin{equation}\label{eq:delta_F_AFM_coefs_intro}
\delta F^{\rm sq, HFA}_{\rm AFM}\left(\Delta^{\rm sq}_{\rm MIT}(\tau)\right) = -f^{\rm eff}_{2,\rm AFM}\left(w_0(\efsq)\right)\tau^2,
\end{equation}
where
\begin{equation}\label{eq:f2_AFM_eff_def}
f^{\rm eff, sq}_{2,\rm AFM}(w) = f^{\rm sq}_{2,\rm AFM}(w) + f^{\rm sq}_{4,\rm AFM}(w)\tau^2(w),
\end{equation}
\begin{eqnarray}\label{eq:f_AFM_coefs.2}
f^{\rm sq}_{2, \rm AFM}(w) &=& \frac{16w - 1 - 2\ln w}{2\pi^2}, \\
\label{eq:f_AFM_coefs.4}
f^{\rm sq}_{4, \rm AFM}(w) &=& \frac{1}{\pi^2}\left(-\gamma(w) w^{-3} + 
w^2  - \frac{w}{16}(11 + 4\ln w) + \frac{\ln^2 w}{64} +  \frac{11}{128}\ln w\right)  + \bar{f}^{\rm sq}_{4, \rm AFM},
\end{eqnarray}
where constant $\bar{f}^{\rm sq}_{4, \rm AFM} = \pi^{-2}\left({61}/{32} 
  - {\ln^22}- {\ln2}\right)/16  + 1/384 - g^{\rm sq}_2/2 = 8.1\cdot10^{-3}$, $\gamma(w)$ is defined by~Eq.~(\ref{eq:gamma_def}). Note that the presence of logarithms in ~Eqs.~(\ref{eq:f_AFM_coefs.2}) and~(\ref{eq:f_AFM_coefs.4}) is a consequence of the van Hove singularity in the spectrum.

\begin{figure}[h!]
\noindent
\includegraphics[angle=-90,width=0.45\textwidth]{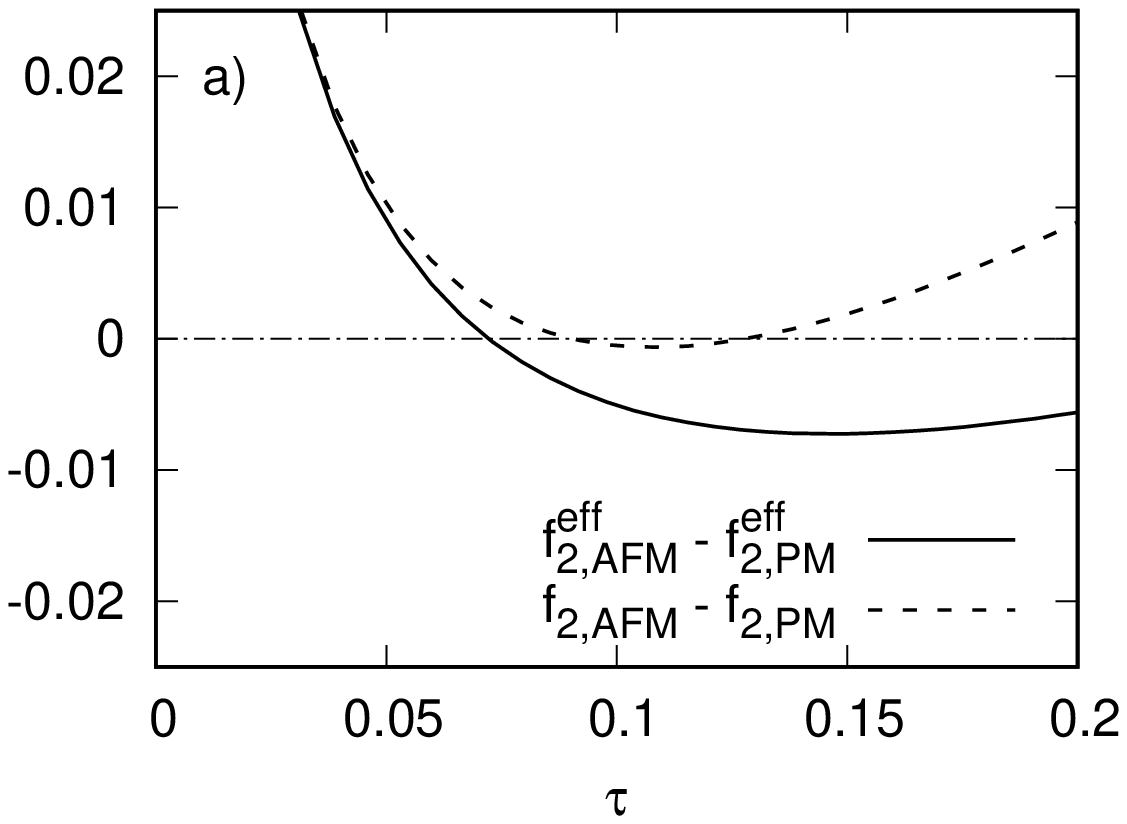}
\includegraphics[angle=-90,width=0.45\textwidth]{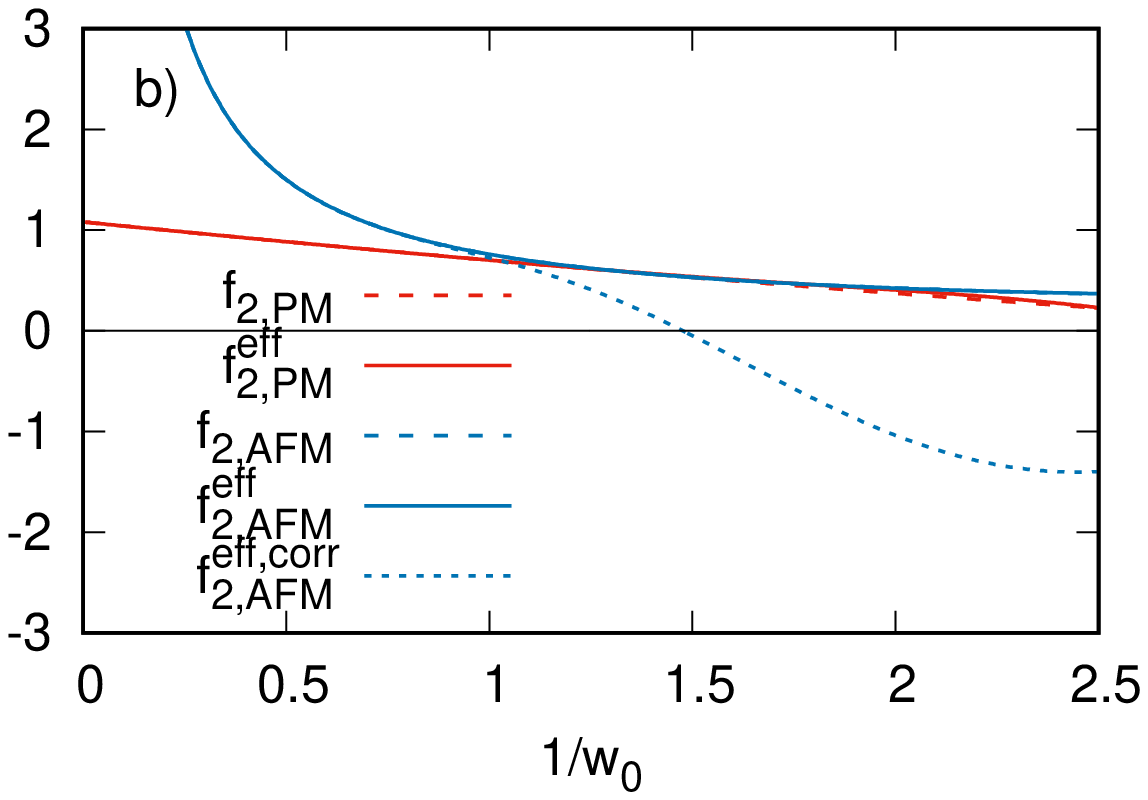}
\caption{
\label{fig:f_coeffs}
(Color online)
Free-energy expansion parameters for the square lattice. (a) The plot of difference of $\tau$-expansion coefficients of the free energy in AFM [$f^{\rm sq}_{2,\rm AFM}(w_0), f^{\rm sq, eff}_{2,\rm AFM}(w_0)$, see Eqs.~(\ref{eq:f2_AFM_eff_def}) and (\ref{eq:f_AFM_coefs.2})] and PM [$f^{\rm sq}_{2,\rm PM}(w_0), f^{\rm sq, eff}_{2,\rm PM}(w_0)$, see~Eqs.~(\ref{eq:f2_PM_eff_def}) and (\ref{eq:f_PM_coefs.2})] phases  as a function of $\tau$. 
(b) The plot of coefficients of $\tau$ expansion $f^{\rm sq}_{2,\rm AFM}(w_0)$, $f^{\rm sq, eff}_{2,\rm AFM}(w_0)$, $f^{\rm sq}_{2,\rm PM}(w_0)$, $f^{\rm sq, eff}_{2,\rm PM}(w_0)$ and $f^{\rm sq, corr}_{2,\rm AFM}(w_0)$ with account of correlation correction $f^{\rm sq, eff}_{4, \rm AFM} \rightarrow f^{\rm sq, corr}_{4, \rm AFM}$, see Eq.~(\ref{eq:f4^corr_def}), as a function of $w_0^{-1}$. 
%Green dotted line shows the giant many-electron renormalization of $f_{4,\rm AFM}$, the factor of $10^{-2}$ being introduced. 
}
\end{figure}

In Fig.~\ref{fig:f_coeffs}a the difference of second-order coefficients $f^{\rm sq}_{2,\rm PM}$ and $f^{\rm sq}_{2,\rm AFM}$ as~functions of~$\tau$ for~PM and~AFM phases without and with account of fourth-order corrections is shown. 
Vanishing of this difference occurs close to the MIT order change point. 
In the relevant region  $0.7\lesssim 1/w_0 \lesssim 2.0$, see Fig.~\ref{fig:w0}, the quadratic coefficients PM and AFM phases $f^{\rm sq}_{2, \rm PM}(w_0), f^{\rm sq}_{2, \rm AFM}(w_0)$ are very close, which explains the closeness of free energies of PM and AFM phases within HFA and implies that logarithmic dependence of the coefficients appears to be very important. 
From  Fig.~\ref{fig:f_coeffs}b one can see that in the relevant interval $f^{\rm sq}_{4,\rm AFM}$ yields only a very small correction to $f^{\rm sq}_{2,\rm AFM}\rightarrow f^{\rm sq, eff}_{2,\rm AFM}$, whereas $f_{4,\rm PM}$ is very important correction to $f^{\rm sq}_{2,\rm PM}\rightarrow f^{\rm sq, eff}_{2,\rm PM}$  removing the second artificial transition point~\cite{2019:Igoshev_JETP_MIT}~[see~Fig.~\ref{fig:f_coeffs}a]. 
%The divergence of $f_{2,\rm AFM}(w_0)$ and $f_{4,\rm AFM}(w_0)$ at small $1/w_0$ originates from vHS at $\tau = 0$ and provides the favoring of AFM phase at small $\tau$. 

%The plots of coefficients $f^{\rm sq}_{2, \rm PM}(w_0),f^{\rm sq}_{4, \rm PM}(w_0)$, see definitions (\ref{eq:f_PM_coefs.2}), (\ref{eq:f_PM_coefs.4}), as functions of $1/w_0$ are shown  in~Fig.~\ref{fig:f_coeffs}b. A domination of $a^{\rm sq}_{kn}$ with senior indices $n$ over $a^{\rm sq}_{kn}$ with elder indices for $k = 2,4$ implies that both $f^{\rm sq}_{2, \rm PM}(w_0)$ and $f^{\rm sq}_{4, \rm PM}(w_0)$ behave almost linearly as a function of $w_0^{-1}$. 
The plot of coefficient $f^{\rm sq}_{2, \rm PM}(w_0)$, see definition (\ref{eq:f_PM_coefs.2}), as functions of $1/w_0$ is shown  in~Fig.~\ref{fig:f_coeffs}b. A domination of $a^{\rm sq}_{2n}$ with senior indices $n$ over $a^{\rm sq}_{2n}$ with elder indices  implies that both $f^{\rm sq}_{2, \rm PM}(w_0)$  behaves almost linearly as a function of $w_0^{-1}$. 
Analogous conclusion is valid for the coefficient $f^{\rm sq}_{4, \rm PM}(w_0)$, see Eq.~(\ref{eq:f_PM_coefs.4}). 
We state that large value of $f^{\rm sq}_{2, \rm PM}(w_0)$ and  the dependencies of $f^{\rm sq}_{2, \rm PM}(w)$, $f^{\rm sq}_{4, \rm PM}(w_0)$ on the logarithmic scale $w_0$ is a direct consequence of the permanent presence of a van Hove singularity in DOS: in the relevant interval of $1/w_0$ (see above) these coefficients fall down by approximately two times. 

We see that logarithmic contributions originating from vHS into the PM phase free energy are of great importance and exhibit a great impact on the 
MIT order change point. 
At the same time, for the AFM phase the account of the second-order term only is sufficient in the relevant interval of $1/w_0$. 
%The effect of permanent presence of van Hove singularity in the electronic spectrum is extreme lowering of the paramagnetic state phase energy.[Ne mogu poyat', tak li eto???]
%the quadratic term of $\Delta F_{\rm MIT}$ is function of $w_0(\efsq)$ due to the presence of van Hove singularities and extremely small, which is concurrence,  and quartic terms as well as  in $f_0(x)$ are important. 

%%%% for sc: d_MIT = tau/3
For sc lattice at small $\tau$, using the Eq.~(\ref{eq:Delta_sc}), we get 
the expansion of $\delta F^{\rm sc, HFA}_{\rm AFM}$ by retaining the leading order contribution,
\begin{equation}\label{eq:delta_F_HFA_sc_leading_contrib}
   \delta F^{\rm sc, HFA, analytic}_{\rm AFM}\left(\Delta^{\rm sq}_{\rm MIT}(\tau)\right) = -a^{\rm sc}_0\tau^2/3,
\end{equation}
$a^{\rm sc}_0 = 0.86$, see Eq.~(\ref{eq:delta_F_AFM_HFA_sc_expansion}) 
%this was Eq. (69)
in~the~Appendix~\ref{appendix:G_expansion}. 
%Eq.~(\ref{eq:delta_F_AFM_HFA_sc_expansion}).  %%%% Reference to Supplemental Material  
%\cite{eq:delta_F_AFM_HFA_sc_expansion}.  %%%% Reference to Supplemental Material  
In contrast to the case of the square lattice, Eq.~(\ref{eq:delta_F_AFM_asymp_square}), we see that leading contribution has no  logarithmic factors, which is directly caused by the regular DOS at $\tau = 0$ in the vicinity of $\epsilon = 0$. Therefore the order of MIT cannot change within small coupling regime (at small $\tau$). 
However, the subleading contribution acquires logarithmic factor despite the fact of the absence of any singularity, 
%Eq.~(\ref{eq:delta_F_AFM_HFA_sc_expansion}) in~Supplemental Material. 
see Eq.~(\ref{eq:delta_F_AFM_HFA_sc_expansion}) 
% this was Eq. (69)
in~the~Appendix~\ref{appendix:PM_expansion}. 
The latter is a common feature.

At the same time, singular  at $\tau = 0$ DOS for bcc lattice results in nonanalytical expansion with logarithmically dependent coefficients of $\tau$ expansion similarly to the square lattice case
%%%%%%%%% bcc: d_MIT = tau/2 
\begin{equation}\label{eq:delta_F_AFM_HFA_bcc_leading_contr}
    \delta F^{\rm bcc, HFA, analytic}_{\rm AFM}\left(\Delta^{\rm bcc}_{\rm MIT}(\tau)\right) = -\frac{2\tau^2}{\pi^3}\left(\ln^2\frac{32}{\tau} + \ln\frac{32}{\tau} + \frac12-\frac{\pi^2}{12}\right),
\end{equation}
see Eq.~(\ref{eq:delta_F_AFM_HFA_bcc_expansion}) 
% this was Eq. (85)
within~the~Appendix~\ref{appendix:G_expansion}.  
%Eq.~(\ref{eq:delta_F_AFM_HFA_bcc_expansion}).    %%%% Reference to Supplemental Material  

For the free energy of PM phase of bcc lattice, we expect substantial logarithmic dependence of $\tau$-expansion coefficients.
%, which should be considered in perspective. 
However, the large value of  $\delta F^{\rm bcc, HFA}_{\rm AFM}\left(\Delta^{\rm bcc}_{\rm MIT}(\tau)\right)$ caused by $\ln^2$-singularity is expected to result in $|\delta F^{\rm bcc, HFA}_{\rm PM}\left(\Delta^{\rm bcc}(\tau)\right)| \ll  |\delta F^{\rm HFA, bcc}_{\rm AFM}\left(\Delta^{\rm bcc}_{\rm MIT}(\tau)\right)|$, which is confirmed by numerical calculations.

From general point of view of phase transition theory, the criterion of 
vanishing $\Delta F_{\rm MIT}(\Delta_\ast)$, which is some polynomial expression with respect to $\Delta_\ast$ (or, in general, spectrum parameters), determining the line of MIT order change is similar to standard statement in Landau theory where 
the equilibrium value of the order parameter is determined by the balance 
between the second- and fourth-order terms. 
However, the balance in this case is determined by the contributions to energy, which contain logarithmic factors, and holds in the region of much 
smaller parameters. 
The expansion coefficients in this case are determined by the coefficients of the singular contributions to the density of states. However, it should be emphasized that $\Delta_\ast$ is not an order parameter since the criterion of transition from the AFM insulator state is generally not determined by vanishing of $\Delta_\ast$ (the role of the order parameter must be played by the spectral weight of quasiparticles in the metal state).

%%%%%%%% Correlation effects in terms of expansion %%%%%%
We consider now an account of correlation effects within SBA modifies the 
expansion found. Despite the fact that correlation correction affects both (\ref{eq:delta_F_AFM}) and~(\ref{eq:delta_F_PM}), it is convenient, with some degree of convention, to cast the correction (\ref{eq:Fc_est}) in the leading order to $\Delta F_{\rm MIT}$  into coefficients of HFA AFM phase  expansion~(\ref{eq:delta_F_AFM_coefs_intro}). 
From Eq.~(\ref{eq:Fc_est}) it is clear that the correlation correction starts from fourth-order coefficient, like Eq.~(\ref{eq:f_AFM_coefs.4}). % of~HFA expansion~(\ref{eq:delta_F_AFM_coefs_intro})\textbf{[???]} by powers of $\Delta$. 
Using the relation between $\tau$ and $\Delta_\ast$ on the MIT line (\ref{eq:MIT_line_at_small_tau}), we get for the subleading term [we have replaced $U$ by its leading term $2/\bar{\Phi}_1\left(\Delta_{\rm MIT}(\tau)\right)$] for arbitrary 
lattice $f_{4, \rm AFM} \rightarrow f^{\rm corr}_{4, \rm AFM}$, where
%It is instructive to consider the renormalization of coefficients of the expansion,     
\begin{equation}\label{eq:f4^corr_def}
	f^{\rm corr}_{4, \rm AFM} = f_{4,\rm AFM} - \frac{\kappa_{\rm MIT}^4}{2U_{\rm BR}}\frac{\left(1 - u_{\rm MIT}(\tau)\right)\left(3 + u_{\rm MIT}(\tau)\right)}{\left(1 +  u_{\rm MIT}(\tau)\right)^4}\bar{\Phi}^2_1\left(\Delta_{\rm MIT}(\tau)\right),
\end{equation}
where $u_{\rm MIT}(\tau) = 2/[U_{\rm BR}\bar{\Phi}_1\left(\Delta_{\rm MIT}(\tau)\right)]$, 
making it in general negative, which enhances the tendency towards first order MIT. Being proportional to $\Phi_1^2(\Delta_{\rm MIT})$ this correction term is strongly sensitive to the type of van Hove singularity.
For the square lattice we get with the use of Eq.~(\ref{eq:w0_def}), see Eq.~(\ref{eq:Phi1_sq_expansion}) 
%this was Eq. (42) in SM
within~the~Appendix~\ref{appendix:G_expansion}
\begin{equation}\label{eq:Phi1_sq_leading}
\bar{\Phi}^{\rm sq}_1(w_0) = \frac1{2\pi^2}\left(
64w_0^2 - 16w_0 + \frac78 + \frac{\ln 2}{2} - 4\ln^22
\right)
 + \frac1{12} + \delta g_0^{\rm sq}.
\end{equation}
%Consider as an expansion an unstable situation for the square lattice: using the below derived leading order expression for $u$ by~Eq.~(\ref{eq:Uc_expansion}), and using Eq.~(\ref{eq:U_HFA_square}) we get an analytical expression for $f_{4,\rm corr, AFM}$, which application is shown in~Fig.~\ref{fig:f_coeffs}b. 
We find that for the square lattice the leading coefficient $f^{\rm sq}_{4,\rm AFM}$ in the free-energy expansion for the AFM phase suffers giant renormalization strongly affecting $f^{\rm sq,eff,corr}_{2, \rm AFM}$ and changing the order of transition, which results in lifting the degeneracy with the corresponding coefficient for the PM phase, see Fig.~\ref{fig:f_coeffs}b. 
This correction  modifies the order of MIT due to the presence of accidental degeneracy within HFA approximation. 
Such strong renormalization substantially shifts the $\tau$ point of MIT order change. 
This circumstance yields the explanation of the change of MIT order at small $\tau$, which is impossible for the system without any van Hove singularities.  
%Below it is shown that the account of many-electron correction within SBA results in giant renormalization of $f^{\rm sq}_{4, \rm AFM}$. 

%\begin{wrapfigure}[25]{l}{0.45\textwidth}
\begin{figure}[h]
\includegraphics[angle=-90,width=0.44\textwidth]{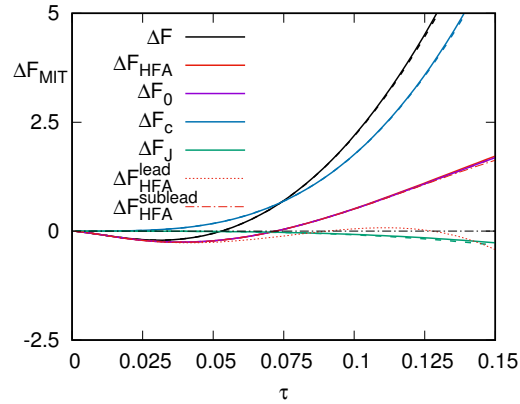}
\caption{
(Color online)
Plots of different contributions to $\Delta F_{\rm MIT}$ (in units of $10^{-4}t$) at $J_{\qq} = 0.5z^2_{\rm A}$ for the square lattice as a function of $\tau$. Two way of calculations is used: Eq.~(\ref{eq:final_delta_F_MIT}), its terms being shown by~solid lines, and Eq.~(\ref{eq:delta_F_MIT_est}), its terms being shown by~dashed lines. The following contributions are shown: $\Delta F_0$ corresponds to first term (originating from HFA), $\Delta F_{\rm c}$  to second term (corresponding to many-electron corrections to HFA), $\Delta F_J$  to third term (originating from many-electron corrections induced by exchange). $\Delta F^{\rm HFA,lead}_{\rm MIT}(\tau)$ is shown by orange dotted line,  $\Delta F^{\rm HFA,lead}_{\rm 
MIT}(\tau)$ is shown by red dotted line, $\Delta F^{\rm sq,HFA}_{\rm MIT}$ is shown by dash-dot-dotted line.  
}
\label{fig:dF_MIT_square}
\end{figure}

It is instructive to apply the estimation (\ref{eq:delta_F_MIT_est}) to the case of square lattice at small~$\tau$, i.e.~in small coupling regime: 
different contributions to $\Delta F_{\rm MIT}$ within the exact, Eq.~(\ref{eq:final_delta_F_MIT}), and approximate splitting, Eq.~(\ref{eq:delta_F_MIT_est}), as well as  total free energy difference as functions of $\tau$ are shown in~Fig.~\ref{fig:dF_MIT_square}. We also show different analytical approximations for $\Delta F^{\rm HFA}_{\rm MIT}$, $\delta F^{\rm HFA,sublead}_{\rm MIT}(\tau) = \delta F^{\rm sq,HFA,analytic}_{\rm AFM}(\tau) - 
\delta F^{\rm sq,HFA,analytic}_{\rm PM}(\tau)$ and $\Delta F^{\rm HFA,sublead}_{\rm MIT}(\tau) = \delta F^{\rm sq,HFA,analytic}_{\rm AFM}(\tau) - \delta F^{\rm sq,HFA,analytic}_{\rm PM}(\tau)$ and its version $\Delta F^{\rm HFA,lead}_{\rm MIT}(\tau)$ with $f_4 = 0$, see Eqs.~(\ref{eq:delta_F_PM_coefs_intro}) and (\ref{eq:delta_F_AFM_coefs_intro}), of second and fourth orders with respect to $\tau$. 
In the case (i) the estimation (\ref{eq:delta_F_MIT_est}) agrees well with exact expression, Eq.~(\ref{eq:final_delta_F_MIT}); (ii) many-electron correction is much larger than the exchange one; and (iii) quadratic (with respect to $\tau$) approximation for $\delta F^{\rm HFA,lead}_{\rm MIT}(\tau)$ cannot reproduce the result of numerical calculation but the inclusion of fourth-order terms, i.e., using $\delta F^{\rm HFA,sublead}_{\rm MIT}(\tau)$ cures the problem. This fact is connected with inapplicability of quadratic approximation (\ref{eq:delta_F_PM_coefs_intro}) for paramagnetic phase $\delta F^{\rm sq,HFA}_{\rm PM}$ due to van Hove singularities. 
The exchange contribution  can increase or decrease $\Delta F_{\rm MIT}$ depending on the sign of $J_{\bf Q}$. We also find that $|\Delta F^{J}_{\rm MIT}|\ll \Delta F^{\rm c}_{\rm MIT}$, which supports the conclusion that many-electron correction term protects the sign of $\Delta F_{\rm MIT}$ from exchange interaction effects (in the regime of weak coupling). 

Furthermore, $|\Delta F^0_{\rm MIT}|$ becomes smaller than $\Delta F^{J}_{\rm MIT}$ at $\tau \gtrsim 0.05 $ and demonstrate an inflection point located at $\tau\approx 0.01$ which is closely connected with logarithmic contributions to the free energy of AFM insulator and PM metal phases. 

Thus, in the context of the MIT order issue, the many-electron corrections to the free energy are crucial and lift the degeneracy induced by two-dimensional vHS for the square lattice. 

\section{Conclusions}

We have revisited in detail the MIT picture connected with the transition 
into AFM phase in the ground state of the nondegenerate Hubbard model within both HFA and SBA. 
Especially interesting is the situation for the square lattice where a beautiful mathematics can be built. 
A simple analytical theory of competition of AFM insulator, AFM metal and 
PM metal phases as a function of crucial next-nearest-neighbor hopping integral is developed. We developed the analytical expansion of the the free energy for PM and AFM phases in the next-nearest-neighbor transfer integral $t'$ and in direct antiferromagnetic gap $\Delta$, respectively. This expansion yields subleading-order nonanalytic contributions in the AFM phase for all the lattices considered. For sc and bcc lattices with strong 
van Hove singularities the nonanalyticity occurs in the leading order.

We highlight a long-standing issue of closeness energy of all these phases for the square lattice at the MIT line in weak-coupling limit. 
We revealed that even in the simplest Hartree-Fock approximation (valid in the latter regime) van Hove singularity of the density of states produces a logarithmic dependence of the coefficients of the expansion of the free energy with respect to $\tau$, which, in turn, produces an  inflection point at very small $\tau$. 
This corollary is a result of a balance of a competition of phase energy of AFM insulator and PM metal phases which both suffer a strong influence 
of van Hove singularity but in very different way: in the limit of small $\tau$ AFM insulator phase generally wins stabilizing the second order MIT, but small increment of $\tau$ makes the paramagnetic phase more energetically favorable, and besides the leading-order expansion cannot reproduce correctly the behaviour of the PM free energy.

Thus for the square lattice we have a first-order transition in a wide parameter region.  
An expansion of the energy in powers of nnn hopping  allows to highlight accidental nature of the energy closeness for the square lattice. 
On the other hand, for the sc lattice, the second-order transition from the AFM insulator to the AFM metal takes place.
For the bcc lattice, the situation is similar, in spite of the presence of van Hove singularities: the second-order transitions occur, since the stability boundary of the insulator AFM phase relative to the PM phase in the  $t'-\Delta$ variables, although being non-linear 
(as well as for a square lattice), does not intersect the line of transition to the AFM metal. 

A possibility of SBA to take into account the difference between the nature of singly and doubly occupied states (which is especially important in the PM phase since its energy is lowered in the Kotliar-Ruckenstein approach) allows highlight such delicate effects as a influence of exchange interaction which is fully inaccessible within HFA as well as a going beyond small coupling regime. 
We obtained that generally the exchange interaction contributes into the relative energy of phases in fourth order with respect to order parameter, 
i.e., is subtle and delicate effect. Nonetheless, it is substantial since the effect is fully missed within the Hartree-Fock approximation and especially important for the case of the square lattice demonstrating the accidental degeneracy of PM metal and AFM insulator energies. 
We found two regimes of MIT within Slater scenario: The first one is a weak-coupling case when the exchange interaction effect is shielded by main 
correlation contribution, so it only weakens changes in the MIT transition line (pure Slater scenario); the second regime corresponds to strong-coupling case when the contribution of exchange energy (being enhanced due to van Hove singularity presence) surpasses the contribution of  main correlation term changing the MIT order (Slater-Heisenberg scenario). 

An investigation of MIT  for bipartite  (sc and bcc) lattices shows that MIT order can be changed to first one in the ``Slater-Heisenberg'' scenario: The AFM insulator phase looks like to be formed by pure local moments, the Hubbard interaction is sufficiently large, and exchange and Coulomb 
energies are comparable, which implies that the competing paramagnetic metal phase is not so far from the Mott transition. Within this scenario, the AFM metal phase is unfavorable since the energy loss due to large intersite ferromagnetic exchange interaction appears to be larger than the gain  reduced by correlation effects: the first-order MIT occurs from AFM insulator to PM metal phase. This correlation effect (i.e., the possibility of ferromagnetic exchange interaction to change the order of MIT transition from the second at $J = 0$ to the first at $J\ne 0$, which occurs only in the regime of strong correlations) was obtained previously within DMFT approximation for Bethe lattice~\cite{1999:Chitra}. However, this result was questioned by some later investigations~\cite{2004:Zitzler} where the antiferromagnetic metal phase was found to be unstable.  

We find that the MIT picture depends strongly  on the lattice geometry through the density of electron states: 
topologically caused van Hove singularity for the square lattice (being stable   at arbitrary values of spectrum parameters) results in strong lowering of the energy of paramagnetic phase. 
It is found for the case of accidental degeneracy that leading quadratic approximation for the free energy is not applicable even qualitatively except for only a very small vicinity of $\tau = 0$, but accounting of total dependence on logarithms of subleading terms allows to obtain a good agreement between numerical and analytical result. 
For other lattices having  van Hove singularities of different strength  (sc and  bcc), the expansion of critical $U$ for MIT transition in the antiferromagnetic state with respect to $\tau$ was developed. 
Analogously, we found a convenient basis of an expansion for AFM insulator phase properties, which is based on full account of  singularities of the density of states.

The results obtained solve the problem of investigation of MIT in a wide parameter region and can be applicable for  compounds like vanadium oxides and other systems, which are described by the Slater MIT scenario or are close to it (see Introduction). However, consideration of nonbipartite lattices and  more complicated magnetic  orderings with different wavevectors  can be required (cf. the treatment  for the fcc lattice in the standard Hubbard model \cite{2016:FCC-Timirgazin}).

\section{Acknowledgments}
The authors are grateful to M.~I.~Katsnelson, M.~A. Timirgazin,
Yu.~N. Skryabin, and A.O. Anokhin for fruitful discussions. 
This work was performed under the State assignment
of the Ministry of Education of the Russian Federation (project ``Quantum'' No.~AAAA-A18-118020190095-4).

\appendix
%\section{Supplemental material for: Metal--insulator transition and antiferromagnetism  in the generalized Hubbard model: a treatment of correlation effects}
\section{Expansion of $G(\Delta)$ at small $\Delta$}\label{appendix:G_expansion}
We consider the expansion of 
%Eq.~(\ref{eq:G_def}) %%%% Reference to main text
Eq.~(\ref{eq:G_def}) %%%% Reference to main text {eq:G_def}
of the main text in powers of $\Delta$, which can be presented through the density of states of the spectrum at $\tau = 0$, $\rho(\varepsilon) \equiv \rho(\varepsilon, \tau = 0)$
\begin{equation}\label{eq:G_expr}
	G(\Delta) = \int_{0}^{+D} \frac{d\varepsilon\,\rho(\varepsilon)}{|\varepsilon| + \sqrt{\Delta^2 + \varepsilon^2}},
\end{equation} 
where the parity of $\rho(\varepsilon)$ for bipartite lattice is taken into account. 
Since direct setting $\Delta = 0$ results in diverging integrals, we conclude that the parameter $\Delta$  cuts the divergence, and leading terms are determined by the contributions of small $\varepsilon$. 
We take $D$ as an energy unit, $\epsilon = \varepsilon/D, d = \Delta/D$,  up to the end of the  Appendix: $G(\Delta) = D^{-1}\mathcal{G}(\Delta/D)$, where
\begin{equation}
	\mathcal{G}(d) = \int_{0}^{+1} \frac{d\epsilon\,\rho_D(\epsilon)}{\epsilon + \sqrt{d^2 + \epsilon^2}},
\end{equation}
where $\rho_D(\epsilon) = D\rho(D\epsilon)$ is rescaled density of states.

We can introduce a ``logarithmic index'' $n_{\rm l}$ for a lattice as a maximal power of logarithms entering  the expansion of the DOS in the vicinity of $\varepsilon = 0$. 
We consider three cases of the DOS behaviour at small $\varepsilon$: (i) double-logarithmic behaviour, $n_{\rm l} = 2$  (an example is bcc lattice), (ii) logarithmic behaviour, $n_{\rm l} = 1$ (an example is square lattice), (iii) regular (analytic) behaviour, $n_{\rm l} = 0$ (an example is sc lattice)~\cite{1969:Jelitto}.  As we will see below, all these case are captured by the following general representation:
\begin{equation}\label{eq:G_general_form}
	\mathcal{G}(d) = \sum_{n = 0}^{n_{\rm l} + 1}a_n\ln^n\frac2{d} + d^2\sum_{n = 0}^{n_{\rm l} + 1}b_n\ln^n\frac2{d} + \mathcal{O}\left(d^4\ln^{n_{\rm l} + 1}\frac2{d}\right),
\end{equation}
where the coefficients $a_n, b_n$ are specified by a concrete lattice.
%\begin{equation}
%	d\cdot\mathcal{G}'(d) = -\sum_{n = 0}^{n_{\rm l}}(n + 1)a_{n + 1}\ln^{n}\frac2{d} + d^2\left(\sum_{n = 0}^{n_{\rm l} + 1}\left(2b_n - (1 - \delta_{n,n_{\rm l}+1})(n + 1)b_{n + 1}\right)\ln^{n}\frac2{d} \right) ,
%\end{equation}

%From the~representations~(\ref{eq:Phi1_in_terms_of_G}),(\ref{eq:Phi2_in_terms_of_G}), %%%% Reference to main text
From the~representations~(\ref{eq:Phi1_in_terms_of_G}),(\ref{eq:Phi2_in_terms_of_G}), %%%% Reference to main text, {eq:Phi1_in_terms_of_G}, {eq:Phi2_in_terms_of_G}
and asymptotic form (\ref{eq:G_general_form}) we get the expansions
\begin{equation}\label{eq:Phi1_through_G}
\Phi_1(d\cdot D) = %2(2\mathcal{G}(d) + d\mathcal{G}'(d)) = 
\frac2{D}\left(\sum_{n = 0}^{n_{\rm l} + 1}(2a_n - (n + 1)a_{n + 1})\ln^n\frac2{d} \\+ d^2\sum_{n = 0}^{n_{\rm l} + 1}\left(4b_n - (n + 1)b_{n + 1}\right)\ln^{n}\frac2{d}\right) + \mathcal{O}\left(d^4\ln^{n_{\rm l} + 1}\frac2{d}\right), 
\end{equation}
\begin{equation}\label{eq:Phi2_through_G}
\delta\Phi_2(d\cdot D) = %-2d^2(\mathcal{G}(d) + d\mathcal{G}'(d)) =
-2D\left(d^2\sum_{n = 0}^{n_{\rm l} + 1}(a_n - (n + 1)a_{n + 1})\ln^n\frac2{d} \\
+ d^4\sum_{n = 0}^{n_{\rm l} + 1}\left(3b_n - (n + 1)b_{n + 1}\right)\ln^{n}\frac2{d}\right) + \mathcal{O}\left(d^6\ln^{n_{\rm l} + 1}\frac2{d}\right),
\end{equation}
where we define $a_{n_{\rm l} + 2} = b_{n_{\rm l} + 2} = 0$. 
%From Eq.~(\ref{eq:F_AFM_HFA_final}) %%%% Reference to main text
From Eq.~(\ref{eq:F_AFM_HFA_final}) %%%% Reference to main text, {eq:F_AFM_HFA_final}
and~asymptotic form (\ref{eq:G_general_form}) we obtain the asymptotics of HFA free energy
\begin{equation}\label{eq:delta_F_HFA_through_G}
    \delta F^{\rm HFA}_{\rm AFM}(d\cdot D) = -D\left(
    d^2\sum_{n = 0}^{n_{\rm l}}(n + 1)a_{n+1}\ln^n\frac2{d} + d^4\sum_{n = 0}^{n_{\rm l} + 1}((n + 1)b_{n+1} - 2b_n)\ln^n\frac2{d} + \mathcal{O}\left(d^6\ln^{n_{\rm l} + 1}\frac2{d}\right)
    \right).
\end{equation}
We see that the expansion of free energy of AFM insulator state always contains singular contributions, but their role dramatically depends on the presence of singularity of DOS at $\tau = 0$: if $\rho(\epsilon)$ is analytic in the vicinity of $\epsilon = 0$, non-analytic contribution enters starting from subleading terms, whereas for singular $\rho(\epsilon)$ already the leading contribution is singular.  

Since at small $d$ main contribution originates from small $\epsilon$, we separate from the $\rho_D(\epsilon)$, which is an even function of $\epsilon$, 
the leading contribution in the vicinity of $\epsilon = 0$:
%We set 
\begin{equation}
	\rho_D(\epsilon) = \rho_{D,\rm s}(\epsilon) + \rho_{D,\rm quad}(\epsilon) + \delta\rho_D(\epsilon),
\end{equation}
where $\rho_{D,\rm s}(\epsilon)$ is leading (double-logarithmic, logarithmic or constant contribution), $\rho_{D,\rm quad}(\epsilon)$ is subleading contribution $\sim \epsilon^2\ln^{n_{\rm l}}(1/\epsilon)$, and $\delta\rho(\epsilon) \sim \epsilon^4\ln^{n_{\rm l}}(1/\epsilon)$ is rest contribution which therefore yields only small corrections. 
\subsection{Asymptotics for $\mathcal{G}(d)$}\label{appendix:G_deriv}
Here we derive a general asymptotics for $\mathcal{G}(d)$ for different forms of the density of states up to $d^2\ln^{n_{\rm l} + 1}\frac1{d}$. 
We determine the singular contributions at $d \rightarrow 0$ and regular contributions up to $o(d)$: $\mathcal{G}(d) = \mathcal{G}_{\rm s}(d) + \mathcal{G}_{\rm quad}(d) +   \delta\mathcal{G}(d)$, where
\begin{equation}\label{eq:G_exp_start}
	\begin{pmatrix}\mathcal{G}_{\text{s}}(d)\\
	\mathcal{G}_{\text{quad}}(d)\\ \delta\mathcal{G}(d)\end{pmatrix} =  \int\limits_0^1 \begin{pmatrix}\rho_{D,\rm s}(\epsilon) \\ 
	\rho_{D,\rm quad}(\epsilon) \\ 
	\delta\rho_D(\epsilon)\end{pmatrix}\frac{d\epsilon}{\sqrt{\epsilon^2 + d^2} + \epsilon}.
\end{equation}
Below we use the identity 
\begin{equation}\label{eq:int_log}
	\int\limits_0^1 dx\,\frac{\ln^n(1/x)}{x + a} = -n! \text{Li}_{n+1}(-1/a),\; a > 0,
\end{equation}
where $\text{Li}_{n}$ is polylogarithm function. 
An idea of derivation of an asymptotic integrals is to choose a simple function having a similar behaviour of the integrand in the regions giving  main contribution to the integral:
%Идея оценки асимптотики интегралов --- выбрать функцию, имитирующую критическое поведение знаменателя в областях, дающих наибольший вклад в интеграл.
%Мы имеем две существенных области:
%, для которых интеграл имеет ??значительное поведение?? содрежит существенный вклад??: область
$\epsilon \ll d$ and  $\epsilon \gg d$:
\begin{equation}
	\sqrt{\epsilon^2 + d^2} + \epsilon \rightarrow 2\epsilon + d.
\end{equation}

We investigate the contribution of double-logarithmic singularity~(dl)$\rho_{D,\rm s}(\epsilon) = \ln^2(1/\epsilon)$:
%Исследуем сначала сингулярный вклад от двойной логарифмической~(dl) особенности $\rho_{D,\rm s}(\epsilon) = \ln^2(1/\epsilon)$:
\begin{equation}\label{eq:G_dl(d)_def}
	\mathcal{G}_{\rm dl}(d) \equiv \int\limits_0^1 d\epsilon\, \frac{\ln^2\left(1/\epsilon\right)}{\sqrt{\epsilon^2 + d^2} + \epsilon} \rightarrow  \bar{\mathcal{G}}_{\rm dl}(d) \equiv \int\limits_0^1 d\epsilon\,\frac{\ln^2\left(1/\epsilon\right)}{2\epsilon + d} = -\mathrm{Li}_3(-2/d).
\end{equation}
Using  Eq.~(\ref{eq:int_log}) and expanding $\mathrm{Li}_3(-2/d)$ at small $d$ we obtain
%\begin{equation}
%	\int\limits_0^1 d\epsilon\,\frac{\ln^2\left[1/\epsilon\right]}{2\epsilon + d} =  -\text{Li}_3\left[-2/d\right].
%\end{equation}
%Разложим это выражение для членов порядка $o(d)$.
%Имеем
%\begin{equation}
%	\text{Li}_3\left[-1/z\right] =  -\frac16\ln^3[1/z]-\frac{\pi^2}6\ln[1/z] + o(1), z\rightarrow 0^+,
%\end{equation}
%\begin{eqnarray}
%	\ln(1 + 1/z) &=& \ln[1/z] + o(1),\\
%	\text{Li}_2\left[-1/z\right] &=&  -\frac12\ln^2[1/z]-\frac{\pi^2}6 + o(1),\\
%	\text{Li}_3\left[-1/z\right] &=&  -\frac16\ln^3[1/z]-\frac{\pi^2}6\ln[1/z] + o(1).
%\end{eqnarray}
\begin{equation}\label{eq:asymp_bar_dl}
\bar{\mathcal{G}}_{\rm dl}(d) = \frac16\left(\ln^3\frac{2}{d} + \pi^2\ln\frac{2}{d}\right) + d/2 - d^2/32+ o(d^2).
\end{equation}
Introducing the variable change $x = \epsilon/d$ we write the difference of $\mathcal{G}_{\rm dl}(d)$ and $\bar{\mathcal{G}}_{\rm dl}(d)$ as 
\begin{equation}
	\mathcal{G}_{\rm dl}(d) - \bar{\mathcal{G}}_{\rm dl}(d) = \int\limits_0^{1/d} dx\, \ln^2\frac{1}{xd}\left(\frac1{\sqrt{x^2 + 1} + x} - \frac1{2x + 1}\right).
\end{equation}
Extending the upper integral limit to infinity 
\begin{equation}
\int\limits_0^{+\infty} dx\, \ln^2\frac{1}{xd}\left(\frac1{\sqrt{x^2 + 1} + x} - \frac1{2x + 1}\right) = \frac{1}4\left(\ln^2\frac2{d} + \left(1-\frac{\pi^2}3\right)\ln\frac2{d} + \left(\frac12 + \zeta(3) + \frac{\pi^2}6\right)\right)
\end{equation}
and estimating an error introduced thereby,  %expanding the integrand of the introduced tail
\begin{equation}
\int\limits_{1/d}^\infty dx\, \ln^2\frac{1}{xd}\left(\frac1{\sqrt{x^2 + 1} + x} - \frac1{2x + 1}\right) = \int\limits_{1/d}^\infty \frac{dx(1 - 1/x + \ldots)}{4x^2}\, \ln^2\frac{1}{xd} = \frac{d(1 - d/8)}2 + o(d^2),
\end{equation}
we have the result 
\begin{equation}\label{eq:diff_dl}
	\mathcal{G}_{\rm dl}(d) - \bar{\mathcal{G}}_{\rm dl}(d) = \frac{1}4\left(\ln^2\frac{2}{d} + \left(1-\frac{\pi^2}3\right)\ln\frac{2}{d} + \left(\frac12 + \zeta(3) + \frac{\pi^2}6\right)\right) - \frac{d(1 - d/8)}2 + o(d^2).
\end{equation}
%Перепишем это выражение по степеням $\ln[2/d]$:
%$$
%	\mathcal{G}_{\rm dl}(d) - \bar{\mathcal{G}}_{\rm dl}(d) = \frac{1}4\left(\ln^2\left[{2}/{d}\right] + 2b^{\rm dl}_1\ln\left[{2}/{d}\right] + b^{\rm dl}_2\right) + o(1).
%$$
Summing the expressions (\ref{eq:asymp_bar_dl}) and (\ref{eq:diff_dl}) we get
\begin{equation}\label{eq:G_asymp_dl}
	\mathcal{G}_{\rm dl}(d) = \frac{1}6\left(\ln^3\frac2{d} + \frac32\ln^2\frac2{d} + \frac12\left(\pi^2 + 3\right) \ln\frac2{d}  + \frac32\left(\frac12 + \zeta(3) + \frac{\pi^2}6\right)\right) + d^2/32 + o(d^2).
\end{equation}
%Запишем полную энергию как функцию параметра $d$
%\begin{equation}
%	\mathcal{F}(\Delta) - \mathcal{F}(0) =  -W d^2\mathcal{G}(d).
%\end{equation}
%\item $\mathcal{G}_{\rm l}:$

Now we consider the contribution from the logarithmic~(l) singularity 
%Рассмотрим теперь сингулярный вклад от логарифмической~(l) особенности 
$\rho_{D,\rm s}(\epsilon) = \ln(1/\epsilon)$:
\begin{equation}\label{eq:G_l(d)_def}
	\mathcal{G}_{\rm l}(d) = \int\limits_0^1 d\epsilon\, \frac{\ln\left(1/\epsilon\right)}{\sqrt{\epsilon^2 + d^2} + \epsilon} \rightarrow  \bar{\mathcal{G}}_{\rm l}(d) = \int\limits_0^1 d\epsilon\,\frac{\ln\left(1/\epsilon\right)}{2\epsilon + d} = -\frac12\text{Li}_{2}(-2/d).
\end{equation}
%Согласно (\ref{eq:int_log}),
%\begin{equation}
%	\bar{\mathcal{G}}_{\rm l}(d) = -\frac12\text{Li}_2\left(-{2}/{d}\right).
%\end{equation}
%Используя тождество Ландена
%\begin{equation}
%	\text{Li}_2(z) + \text{Li}_2(1/z) = -\frac{\pi^2}6 - \frac12\ln^2(-z).
%\end{equation}
%и разложение
%\begin{equation}
%	\text{Li}_2(z) = \sum_{k = 1}^\infty \frac{z^k}{k^2} = z + \frac14z^2 + o(z^2),
%\end{equation}
%имеем
%\begin{equation}
%	\text{Li}_2\left(-\frac{2}{d}\right) = -\text{Li}_2\left(-\frac{d}{2}\right) -\frac{\pi^2}6 - \frac12\ln^2\left(\frac{2}{d}\right).
%\end{equation}
%Имеем
%\begin{equation}
%	\text{Li}_2\left[-1/z\right] =  -\frac12\ln^2[1/z]-\frac{\pi^2}6 + o(1), z\rightarrow 0^+\\
%\end{equation}
%Отсюда получаем асимптотику при $d\rightarrow 0$
%\begin{equation}
%	\text{Li}_2\left(-\frac{2}{d}\right) = - \frac12\ln^2\left(\frac{2}{d}\right) -\frac{\pi^2}6 + o(d), d\rightarrow 0^+.
%\end{equation}
%Из уравнения (\ref{eq:int_log}) имеем асимптотику
From Eq.~(\ref{eq:int_log}) we obtain the asymptotics
 %$\bar{\mathcal{G}}_{\rm s}(d)$
%\begin{equation}
%	\bar{\mathcal{G}}_{\rm s}(d) = \frac14\ln^2\left[{2}/{d}\right] +\frac{\pi^2}{12} + o(d).
%\end{equation}
%или, разлагая логарифм
%\begin{equation}
%	\ln\left[1 + \frac{2}{d}\right] = \ln\left[\frac{2}{d}\right] +\frac{d}{2} - \frac12\left(\frac{d}{2}\right)^2 + \mathcal{O} \left(d^3\right).
%\end{equation}
\begin{equation}\label{eq:barG_2s_asymp}
	\bar{\mathcal{G}}_{\rm l}(d) = \frac14\ln^2\frac{2}{d}
	+ \frac{\pi^2}{12} - d/4 + d^2/32 + o(d).
\end{equation}
As above, we introduce the variable change $x = \epsilon/d$ to obtain
\begin{equation}
	\mathcal{G}_{\rm l}(d) - \bar{\mathcal{G}}_{\rm l}(d) = \int\limits_0^{1/d} dx\, \ln\frac{1}{xd}\left(\frac1{\sqrt{x^2 + 1} + x} - \frac1{2x + 1}\right).
\end{equation}
Similar to above consideration, we extend the upper limit to infinity,
\begin{equation}
\int\limits_0^{+\infty} dx\, \ln\frac{1}{xd}\left(\frac1{\sqrt{x^2 + 1} + x} - \frac1{2x + 1}\right) = \frac14\left(\ln\frac2{d} + \frac{3-\pi^2}6\right)
\end{equation}
and 
\begin{equation}
\int\limits_{1/d}^\infty dx\, \ln\frac{1}{xd}\left(\frac1{\sqrt{x^2 + 1} + x} - \frac1{2x + 1}\right) = \int\limits_{1/d}^\infty \frac{dx(1 - 1/x + \ldots)}{4x^2}\, \ln^2\frac{1}{xd} = -\frac{d}{4}(1 - d/4) + o(d^2)
\end{equation}
\begin{equation}\label{eq:diff_l}
	\mathcal{G}_{\rm l}(d) - \bar{\mathcal{G}}_{\rm l}(d) = \frac14 \left(\ln\frac2{d} +  \frac{3 -\pi^2}{6}\right) + \frac{d}{4}(1 - d/4) + o(d^2).
\end{equation}
Summing both the contributions (\ref{eq:barG_2s_asymp}) and (\ref{eq:diff_l}), we obtain
%\begin{multline}
%	\mathcal{G}_{\rm s}(\Delta) = \frac14\ln^2\left(\frac{W}{\Delta}\right)+ \frac12\lnB{B}\ln\left(1 + \frac{W}{\Delta}\right) +\frac{\pi^2}{12} + \frac12\left(\frac{\Delta}{W}\right)^2 +\\
%	+ \frac14 \ln\left(\frac{B}{\Delta}\right) + \frac{6\ln(2) + 3 -\pi^2}{24} - \frac{\Delta}{2W}\left[\lnB{B} - 1\right] +\\
%	+ \left(\frac{\Delta}{2W}\right)^2\left[2\lnB{B} - 1\right].
%\end{multline}
%\begin{multline}
%	\mathcal{G}_{\rm s}(\Delta) = \frac14\ln^2\left(\frac{W}{\Delta}\right)+ \frac12\lnB{B}\ln\left(1 + \frac{W}{\Delta}\right) + \frac14 \ln\left(\frac{B}{\Delta}\right)  +\\
%	 + \frac{6\ln(2) + 3 +\pi^2}{24} - \frac{\Delta}{2W}\left[\lnB{B} - 1\right]
%	+ \left(\frac{\Delta}{2W}\right)^2\left[2\lnB{B} + 1\right].
%\end{multline}
%\begin{multline}
%	\mathcal{G}_{\rm s}(\Delta) = \frac14\ln^2\left(\frac{W}{\Delta}\right)+ \frac12\lnB{B}\left[\ln\left(\frac{W}{\Delta}\right) +\frac{\Delta}{W} - \left(\frac{\Delta}{W}\right)^2\right] + \frac14 \ln\left(\frac{B}{\Delta}\right)  +\\
%	 + \frac{6\ln(2) + 3 +\pi^2}{24} - \frac{\Delta}{2W}\left[\lnB{B} - 1\right]
%	+ \left(\frac{\Delta}{2W}\right)^2\left[2\lnB{B} + 1\right].
%\end{multline}
%\begin{multline}
%	\mathcal{G}_{\rm s}(\Delta) = \frac14\ln^2\left(\frac{W}{\Delta}\right)+ \frac12\lnB{B}\ln\left(\frac{W}{\Delta}\right)  + \frac14 \ln\left(\frac{W}{\Delta} \right)+ \frac14 \ln\left(\frac{B}{W}\right)  +\\
%	 + \frac{6\ln(2) + 3 +\pi^2}{24} + \frac{\Delta}{2W}
%	+ \left(\frac{\Delta}{2W}\right)^2.
%\end{multline}
\begin{equation}\label{eq:G_asymp_l}
	\mathcal{G}_{\rm l}(d) = \frac14\left(\ln^2\frac{2}{d} + \ln\frac{2}{d} +  \frac{3 +\pi^2}{6}\right) - d^2/32 + o(d^2).
\end{equation}
%\item $\mathcal{G}_0:$

The contribution from $\rho_{D}(\epsilon) = 1$  is
%Вклад от постоянной в окрестности $\epsilon = 0$ плотности состояний равен
%\begin{equation}
%	\mathcal{G}_{0}(d) = \int\limits_0^1 \frac{d\epsilon}{\sqrt{\epsilon^2 + d^2} + \epsilon}  = \frac12\left(\frac1{\sqrt{d^2+1}+1} + \ln
%   \left(\frac{\sqrt{d^2+1}+1}{d}\right)\right).
%\end{equation}
\begin{equation}\label{eq:G_0(d)_def}
	\mathcal{G}_0(d) = \frac12\left(\frac12 + \ln\frac{2}{d}\right)   + d^2/16+ o(d^2).
\end{equation}

%\item $\mathcal{G}_2:$
%
%\begin{multline}
%	\mathcal{G}_2(\Delta) = \frac{\alpha_2(W/2)^2}8\left[
%	(2\Delta/W)^2\ln\left[\frac{1 + \sqrt{1 + (2\Delta/W)^2}}{2\Delta/W} \right] + \right. \\ \left. + \sqrt{1 + (2\Delta/W)^2} + \frac2{1 + \sqrt{1 + (2\Delta/W)^2}}
%	\right].
%\end{multline}
%Асимптотика имеем вид
%\begin{equation}
%		\mathcal{G}_2(\Delta) = \frac{\alpha_2}4(W/2)^2\left(1 + \frac{(2\Delta/W)^2}{8}\left(1 - 4\ln\left[\frac{W}{\Delta}\right]\right)\right) + o(\Delta^2).
%\end{equation}
%

Now we consider the contribution from quadratic (possibly with logarithmic corrections) term in the density of state $\rho_{D;\rm quad}(\epsilon) = \epsilon^2$:
\begin{equation}\label{eq:G_quad(d)_def}
	\mathcal{G}_{\rm quad}(d) = \int\limits_0^1 d\epsilon\, \frac{\epsilon^2}{\sqrt{\epsilon^2 + d^2} + \epsilon} = \frac14 - \frac{d^2}{32}\left(4\ln\frac2{d} - 1\right) + o(d^3),
\end{equation}
and for $\rho_{D;\rm quad,l}(\epsilon) = \epsilon^2\ln(1/|\epsilon|)$
\begin{equation}\label{eq:G_quad,l(d)_def}
	\mathcal{G}_{\rm quad,l}(d) = \int\limits_0^1 d\epsilon\, \frac{\epsilon^2\ln(1/\epsilon)}{\sqrt{\epsilon^2 + d^2} + \epsilon} = \frac18 - \frac{d^2}{16}\left(\ln^2\frac2{d} - \frac12\ln\frac2{d} + \frac{4\pi^2 - 15}{24}\right) + o(d^3).
\end{equation}
For $\rho_{D;\rm quad,2l}(\epsilon) = \epsilon^2\ln^2(1/|\epsilon|)$
\begin{equation}\label{eq:G_quad,dl(d)_def}
	\mathcal{G}_{\rm quad,dl}(d) = \int\limits_0^1 d\epsilon\, \frac{\epsilon^2\ln^2(1/\epsilon)}{\sqrt{\epsilon^2 + d^2} + \epsilon} = \frac18 
	-d^2\sum_{n = 0}^3g^{\rm quad,2l}_n\ln^n\frac2{d} + o(d^3),
\end{equation}
where
$g^{\rm quad,2l}_0 =  \left(48 \zeta (3) - 4 \pi ^2 - 39\right)/{768}$,   $g^{\rm quad,2l}_1 = (4 \pi ^2-15)/{192}$,
   $g^{\rm quad,2l}_2 = -1/{32}$,    $g^{\rm quad,2l}_3 = 1/{24}$.

Finally, we take into account the  contribution $\delta\mathcal{G}(d)$ from the rest part of the density of states presenting $\rho_D(\epsilon) = \rho_{D, \rm leading}(\epsilon) + \delta\rho_D(\epsilon)$, where $\rho_{D, \rm leading}(\epsilon)$ is some linear combination of the above-considered logarithmic and quadratic contributions, and $\delta\rho_D(\epsilon) = o(\epsilon^3)$.
%Наконец, учтем вклад $\delta\mathcal{G}$ от регулярной части плотности состояний 
%$\delta \rho_D(\epsilon) = \rho_D(\epsilon) - \rho_{D,\rm s}(\epsilon) = a_0\epsilon^2 + o(\epsilon^3)$ or $a_{\rm l}\epsilon^2\ln(1/|\epsilon|) + o(\epsilon^3)$. 
%Мы предполагаем, что $\int\limits_0^1\,d\epsilon\delta \rho_D(\epsilon)/\epsilon$ сходится, так как сингулярный вклад $\rho_{D,\rm s}$ вычтен. 
Since the integral $\int\limits_0^1\,d\epsilon\delta\rho_D(\epsilon)/\epsilon^3$ converges,  we apply the expansion for the contribution of $\delta_D\rho(\epsilon)$
\begin{equation}\label{eq:trivial_expansion}
	\frac1{\sqrt{\epsilon^2 + d^2} + \epsilon} = \frac1{2\epsilon} - \frac{d^2}{8\epsilon^3} + \mathcal{O}(d^4).
\end{equation}
We directly get
\begin{equation}\label{eq:delta_G(d)_def}
\delta\mathcal{G}(d) = \frac12\int\limits_0^1 d\epsilon\, \frac{\delta\rho_D(\epsilon)}{\epsilon} - \frac{d^2}8\int\limits_0^1 d\epsilon\, \frac{\delta\rho_D(\epsilon)}{\epsilon^3} + o(d^3).
\end{equation}
Contrary to the above  contributions being determined by the asymptotic behaviour in the center of the band, the rest contribution $\delta\mathcal{G}(d)$ depends on the spectrum over all the band and can be directly calculated numerically.
%Therefore $d$ independent contribution for each lattice can be found numerically:
%%Таким образом, постоянный вклад для каждой решетки можно определить численно:
%\begin{equation}
%	\delta\mathcal{G}(d) = \delta\mathcal{G} + o(d),
%\end{equation}
%%где
%where
%\begin{equation}\label{eq:G_asymp_delta}
%	\delta\mathcal{G} = \frac12\int\limits_0^1 d\epsilon\, \frac{\delta \rho_D(\epsilon)}{\epsilon}.
%\end{equation}
%The result for 
%%Результат для 
%$\delta\rho_D(\epsilon)$ does not contain singular contributions into $\mathcal{G}(d)$.
%%не содержит сингулярных вкладов в $\mathcal{G}(d)$.

\subsection{Application for lattices}\label{appendix:lattice_appl}
Here we apply the method for the following lattices.
%Применим полученные асимптотики для следующих решеток.

\textbf{Square lattice.} $D = D_{\rm sq} = 4$.
\begin{figure}[h]
\includegraphics[angle=-90,width=0.45\textwidth]{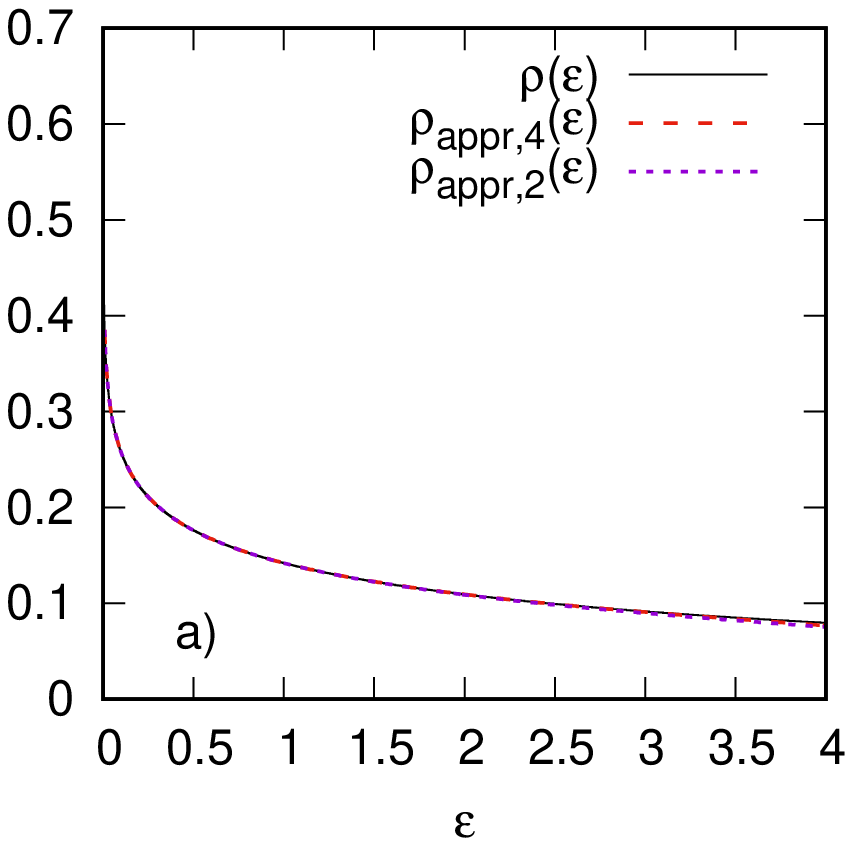}
\includegraphics[angle=-90,width=0.45\textwidth]{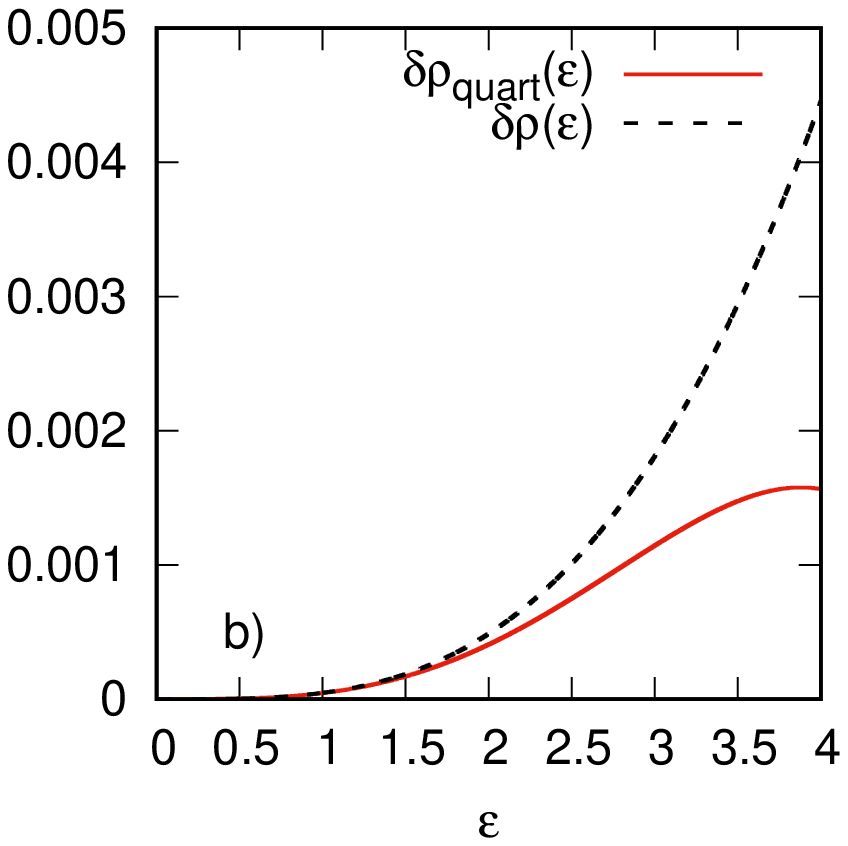}
\caption{
(a) The density of states of the square lattice (see Eq.~(\ref{eq:DOS_square})) and different approximations for it (\ref{eq:DOS_square_appr2}),(\ref{eq:DOS_square_appr4}). (b) The rest of $\rho^{\rm sq}_{\rm appr,2}(\varepsilon)$ approximation $\delta\rho^{\rm sq}(\varepsilon)$ and its leading contribution in $\delta\rho^{\rm sq}_{\rm quart}(\varepsilon)$, see  Eq.~(\ref{eq:DOS_square_contrib_quart}). Due to symmetry only $\varepsilon > 0$ region is shown.
}
\label{fig:dos_sq}
\end{figure}
According to above considered scheme we present different approximations for the density of states of the square lattice
\begin{equation}\label{eq:DOS_square}
	\rho^{\rm sq}_D(\epsilon) = \frac{2}{\pi^2}\mathbb{K}(1 - \epsilon^2):
\end{equation}
%which can be presented as 
\begin{eqnarray}
\label{eq:DOS_square_appr2}
	\rho^{\rm sq}_{D, \rm appr,2} &=& \rho^{\rm sq}_{D, \rm s}(\epsilon)+ \rho^{\rm sq}_{D, \rm quad}(\epsilon),\\
\label{eq:DOS_square_appr4}
	\rho^{\rm sq}_{D, \rm appr,4} &=& \rho^{\rm sq}_{D, \rm s}(\epsilon)+ \rho^{\rm sq}_{D, \rm quad}(\epsilon)+ \rho^{\rm sq}_{D, \rm quart}(\epsilon),
\end{eqnarray}
where
\begin{eqnarray}\label{eq:DOS_square_contrib_main}
\rho^{\rm sq}_{D, \rm s}(\epsilon) &=& \frac2{\pi^2}\ln\frac4{|\epsilon|},\\
\label{eq:DOS_square_contrib_quad}
\rho^{\rm sq}_{D, \rm quad}(\epsilon) &=& \frac{\epsilon^2}{2\pi^2}\left(\ln\frac4{|\epsilon|} - 1\right), \\
\label{eq:DOS_square_contrib_quart}
\rho^{\rm sq}_{D, \rm quart}(\epsilon) &=& \frac{9\epsilon^4}{32\pi^2}\left(\ln\frac4{\epsilon}- \frac76\right),
\end{eqnarray}
so that the representation 
\begin{equation}
	\rho^{\rm sq}_{D}(\epsilon) =  \rho^{\rm sq}_{D, \rm appr,2} + \delta\rho^{\rm sq}_{D}(\epsilon),
\end{equation}
holds and (\ref{eq:DOS_square_contrib_quart}) can be used as a leading contribution to the rest $\delta\rho^{\rm sq}_{D}(\epsilon)$. 
We directly obtain from the results of subsection~\ref{appendix:G_deriv}
\begin{equation}
	\mathcal{G}_{\rm sq}(d) = \frac2{\pi^2}\mathcal{G}_{\rm l}(d) + \frac{4\ln 2}{\pi^2} \mathcal{G}_0(d) + \frac{2\ln2 - 1}{2\pi^2}\mathcal{G}_{\rm quad}(d)  + \frac{1}{2\pi^2}\mathcal{G}_{\rm quad, l}(d)+   \delta\mathcal{G}_{\rm sq}(d)   + o(d^3),
\end{equation}
where $\delta\mathcal{G}_{\rm sq}(d) = \delta g^{\rm sq}_0 + \delta g^{\rm sq}_2d^2$ and numerical calculation yields, 
\begin{eqnarray}
\label{eq:delta_g0_sq_val}
	\delta g^{\rm sq}_0 &=& \frac12\int\limits_0^1d\epsilon\,\frac{\delta\rho^{\rm sq}_D(\epsilon)}{\epsilon} = 2.8017535604627\cdot10^{-3},\\
	\delta g^{\rm sq}_2 &=& -\frac18\int\limits_0^1d\epsilon\,\frac{\delta\rho^{\rm sq}_D(\epsilon)}{\epsilon^3} = -1.7397692893940\cdot10^{-3}.
\end{eqnarray} 
Substituting all $\mathcal{G}$ contributions we get
%\begin{multline}
%	\mathcal{G}_{\rm sq}(d) = \frac1{2\pi^2}\left(\ln^2\frac{2}{d} + \ln\frac{2}{d} +  \frac{3 +\pi^2}{6}\right) + \frac{4\ln 2}{2\pi^2} \left(\frac12 + \ln\frac{2}{d}\right) + \frac{2\ln2 - 1}{2\pi^2}\frac14  + \frac{1}{2\pi^2}\frac18 + g^{\rm sq}_0 \\
%	d^2\left(
%	 \frac{4\ln 2-1}{16\pi^2}  -\frac{2\ln2 - 1}{64\pi^2} \left(4\ln\frac2{d} - 1\right)
%	-\frac{1}{32\pi^2}\left(\ln^2\frac2{d} - \frac12\ln\frac2{d} + \frac{4\pi^2 - 15}{24}\right) + g^{\rm sq}_2
%\right)	
%	+ o(d^3),
%\end{multline}
%\begin{multline}
%	\mathcal{G}_{\rm sq}(d) = \frac1{2\pi^2}\left(\ln^2\frac{2}{d} + (1 + 4\ln2)\ln\frac{2}{d} + \frac{\pi^2}6 + \frac52\ln2 + \frac38  + 2\pi^2g^{\rm sq}_0\right) \\
%	+ d^2\left(
%	 \frac{4\ln 2-1}{16\pi^2}  -\frac{2\ln2 - 1}{64\pi^2} \left(4\ln\frac2{d} - 1\right)
%	-\frac{1}{32\pi^2}\left(\ln^2\frac2{d} - \frac12\ln\frac2{d} + \frac{4\pi^2 - 15}{24}\right) + g^{\rm sq}_2
%\right)	
%	+ o(d^3),
%\end{multline}
\begin{multline}\label{eq:G_sq_result}
	\mathcal{G}_{\rm sq}(d) = \frac1{2\pi^2}\left(\ln^2\frac{2}{d} + (1 + 4\ln2)\ln\frac{2}{d} + \frac{\pi^2}6 + \frac52\ln2 + \frac38  + 2\pi^2\delta g^{\rm sq}_0\right) \\
	+ \frac{d^2}{64\pi^2}\left(-2\ln^2\frac2{d} + (5 - 8\ln2)\ln\frac2{d} + 18\ln2 - 15/4 - \pi^2/3
	  + 64\pi^2\delta g^{\rm sq}_2
\right)	
	+ o(d^3).
\end{multline}
%It is useful to return to old variable $\varepsilon$ in the integral 
%\begin{eqnarray}
%\label{eq:delta_g0_sq_val}
%	\delta g^{\rm sq}_0 &=& \frac{D}2\int\limits_0^Dd\varepsilon\,\frac{\delta\rho^{\rm sq}(\varepsilon)}{\varepsilon} = 2.8017535604627\cdot10^{-3},\\
%	\delta g^{\rm sq}_2 &=& -\frac{D^3}8\int\limits_0^Dd\varepsilon\,\frac{\delta\rho^{\rm sq}(\varepsilon)}{\varepsilon^3} = -1.7397692893940\cdot10^{-3}.
%\end{eqnarray} 

We apply Eq.~(\ref{eq:Phi1_through_G}) 
%\begin{multline}
%\Phi_1(d) = 4\left(a_2\ln^2\frac2{d} + (a_1 - a_2)\ln\frac2{d} + a_0 - a_1/2  \right) \\
%+ 2d^2\left(
%4b_2\ln^2\frac2{d} + 2(2b_1 - b_2)\ln\frac2{d} + 4b_0 - b_1
%\right)
%\end{multline}
%\begin{multline}
%\Phi_1(d) = \frac{4}{2\pi^2}\left(\ln^2\frac8{d} + \frac{\pi^2}6 + \frac12\ln2 - 4\ln^22- \frac18  + 2\pi^2\delta g^{\rm sq}_0\right) \\
%+ \frac{2d^2}{64\pi^2}\left(
%-8\ln^2\frac2{d} + 2(12 - 16\ln2)\ln\frac2{d} + 80\ln2 - 20 - 4\pi^2/3
%	  + 256\pi^2\delta g^{\rm sq}_2
%\right)
%\end{multline}
\begin{multline}\label{eq:Phi1_sq_expansion}
\Phi^{\rm sq}_1(d\cdot D_{\rm sq}) = \frac{1}{2\pi^2}\left(\ln^2\frac8{d} + \frac{\pi^2}6 + \frac12\ln2 - 4\ln^22- \frac18  + 2\pi^2\delta g^{\rm sq}_0\right) \\
- \frac{d^2}{16\pi^2}\left(
\ln^2\frac8{d} - 3\ln\frac8{d} - 4\ln^22 - 7\ln2 + 5/2 + \pi^2/6
	  - 32\pi^2\delta g^{\rm sq}_2
\right).
\end{multline}
and (\ref{eq:delta_F_HFA_through_G})
\begin{multline}\label{eq:delta_F_HFA_sq_expansion}
    \delta F^{\rm HFA, sq}_{\rm AFM}(d\cdot D_{\rm sq}) \\
    = -\frac{2d^2}{\pi^2}\left(
    2\ln\frac8{d} + 1 \right)
      - \frac{d^4}{4\pi^2}\left(
    \ln^2\frac2{d} - \frac12(7 - 8\ln2)\ln\frac2{d}  + \frac14(25/2 - 44\ln 2 + 2\pi^2/3 - 128\pi^2\delta g^{\rm sq}_2)
    \right).
\end{multline}

%\begin{equation}\label{eq:delta_G_sq}
%\delta\mathcal{G}_{\rm sq} = \frac12\int\limits_0^1d\epsilon(\delta\rho^{\rm sq}_D(\epsilon)/\epsilon) = 0.0140268.
%\end{equation}

\textbf{SC lattice.} $D = D_{\rm sc} = 6$.
\begin{figure}[h]
\includegraphics[angle=-90,width=0.45\textwidth]{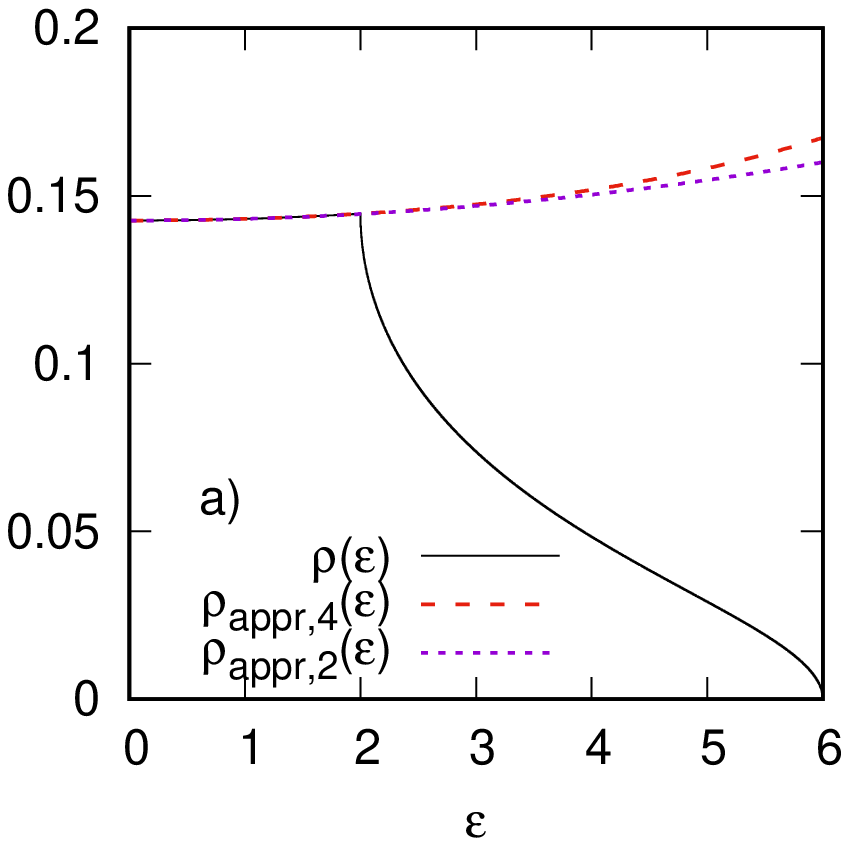}
\includegraphics[angle=-90,width=0.45\textwidth]{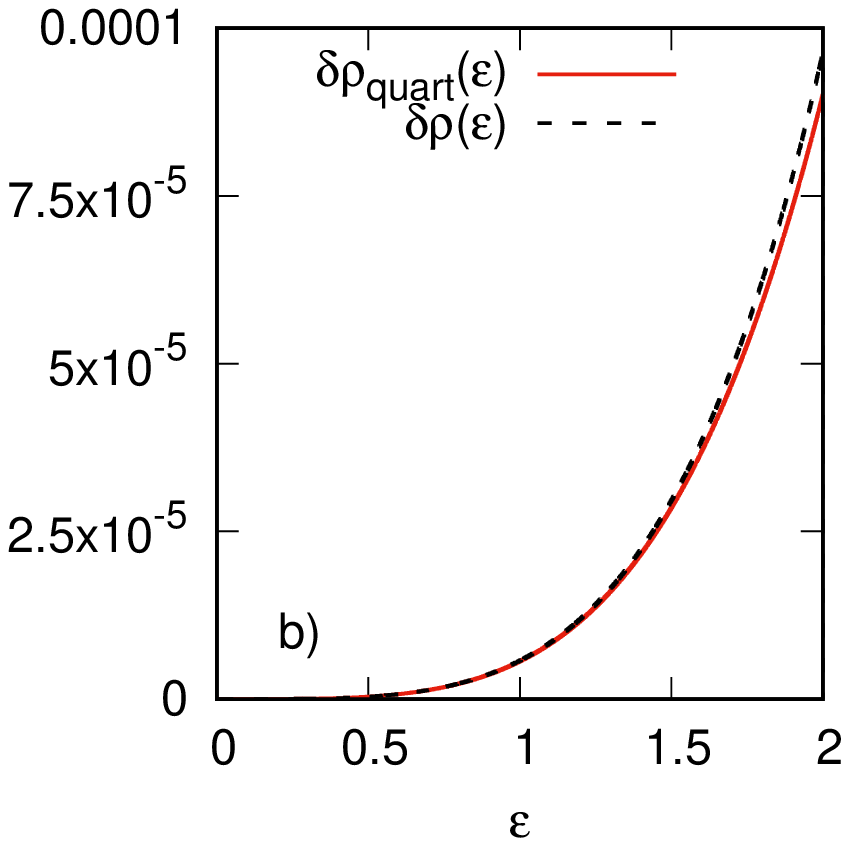}
\caption{
(a) The densify of states of the sc lattice (see  Eq.~(\ref{eq:DOS_sc})) and different approximations for it (\ref{eq:DOS_SC_appr2}),(\ref{eq:DOS_SC_appr4}). (b) The rest of $\rho^{\rm sc}_{\rm appr,2}(\varepsilon)$ approximation $\delta\rho^{\rm sq}(\varepsilon)$ and the leading contribution in $\delta\rho^{\rm sc}(\varepsilon)$, Eq.~(\ref{eq:DOS_SC_appr4}) is shown at $\varepsilon$ up to the kink point. Due to symmetry, only $\varepsilon > 0$ region is shown.
}
\label{fig:dos_sc}
\end{figure}
For this lattice the density of state is analytical in the center of the band.  
From Ref.~\onlinecite{1969:Jelitto} we get
\begin{equation}\label{eq:DOS_sc}
	\rho^{\rm sc}_D(\epsilon) = \frac{3}{\pi^3} 
\begin{cases}	
	R_{\rm sc}(3\epsilon) + R_{\rm sc}(-3\epsilon),& 0 \le |\epsilon| < 1/3,\\
	\lint_{3|\epsilon|-2}^{+1}\frac{dx\,K'\left(\frac{x - 3|\epsilon|}2\right)}{\sqrt{1 - x^2}},& 1/3 < |\epsilon| < 1.
	\end{cases}
\end{equation}
%%%%%% Old version %%%%%%%%%
%\begin{equation}\label{eq:DOS_sc}
%	g(E) = \frac{1}{\pi^3} 
%\begin{cases}	
%	R_{\rm sc}(E) + R_{\rm sc}(-E),& 0 \le |E| < 1,\\
%	\lint_{|E|-2}^{+1}\frac{dx\,K'\left(\frac{x - |E|}2\right)}{\sqrt{1 - x^2}},& 1 < |E| < 3.
%	\end{cases}
%\end{equation}
%%%%%%%%%%%%%%%%%%%%%%%%%%%%%%%%
where an auxiliary function is introduced,
\begin{equation}\label{eq:f_def}
	R_{\rm sc}(E) = \lint_E^{+1}\frac{dx\,K'\left(\frac{x - E}2\right)}{\sqrt{1 - x^2}},
\end{equation}
%%%%%% Old version %%%%%%%%%%%%%
%\begin{equation}\label{eq:f_def}
%	R_{\rm sc}(E) = \lint_E^{+1}\frac{dx\,K'\left(\frac{x - E}2\right)}{\sqrt{1 - x^2}},
%\end{equation}
%%%%%%%%%%%%%%%%%%%%%%%%%%%%%%%%
and $K'(x) = \mathbb{K}(1 - x^2)$, where we set
$\mathbb{K}(m) = \int_0^1 {dx}/{\sqrt{(1 - x^2)(1 - mx^2)}}$. 
To expand Eq.~(\ref{eq:f_def}) in $E$, we change variable $x = E + (1 - E)t$
\begin{equation}
	R_{\rm sc}(E) = (1 - E)\int_0^1\frac{dt K'((1 - E)t/2)}{\sqrt{1 - (E + (1 - E)t)^2}}.
\end{equation}
%We use the asymptotics
%\begin{equation}
%	K'(x) = \ln\frac4{x} + \frac{\ln\frac4{x} - 1}4x^2 + o(x^3).
%\end{equation}
We differentiate directly to derive at small $\epsilon$
\begin{equation}
	\rho^{\rm sc}_D(\epsilon) = \frac{6}{\pi^3}(R_{\rm sc}(0) + 9R''_{\rm sc}(0)\epsilon^2/2 + 27R^{(4)}_{\rm sc}(0)\epsilon^4/8 \ldots),
\end{equation}
%%%%%% Old version %%%%%%%%%
%\begin{equation}
%	g(E) = \frac{2}{\pi^3}(f(0) + f''(0)E^2/2 + f^{(4)}(0)E^4/24 \ldots),
%\end{equation}
%%%%%%%%%%%%%%%%%%%%%%%%%%%%%%%%
where 
\begin{equation}
	R^{(n)}_{\rm sc}(0) = \int\limits_0^1\frac{dx}{\sqrt{1-x^2}}\frac{ p_{n,K}(x) K'\left(x/2\right)-p_{n,E}(x)
   E'\left(x/2\right)}{(x+1)^n 
   \left(4-x^2\right)^n},
\end{equation}
with $n = 0,2,4 $ and $p_{0,K}(t) = 1$, $p_{0,E}(t) = 0$,
\begin{eqnarray*}
p_{2,K}(x) &=& 16 - 32x + 24x^3 + x^4 + 2x^6,\\
p_{2,E}(x) &=&  4(4 - 3x^2 + 4x^3 + 3x^4),\\
p_{4,K}(x) &=& 24 x^{12}+124 x^{10}+1152 x^9+497 x^8-2064 x^7+4068 x^6+5184 x^5-8896
   x^4\\
   &+& 2304 x^3 +7744 x^2  - 9216 x+2304,\\
p_{4,E}(x) &=& 8(25 x^{10}+48 x^9+21 x^8+336 x^7+602 x^6-408 x^5-448 x^4+960 x^3
   +96x^2-384 x+256),
\end{eqnarray*}
$E'(x) = \mathbb{E}(1 - x^2)$, where
$\mathbb{E}(m) = \int_0^1 {dx\sqrt{1 - mx^2}}/{\sqrt{1 - x^2}}$.
 
We therefore obtain the expansion
\begin{equation}\label{eq:g_exp}
	\rho^{\rm sc}_D(\epsilon) = a^{\rm sc}_0 + a^{\rm sc}_2\epsilon^2 + a^{\rm sc}_4\epsilon^4 + \mathcal{O}(\epsilon^6), 
\end{equation}
where 
$$a^{\rm sc}_n = \frac{2\cdot3^{n+1}}{\pi^3n!}\int\limits_0^1\frac{dt}{\sqrt{1- t^2}}\frac{p_{n,K}(t)K'(t/2) - p_{n,E}(t)E'(t/2)}{(1+t)^n(4- t^2)^n}$$.

%%%%%% Old version %%%%%%%%%
%\begin{eqnarray}
%a^{\rm sc}_0 &=& \frac{2}{\pi^3}\int\limits_0^1\frac{dt K'(t/2)}{\sqrt{1 - t^2}},\\
%a^{\rm sc}_n &=& \frac{2}{\pi^3n!}\int\limits_0^1\frac{dt}{\sqrt{1- t^2}}\frac{p_{n,K}(t)K'(t/2) - p_{n,E}(t)E'(t/2)}{(1+t)^n(4- t^2)^n}.
%\end{eqnarray}
%%%%%%%%%%%%%%%%%%%%%%%%%%%%%%%%
Numerical calculation yields 
\begin{eqnarray}
\label{eq:a0_sc_def}
a^{\rm sc}_0  &=& 0.856038, \\
a^{\rm sc}_2 &=& 0.104223, \\
\label{eq:a4_sc_def}
a^{\rm sc}_4 &=& 0.0437667.
\end{eqnarray} 
%%%%%% Old version %%%%%%%%%
%\begin{eqnarray}
%\label{eq:a0_sc_def}
%a^{\rm sc}_0  &=& 0.28534597, \\
%a^{\rm sc}_2 &=& 0.38601\cdot10^{-2}, \\
%\label{eq:a4_sc_def}
%a^{\rm sc}_4 &=& 0.18011\cdot10^{-3}.
%\end{eqnarray} 
%%%%%%%%%%%%%%%%%%%%%%%%%%%%%%%%

Different approximations for $\rho^{\rm sc}(\epsilon)$
\begin{eqnarray}
\label{eq:DOS_SC_appr2}
	\rho^{\rm sc}_{D,\rm appr,2}(\epsilon) &=& a^{\rm sc}_0 + a^{\rm sc}_2\epsilon^2,\\ 
\label{eq:DOS_SC_appr4}
	\rho^{\rm sc}_{D,\rm appr,4}(\epsilon) &=& a^{\rm sc}_0 + a^{\rm sc}_2\epsilon^2 + a^{\rm sc}_4\epsilon^4.
\end{eqnarray}
%%%%%% Old version %%%%%%%%%
%\begin{eqnarray}
%\label{eq:DOS_SC_appr2}
%	\rho^{\rm sc}_{D,\rm appr,2}(\epsilon) &=& 3(a^{\rm sc}_0 + 9a^{\rm sc}_2\epsilon^2),\\ 
%\label{eq:DOS_SC_appr4}
%	\rho^{\rm sc}_{D,\rm appr,4}(\epsilon) &=& 3(a^{\rm sc}_0 + 9a^{\rm sc}_2\epsilon^2 + 81a^{\rm sc}_4\epsilon^4).
%\end{eqnarray}
%%%%%%%%%%%%%%%%%%%%%%%%%%%%%%%%
are valid in the vicinity of $\epsilon = 0$.  
We can split the density of states
\begin{equation}\label{eq:SC_DOS_expansion}
	\rho^{\rm sc}_D(\epsilon) = \rho^{\rm sc}_{D,\rm appr,2}(\epsilon) + \delta\rho^{\rm sc}_{D}(\epsilon).
\end{equation}
The leading contribution to the rest $\delta\rho^{\rm sc}_{D}$ is 
\begin{equation}\label{eq:DOS_SC_contrib_quart}
\rho^{\rm sc}_{D, \rm quart}(\epsilon) = a^{\rm sc}_4\epsilon^4.
\end{equation} 
To treat accurately the kink of $\rho^{\rm sc}_D(\epsilon)$ at $\epsilon = 1/3$ we split
\begin{equation}
\mathcal{G}_{\rm sc}(d) = \mathcal{G}_{\rm sc,1}(d) + \mathcal{G}_{\rm sc,2}(d),
\end{equation}
where $$
\mathcal{G}_{\rm sc,1}(d) = \int_0^{1/3} \frac{d\epsilon\rho_D(\epsilon)}{\epsilon + \sqrt{\epsilon^2 + d^2}},
\mathcal{G}_{\rm sc,2}(d) = \int_{1/3}^1 \frac{d\epsilon\rho_D(\epsilon)}{\epsilon + \sqrt{\epsilon^2 + d^2}}$$. 
To use Eq.~(\ref{eq:G_exp_start}) for $\mathcal{G}_{\rm sc,1}(d)$, we pass to effective half-bandwidth $D' = 1/3$: $\epsilon = \epsilon'/3$ 
\begin{equation}
\mathcal{G}_{\rm sc,1}(d) = \int_0^{1} \frac{d\epsilon'\tilde\rho_D(\epsilon')}{\epsilon' + \sqrt{\epsilon'^2 + (d')^2}},	
\end{equation}
where $d' = 3d$, $\tilde\rho_D(\epsilon') = \rho_D(\epsilon'/3)$ and the expansion (\ref{eq:DOS_SC_appr4}) can be rewritten as
$\tilde\rho_D(\epsilon') = a^{\rm sc}_0 + a^{\rm sc}_2\left(\epsilon'\right)^2/9 + a^{\rm sc}_4\left(\epsilon'\right)^4/81 + \mathcal{O}(\left(\epsilon'\right)^6)$.
%%%%%  OLD version %%%%%%%%%
%\begin{equation}\label{eq:SC_DOS_expansion1}
%	\tilde\rho_D(\epsilon) = 3(a^{\rm sc}_0 + a^{\rm sc}_2\epsilon^2 + a^{\rm sc}_4\epsilon^4) + \mathcal{O}(\epsilon^6).
%\end{equation}
%%%%%%%%%%%%%%%%%%%%%%%%%%%%

%\begin{equation}
%	\rho^{\rm sc}_D(\epsilon) = \rho^{\rm sc}_D(0) + \delta\rho_D^{\rm sc}(\epsilon),
%\end{equation}
%where $\rho^{\rm sc}_D(0) = 0.856038$~\cite{1969:Jelitto}, $\delta\rho_D^{\rm sc}(\epsilon) = o(\epsilon)$. 

Then, according to equations (\ref{eq:SC_DOS_expansion}),(\ref{eq:G_0(d)_def}), (\ref{eq:G_quad(d)_def}) and (\ref{eq:delta_G(d)_def}), we get
%Тогда согласно уравнениям (\ref{eq:G_0(d)_def}) и (\ref{eq:G_asymp_delta})
\begin{equation}
\mathcal{G}_{\rm sc,1}(d) = a^{\rm sc}_0\mathcal{G}_0(3d) + \frac{a^{\rm sc}_2}9\mathcal{G}_{\rm quad}(3d)+\delta\mathcal{G}_{\rm sc,1}(3d) + o(d^3),
\end{equation}
%and we obtain the coefficients of representation~(\ref{eq:a_def})
%и мы получим коэффициенты представления 
%\begin{eqnarray}
%\label{eq:G_sc}
%	a^{\rm sc}_0 &=& \rho_D(0)/4 + \delta\mathcal{G}_{\rm sc},\\
%	a^{\rm sc}_1 &=& \rho_D(0)/2,
%\end{eqnarray}
where $\delta\mathcal{G}_{\rm sc,1}(d) = \delta g^{\rm sc,1}_0 + \delta g^{\rm sc,1}_2d^2$, where
%\begin{equation}\label{eq:delta_G_sc}
%\delta\mathcal{G}_{\rm sc} = \frac12\int\limits_0^1d\epsilon(\rho_D^{\rm sc}(\epsilon)/\epsilon) = -0.173.
%\end{equation}
\begin{eqnarray*}
\delta g^{\rm sc,1}_0 &=& \frac12\lint_0^1\frac{d\epsilon'}{\epsilon'}\left(\tilde\rho_D(\epsilon') - a^{\rm sc}_0 - a^{\rm sc}_2\left(\epsilon'\right)^2/9\right), \\
\delta g^{\rm sc,2}_2 &=& -\frac18\lint_0^1\frac{d\epsilon'}{{\epsilon'}^3}\left(\tilde\rho_D(\epsilon') - a^{\rm sc}_0 - a^{\rm sc}_2\left(\epsilon'\right)^2/9\right). 
\end{eqnarray*}
It is convenient to return to $\epsilon$ in two latter integrals:
\begin{equation}\label{eq:delta_g_sc_1}
	\delta g^{\rm sc,1}_0 = \frac12\lint_0^{1/3}\frac{d\epsilon}{\epsilon}\left(\rho_D(\epsilon) - \rho_{D,\rm appr,2}(\epsilon)\right), 
	\delta g^{\rm sc,2}_2 = -\frac1{72}\lint_0^{1/3}\frac{d\epsilon}{\epsilon^3}\left(\rho_D(\epsilon) - \rho_{D,\rm appr,2}(\epsilon)\right),
\end{equation}
where $\rho_{D,\rm appr,2}(\epsilon)$ is defined by~Eq.~(\ref{eq:DOS_SC_appr2}). 

%Numerical calculation using the asymptotics (\ref{eq:g_exp}) yields $\delta g^{\rm sc}_0 = ???$, $ \delta g^{\rm sc}_2 = ???$.  
For $\mathcal{G}_{\rm sc, 2}(d)$ we directly use the expansion (\ref{eq:trivial_expansion}) and obtain
\begin{equation}
	\delta\mathcal{G}_{\rm sc,2}(d) = \delta g^{\rm sc}_{0,2} + \delta g^{\rm sc,2}_{2}d^2 + \mathcal{O}(d^4), 
\end{equation}
where
\begin{equation}\label{eq:delta_g_sc_2}
	\delta g^{\rm sc,2}_0 = \frac12\lint_{1/3}^1\frac{d\epsilon}{\epsilon}\rho_D(\epsilon), 
	\delta g^{\rm sc,2}_2 = -\frac18\lint_{1/3}^1\frac{d\epsilon}{\epsilon^3}\rho_D(\epsilon).
\end{equation}
Using Eqs.~(\ref{eq:delta_g_sc_1}), (\ref{eq:delta_g_sc_2}) we get
\begin{equation}
\mathcal{G}_{\rm sc}(d) = a^{\rm sc}_0\mathcal{G}_0(3d) + \frac{a^{\rm sc}_2}9\mathcal{G}_{\rm quad}(3d)+\delta\mathcal{G}_{\rm sc}(d) + o(d^3),
\end{equation}
where 
\begin{equation}
\delta\mathcal{G}_{\rm sc}(d) = \delta g^{\rm sc}_0 + \delta g^{\rm sc}_2d^2,
\end{equation}
and numerically calculated integrals are 
\begin{eqnarray}
\label{eq:delta_g0_sc_val}
	\delta g^{\rm sc}_0 &=& \frac12\lint_{0}^1\frac{d\epsilon}{\epsilon}(\rho^{\rm sc}_D(\epsilon) - \theta(1/3 - \epsilon)\rho^{\rm sc}_{D,\rm appr,2}(\epsilon)) = 0.208275, \\ %0.20827508785460
\label{eq:delta_g2_sc_val}
	\delta g^{\rm sc}_2 &=& -\frac18\lint_{0}^1\frac{d\epsilon}{\epsilon^3}(\rho^{\rm sc}_D(\epsilon) - \theta(1/3 - \epsilon)\rho^{\rm sc}_{D,\rm appr,2}(\epsilon)) = -0.247755. %-0.24775498366164
\end{eqnarray}
%Here we, as above, also return to old variable $\epsilon = \varepsilon/D$: 
%\begin{eqnarray}
%\label{eq:delta_g0_sc_val}
%	\delta g^{\rm sc}_0 &=& \frac{D}2\lint_{0}^D\frac{d\varepsilon}{\varepsilon}(\rho^{\rm sc}(\varepsilon) - \theta(D/3 - \varepsilon)\rho^{\rm sc}_{\rm appr}(\varepsilon)) = 0.20827508785460, \\
%	\delta g^{\rm sc}_2 &=& -\frac{D^3}8\lint_{0}^{D}\frac{d\varepsilon}{\varepsilon^3}(\rho^{\rm sc}(\varepsilon) - \theta(D/3 - \varepsilon)\rho^{\rm sc}_{\rm appr}(\epsilon)) = -0.24775498366164.
%\end{eqnarray}
We finally have 
\begin{equation}\label{eq:G_sc_result}
\mathcal{G}_{\rm sc}(d) = \frac{a^{\rm sc}_0}2\ln\frac{2}{3d} +  \frac{a^{\rm sc}_0}4 + \frac{a^{\rm sc}_2}{36} + \delta g^{\rm sc}_0 -  d^2\left(\frac{a^{\rm sc}_2}8\ln\frac{2}{3d} - \frac{9a^{\rm sc}_0}{16} - \frac{a^{\rm sc}_2}{32} - \delta g^{\rm sc}_2\right) + o(d^3),
\end{equation}
and from~Eq.~(\ref{eq:Phi1_through_G})
%\begin{equation}
%\Phi_1(d) = 4\left( a_1\ln\frac2{3d} + a_0 - a_1/2  \right) 
%+ 2d^2\left(
% 4b_1\ln\frac2{3d} + 4b_0 - b_1
%\right)
%\end{equation}
%\begin{equation}
%\Phi_1(d) = 4\left( \frac{3a^{\rm sc}_0}2\ln\frac2{3d} + \frac{3a^{\rm sc}_2}4  \right) 
%+ 2d^2\left(
% 4b_1\ln\frac2{3d} + 4b_0 - b_1
%\right)
%\end{equation}
%\begin{equation}
%\Phi_1(d) = 6\left( a^{\rm sc}_0\ln\frac2{3d} + \frac{a^{\rm sc}_2}{2}  \right) 
% + 4\delta g^{\rm sc}_0 + 2d^2\left(
% 4b_1\ln\frac2{3d} + 4b_0 - b_1
%\right)
%\end{equation}
\begin{equation}\label{eq:Phi1_sc_expansion}
\Phi^{\rm sc}_1(D_{\rm sc}\cdot d) =  \frac{a^{\rm sc}_0}3\ln\frac2{3d} + \frac{a^{\rm sc}_2}{54}
 + \frac{2\delta g^{\rm sc}_0}3 - \frac{d^2}6\left(
 a^{\rm sc}_2\ln\frac2{3d} - \frac{9a^{\rm sc}_0}{2}-  \frac{a^{\rm sc}_2}2 - 8\delta g^{\rm sc}_2 
\right).
\end{equation}
Doing the same procedure for the free energy, see Eq.~(\ref{eq:delta_F_HFA_through_G}),
\begin{equation}\label{eq:delta_F_AFM_HFA_sc_expansion}
\delta F^{\rm HFA, sc}_{\rm AFM}(D_{\rm sc}\cdot d) = -3d^2\left(
a^{\rm sc}_0 + \frac{9d^2}{2}\left(\ln\frac{2}{3d} - \frac{a^{\rm sc}_0}6 - \frac{a^{\rm sc}_2}{36} - \frac{8\delta g^{\rm sc}_2}9 \right)
\right) + \mathcal{O}\left(d^6\ln\frac2{d}\right).
\end{equation}

\textbf{BCC lattice.} $D  = D_{\rm bcc}= 8$.
\begin{figure}[h]
\includegraphics[angle=-90,width=0.45\textwidth]{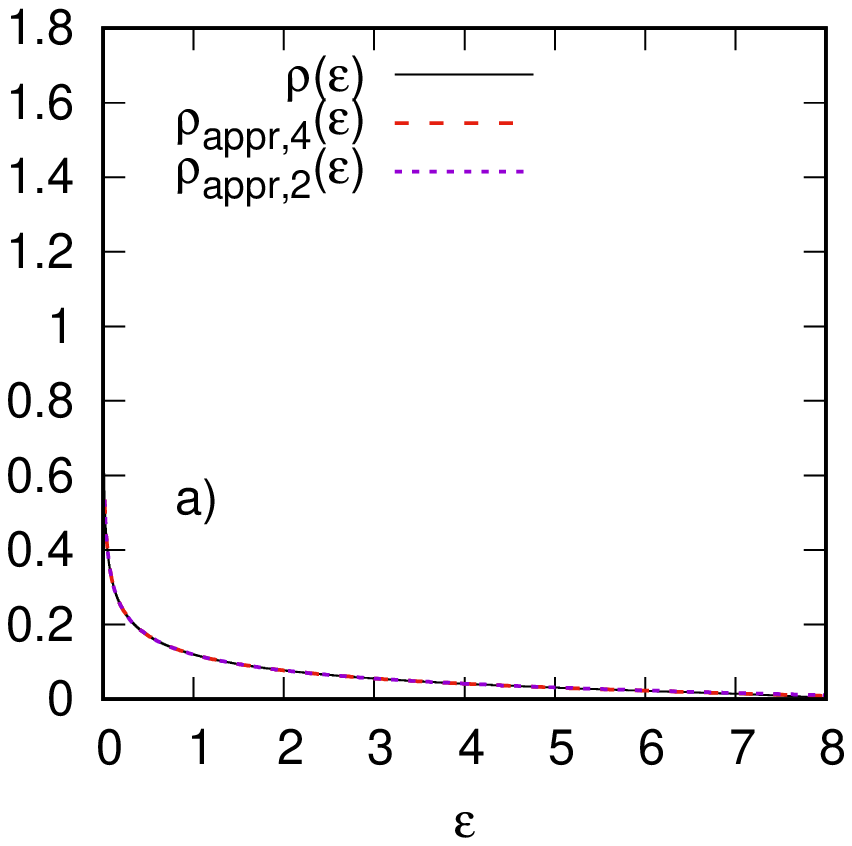}
\includegraphics[angle=-90,width=0.45\textwidth]{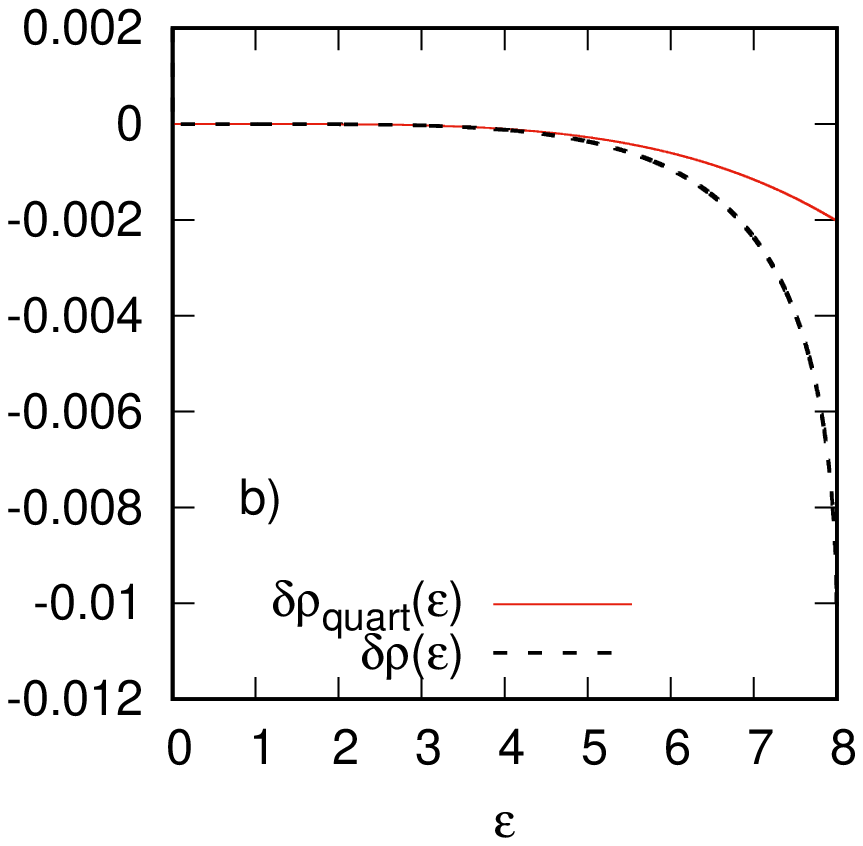}
\caption{
(a) The density of states of the sc lattice (see Eq.~(\ref{eq:DOS_bcc})) and different approximations for it (\ref{eq:DOS_BCC_appr2}),(\ref{eq:DOS_BCC_appr2}). (b)The rest of $\rho^{\rm bcc}_{\rm appr,2}(\varepsilon)$ approximation $\delta\rho^{\rm bcc}(\varepsilon)$ and its leading contribution in $\delta\rho^{\rm bcc}_{\rm quart}(\varepsilon)$, see Eq.~(\ref{eq:DOS_bcc_contrib_quart}). Due to symmetry only $\varepsilon > 0$ region is shown.
}
\label{fig:dos_bcc}
\end{figure}
From Ref.~\onlinecite{1971:Katsura} we get the following expression for the density of states expressed through the Gaussian hypergeometric function
\begin{equation}\label{eq:DOS_bcc}
\rho^{\rm bcc}_D(\epsilon) = -\frac1{\pi|\epsilon|}{\rm Im}\,{}_2\mathbb{F}_1\left(\frac12,\frac12,1; \frac12\left(1-\sqrt{1 - \epsilon^{-2}}\right)\right).
\end{equation}
As above, we introduce the representation for $\rho^{\rm bcc}_D(\epsilon)$ separating a leading contribution in the limit $\epsilon\rightarrow0$
\begin{eqnarray}
\label{eq:DOS_BCC_appr2}
	\rho^{\rm bcc}_{D, \rm appr,2}(\epsilon) &=& \rho^{\rm bcc}_{D, \rm s}(\epsilon) + \rho^{\rm bcc}_{D, \rm quad}(\epsilon),\\
\label{eq:DOS_BCC_appr4}
	\rho^{\rm bcc}_{D, \rm appr,4}(\epsilon) &=& \rho^{\rm bcc}_{D, \rm s}(\epsilon) + \rho^{\rm bcc}_{D, \rm quad}(\epsilon)+ \rho^{\rm bcc}_{D, \rm quart}(\epsilon),
\end{eqnarray}
where (see Ref.~\onlinecite{1971:Katsura})
\begin{eqnarray}\label{eq:DOS_bcc_contrib_main}
\rho^{\rm bcc}_{D, \rm s}(\epsilon) &=& \frac{2}{\pi^3}
\left(
\ln^2\frac8{\epsilon} - \frac{\pi^2}4\right),\\
\label{eq:DOS_bcc_contrib_quad}
\rho^{\rm bcc}_{D, \rm quad}(\epsilon) &=&  \frac{\epsilon^2}{4\pi^3}
\left(
\ln^2\frac8{\epsilon} - 3\ln\frac8{\epsilon} 
- \frac{\pi^2}4
\right), \\
\label{eq:DOS_bcc_contrib_quart}
\rho^{\rm bcc}_{D, \rm quart}(\epsilon) &=& \frac{27\epsilon^4}{256\pi^3}
\left(
\ln^2\frac8{\epsilon} - \frac72\ln\frac8{\epsilon} + \frac23 - \frac{\pi^2}{4}\right),
\end{eqnarray}
so that the representation 
\begin{equation}
	\rho^{\rm bcc}_{D}(\epsilon) =  \rho^{\rm bcc}_{D, \rm appr,2} + \delta\rho^{\rm bcc}_{D}(\epsilon),
\end{equation}
holds and (\ref{eq:DOS_square_contrib_quart}) can be used as leading contribution to the rest $\delta\rho^{\rm bcc}_{D}(\epsilon)$.

%
%and asymptotics for density of states
%%\begin{equation}
%%	\rho^{\rm bcc}_D(\epsilon) = \frac2{\pi^3}\left(\ln^2\frac{8}{|\epsilon|} - \left(\frac{\pi}4\right)^2\right) + \delta\rho_{\rm bcc}(\epsilon),
%%\end{equation}
%\begin{equation}\label{eq:dos_bcc_contributions}
%	\rho^{\rm bcc}_{D}(\epsilon) =  \rho^{\rm bcc}_{D, \rm s}(\epsilon)+ \rho^{\rm bcc}_{D, \rm quad}(\epsilon) + \delta\rho^{\rm bcc}_{D}(\varepsilon),
%\end{equation}
%where
%\begin{eqnarray}
%\rho^{\rm bcc}_{D, \rm s}(\epsilon) &=& \frac{2}{\pi^3}
%\left(
%\ln^2\frac8{\epsilon} - \pi^2/4\right),\\
%\rho^{\rm bcc}_{D, \rm quad}(\epsilon) &=& \frac{\epsilon^2}{4\pi^3}
%\left(
%\ln^2\frac8{\epsilon} - 3\ln\frac8{\epsilon} 
%- \frac{\pi^2}4
%\right),\\
%\delta\rho_D(\epsilon)&=& \frac{27\epsilon^4}{256\pi^3}
%\left(
%\ln^2\frac8{\epsilon} - \frac72\ln\frac8{\epsilon} + \frac23 - \frac{\pi^2}{4}\right) + \mathcal{O}(\epsilon^6\ln^2\epsilon).
%\end{eqnarray}
From  Eqs.~(\ref{eq:G_asymp_dl}), (\ref{eq:G_0(d)_def}), (\ref{eq:G_quad,dl(d)_def}),  (\ref{eq:delta_G(d)_def}) we have
\begin{multline}\label{eq:G_bcc_result}
	\mathcal{G}_{\rm bcc}(d) = \frac2{\pi^3}\left[\mathcal{G}_{\rm dl}(d) + 2\ln 8\cdot\mathcal{G}_{\rm l}(d) + (\ln^2  8 - \pi^2/4)\mathcal{G}_{0}(d)\right] \\
	+ \frac1{4\pi^3}\left[\mathcal{G}_{\rm quad,dl}(d) + 3(2\ln2 - 1)\mathcal{G}_{\rm quad,l}(d) 
+ \left(9\ln^22 - 9\ln2 -\frac{\pi^2}{4}\right)\mathcal{G}_{\rm quad}(d)	
	\right]
	+ \delta\mathcal{G}_{\rm bcc}(d) + o(d^3),
\end{multline}
where $\delta\mathcal{G}_{\rm bcc}(d) = \delta g_0^{\rm bcc} + \delta g_2^{\rm bcc}d^2$,  
\begin{eqnarray}
	\delta g^{\rm bcc}_0 &=& \frac12\int\limits_0^1d\epsilon\,\frac{\delta\rho^{\rm bcc}_D(\epsilon)}{\epsilon} = -4.0031\cdot10^{-3},\\ % -4.0030986226766\cdot10^{-3}
	\delta g^{\rm bcc}_2 &=& -\frac18\int\limits_0^1d\epsilon\,\frac{\delta\rho^{\rm bcc}_D(\epsilon)}{\epsilon^3} = 1.5184\cdot10^{-3}. %1.5184276826778\cdot10^{-3}
\end{eqnarray}

\begin{figure}[h]
\includegraphics[angle=-90,width=0.45\textwidth]{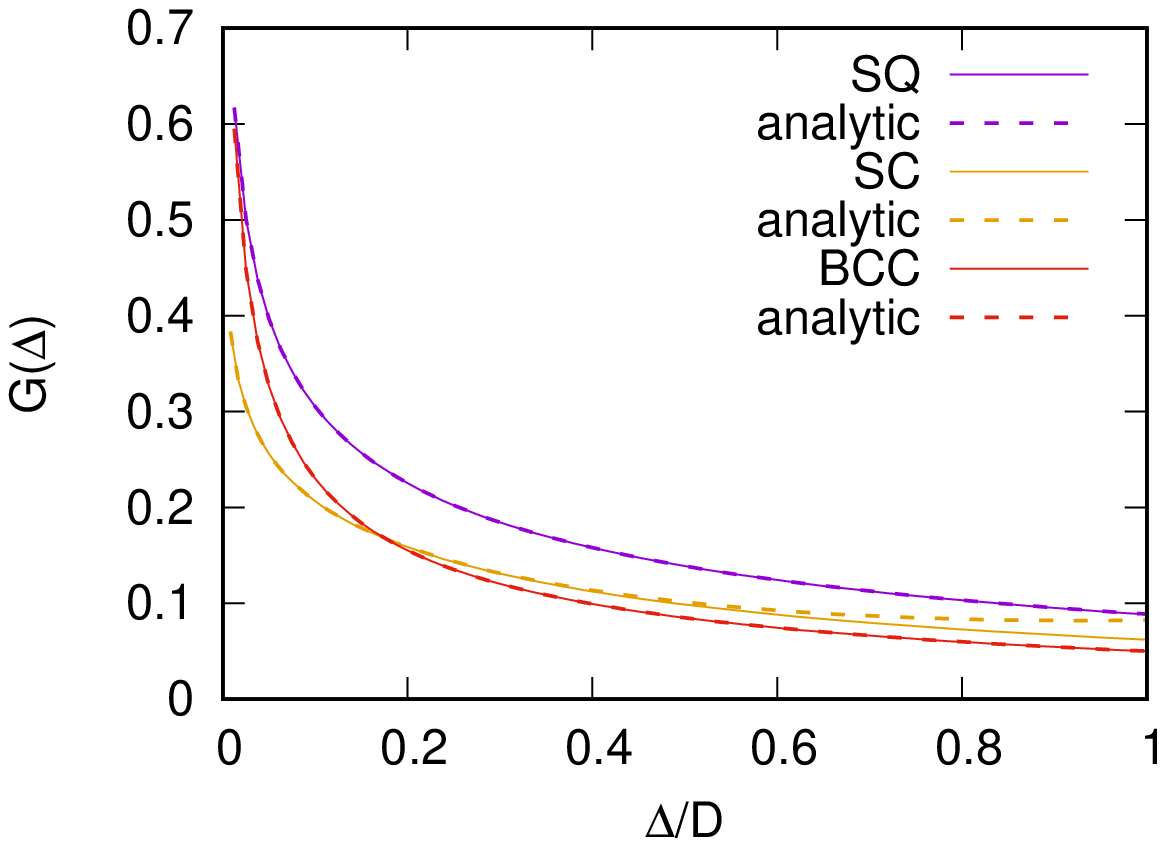}
\includegraphics[angle=-90,width=0.45\textwidth]{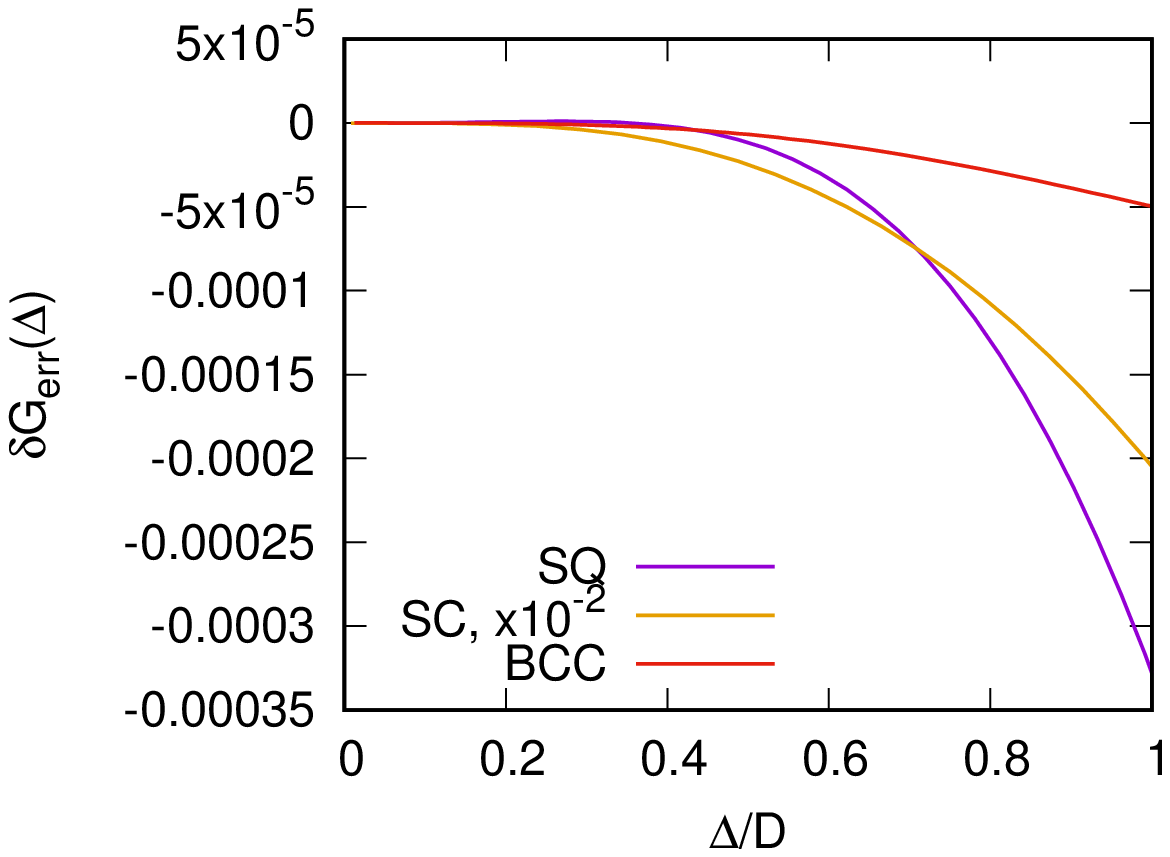}
\caption{
(a) The numerical calculation (solid line) $G(\Delta)$ (see Eq.~(\ref{eq:G_expr})) and its asymptotics (dashed line) $G_{\rm asymp}(\Delta)$  for square~(Eq.~(\ref{eq:G_sq_result})), sc~(Eq.~(\ref{eq:G_sc_result})), bcc ~(Eq.~(\ref{eq:G_bcc_result})) lattices, (b) the difference  $G(\Delta) - G_{\rm asymp}(\Delta)$.
}
\label{fig:G}
\end{figure}
We find the coefficients of the representation (\ref{eq:G_general_form}):
\begin{eqnarray*}
a^{\rm bcc}_3 = (3\pi^3)^{-1},\\ 
a^{\rm bcc}_2 = (1+6\ln2)/(2\pi^3), \\
a^{\rm bcc}_1 = (6-\pi ^2+108 \ln^22+36 \ln2)/(12\pi^3),\\
a^{\rm bcc}_0 = (96 \zeta (3)+36-11 \pi ^2+12\cdot9^2 \ln^22+ 24(9  +4 \pi^2)\ln2)/(3\cdot64\pi^3) + \delta g^{\rm bcc}_0,\\
b^{\rm bcc}_3 = -(96\pi^3)^{-1},\\
b^{\rm bcc}_2 = (7 - 12\ln 2)/(128\pi^3), \\
b^{\rm bcc}_1 = (-12+8 \pi ^2+1008 \ln2-144\cdot6 \ln^22)/(3072 \pi ^3),\\
b^{\rm bcc}_0 = (-48 \zeta (3)+141-74 \pi ^2+3672 \ln^22- 12(99 + 4 \pi ^2)\ln2)/(3072 \pi ^3) + \delta g^{\rm bcc}_2.
\end{eqnarray*}

Using Eq.~(\ref{eq:Phi1_through_G}) we get
\begin{equation}\label{eq:Phi1_bcc_expansion}
    \Phi_{\rm bcc}(d\cdot D_{\rm bcc}) = \sum_{n = 0}^3\phi^{\rm bcc}_{0,n}\ln^n\frac2{d} + d^2\sum_{n = 0}^3\phi^{\rm bcc}_{2,n}\ln^n\frac2{d},
\end{equation}
where the coefficients read
\begin{eqnarray}
\label{eq:bcc_phi_0,3}
    \phi^{\rm bcc}_{0,3} &=& \frac1{6\pi^3},
    \phi^{\rm bcc}_{0,2} = \frac{3\ln 2}{2\pi^3},
    \phi^{\rm bcc}_{0,1} = \frac{108\ln^22 - \pi^2}{24\pi^3},\\
\label{eq:bcc_phi_0,0}    
    \phi^{\rm bcc}_{0,0} &=& \frac1{128\pi^3}(32\zeta(3) - 4 - \pi^2 + 36\ln^22 + 8(4\pi^2 - 3)\ln 2) + \delta g^{\rm bcc}_0/2
\end{eqnarray}
and
\begin{eqnarray}    
    \phi^{\rm bcc}_{2,3} &=& -\frac1{96\pi^3},
    \phi^{\rm bcc}_{2,2} = \frac{2-3\ln2}{32\pi^3},
    \phi^{\rm bcc}_{2,1} = \frac1{384\pi^3}(-108\ln^22 + 144\ln2 + \pi^2 - 12), \\   
    \phi^{\rm bcc}_{2,0} &=& \delta g^{\rm bcc}_2 + \frac1{768\pi^3}(972\ln^22 - 3(127 + 4\pi^2)\ln2 + 36 - 12\zeta(3)  - 19\pi^2) + \delta g^{\rm bcc}_2.
\end{eqnarray}
Doing the same procedure for the free energy, see Eq.~(\ref{eq:delta_F_HFA_through_G}), we have
\begin{equation}\label{eq:delta_F_AFM_HFA_bcc_expansion}
    \delta F^{\rm HFA, bcc}_{\rm AFM}(d\cdot D_{\rm bcc}) = -8d^2\left(\sum_{n = 0}^2\mathcal{F}^{\rm bcc}_{0,n}\ln^n\frac2{d} + d^2\sum_{n = 0}^3\mathcal{F}^{\rm bcc}_{2,n}\ln^n\frac2{d}\right),
\end{equation}
where 
\begin{equation}
	\mathcal{F}^{\rm bcc}_{0,2} = \frac1{\pi^3},
	\mathcal{F}^{\rm bcc}_{0,1} = \frac{1+6\ln2}{\pi^3},
	\mathcal{F}^{\rm bcc}_{0,0} = \frac{6-\pi ^2+108 \ln^22+36 \ln2}{12\pi^3}
\end{equation}
and 
\begin{eqnarray}
	\mathcal{F}^{\rm bcc}_{2,3} &=& \frac1{48\pi^3},
	\mathcal{F}^{\rm bcc}_{2,2} = -\frac{9 - 12\ln 2}{64\pi^3},
	\mathcal{F}^{\rm bcc}_{2,1} = \frac{180 - 8 \pi ^2 - 1296 \ln2 + 864\ln^22}{1536\pi ^3},\\
	\mathcal{F}^{\rm bcc}_{2,0} &=& \frac{48 \zeta (3) - 147 + 78 \pi ^2  - 4104 \ln^22 + 12(141 + 4 \pi ^2)\ln2}{1536 \pi ^3} - 2\delta g^{\rm bcc}_2.
\end{eqnarray}

\section{Expansion of free energy of paramagnetic phase}\label{appendix:PM_expansion}
In this Appendix we consider in detail the dependence of the band energy of non-interacting electrons per one spin projection 
\begin{equation}\label{eq:E_def}
    \mathcal{E}(\tau, E_{\rm F}) = \frac1{N}\sum_{\mathbf{k}}\ek f(\ek),
\end{equation}
on $\tau$. The Fermi energy $E_{\rm F}$ is set by equation on electron filling 
\begin{equation}\label{eq:appendix:ef_equation}
    \mathcal{I}(\tau, E_{\rm F}) \equiv \frac1{N}\sum_{\mathbf{k}} f(\ek) = 1/2.
\end{equation}
%derived from~Eq.~(\ref{eq:PM_n}). 
We focus attention on the case of DOS van Hove singularities impact (the square lattice). 

\textbf{Square lattice.} 

For convenience we count the spectrum from the position of van Hove singularity, $\varepsilon = \ek + 4\tau$, and rewrite Eq.~(\ref{eq:E_def}) through a density of states  for the square lattice
\begin{equation}
\rho_{\rm sq}(\varepsilon, \tau) = \frac{\mathbb{K}\left(1 - \frac{\varepsilon^2/16}{1 + \varepsilon\tau - 4\tau^2}\right)}{2\pi^2\sqrt{1 + \varepsilon\tau - 4\tau^2}}.
\end{equation}
\begin{equation}\label{eq:appendix:E_def}
	\mathcal{E}_{\rm sq}(\tau, \efsq) = \int\limits_{\varepsilon_1(\tau)}^\efsq\varepsilon\rho_{\rm sq}(\varepsilon,\tau)d\varepsilon,
\end{equation}
where $\varepsilon_1(\tau) = -4 + 8\tau$ and the Fermi level $\efsq = \efsq(\tau)$ is determined by the equation (\ref{eq:appendix:ef_equation}), for sqaure lattice we write $I$ as
\begin{equation}\label{eq:appendix:I_def}
I_{\rm sq}(\tau, \efsq) = \int\limits_{\varepsilon_1(\tau)}^\efsq\rho(\varepsilon,\tau)d\varepsilon ,
\end{equation}
%Here the reference point for $\varepsilon$ count is taken in such a way to combine the van Hove singularity and zero. 
Both the integrals (\ref{eq:appendix:E_def}),(\ref{eq:appendix:I_def}) have common form and are presented  as
\begin{equation}
\int\limits_{\varepsilon_1(\tau)}^\efsq = \int\limits_{\varepsilon_1(\tau)}^{\varepsilon_1} + \int\limits_{\varepsilon_1}^0 + \int\limits_{0}^\efsq,
\end{equation}
where $\varepsilon_1 = \varepsilon_1(0)$. 
We split the integral in (\ref{eq:appendix:I_def}):
\begin{equation}\label{eq:I_split}
	I_{\rm sq}(\tau, \efsq) = I^{\rm sq}_{\rm bnd}(\tau) + I^{\rm sq}_0(\tau) + I^{\rm sq}_{\rm sing}(\tau, \efsq),
\end{equation}
where only $I^{\rm sq}_{\rm sing}(\tau, \efsq)$ contains non-analytic dependence on $\efsq$ and
\begin{equation}
	I^{\rm sq}_{\rm bnd}(\tau) = \int\limits_{\varepsilon_1(\tau)}^{\varepsilon_1}\rho_{\rm sq}(\varepsilon, \tau)d\varepsilon,
	I^{\rm sq}_0(\tau) =  \int\limits_{\varepsilon_1}^0\rho_{\rm sq}(\varepsilon,\tau)d\varepsilon,
	I^{\rm sq}_{\rm sing}(\tau, \efsq) = \int\limits_0^{\efsq}\rho_{\rm sq}(\varepsilon, \tau)d\varepsilon.
\end{equation}
To treat the first term we use the fact that $\rho_{\rm sq}(\ebt, \tau)$ is analytic function in the vicinity of $\varepsilon = \ebt$, so that we  directly obtain %for the integrand
\begin{equation}\label{eq:I_bnd_preliminary_exp}
I^{\rm sq}_{\rm bnd}(\tau) = \int\limits_{\ebt}^{\varepsilon_1}\left(\rho_{\rm sq}(\ebt, \tau) + \partial_\varepsilon\rho_{\rm sq}(\ebt, \tau)(\varepsilon - \ebt) + \cdots\right)d\varepsilon.
\end{equation}
Expanding this up to fourth-order terms with respect to $\tau$ we have
\begin{equation}\label{eq:I_bnd_exp}
I^{\rm sq}_{\rm bnd}(\tau) = -\frac{2 \tau }{\pi }-\frac{3 \tau ^2}{\pi }-\frac{41 \tau ^3}{6 \pi } + \mathcal{O}(\tau^4).
\end{equation}
The direct Taylor expansion of $I_0$ in $\tau$ yields
\begin{equation}\label{eq:I_0_exp}
I^{\rm sq}_0(\tau) = 1/2 + \sum_{n = 1}^\infty A_{In}\tau^n . 
\end{equation}
where
\begin{equation}\label{eq:appendix:AI_def}
A_{In} = \frac{1}{n!}\int_{\varepsilon_1}^0\left.\frac{\partial^n\rho_{\rm sq}(\varepsilon,\tau)}{\partial\tau^n}\right|_{\tau = 0}d\varepsilon.
\end{equation}
We hold only  few of coefficients:
\begin{eqnarray}
	A_{I1} &=& (2/\pi^2)(\pi - 2),\\
	A_{I2} &=& 3/\pi,\\
\label{eq:appendix:AI3_value}
	A_{I3} &=& 1.7248.
\end{eqnarray} 
%Summing the Eqs.~(\ref{eq:I_0_exp}), (\ref{eq:I_bnd_exp}) we get
%$$
%I_0(\tau) + I_{\rm bnd}(\tau) = \frac12 -\frac{4\tau}{\pi^2}  + \left(A_{I3} -\frac{41}{6 \pi }\right)\tau^3 + \mathcal{O}(\tau^4).
%$$

To investigate $I_{\rm sing}$, we present $\rho_{\rm sq}(\varepsilon, \tau)$ in the following form
\begin{equation}
\rho_{\rm sq}(\varepsilon, \tau) = (2\pi^2)^{-1}F(\varepsilon^2/16, \varepsilon\tau - 4\tau^2),
\end{equation} 
where
$F(u, v) = \frac{\mathbb{K}\left(1 - \frac{u}{1 + v}\right)}{\sqrt{1 + v}}$, 
which is convenient for the expansion with respect to small $u,v$ being quadratic in small parameters $\varepsilon$ and $\tau$. We obtain singular (logarithmic) terms with respect to $u$ and an analytic dependence on $v$. We expand the integrand up to second order in $u$ and $v$ and integrate the result with respect to $\varepsilon$
\begin{multline}\label{eq:I_sing_exp}
I_{\rm sing}(\tau,\efsq) = \frac{\efsq}{2\pi^2}\left(1 + \ln\frac{16}{\efsq} 
\right.\\
\left.
+ \frac{(\efsq)^2}{576}\left(-2 + 3\ln\frac{16}{\efsq}\right) + \frac{\tau\efsq}{8}\left(1 - 2\ln\frac{16}{\efsq}\right) + 2\tau^2\ln\frac{16}{\efsq}\right) + o(|\efsq|(\tau + |\efsq|)^3).
\end{multline}

Sutstituting  Eqs.~(\ref{eq:I_bnd_exp}),(\ref{eq:I_0_exp}),(\ref{eq:I_sing_exp})  %We get from the Eqs.~(\ref{eq:Ef_eq}), into the Eq.~(\ref{eq:I_split}) 
we obtain
%We get the result for (\ref{eq:I_split})
\begin{multline}\label{eq:I_final}
I_{\rm sq}(\tau, \efsq) = \frac12 -\frac{4\tau}{\pi^2} + \frac{\efsq}{2\pi^2}\left(1 + \ln\frac{16}{\efsq}\right)   + \\
+ \left(A_{I3} -\frac{41}{6 \pi }\right)\tau^3 + \frac{\efsq}{2\pi^2}\left(\frac{\left(\efsq\right)^2}{576}\left(-2 + 3\ln\frac{16}{\efsq}\right) + \frac{\tau\efsq}{8}\left(1 - 2\ln\frac{16}{\efsq}\right) + 2\tau^2\ln\frac{16}{\efsq}\right) + o((|\tau|+|\efsq|)^4).
\end{multline}
We substitute this result in  Eq.~(\ref{eq:appendix:ef_equation}), where, for convenience, the parameter  $w = \tau/\efsq$ is introduced, which results in the equation
%\begin{multline}
%{\efsq}\left(1 + \ln\frac{16}{\efsq} + \frac{\left(\efsq\right)^2}{576}\left(-2 + 3\ln\frac{16}{\efsq}\right) + \frac{\tau\efsq}{8}\left(1 - 2\ln\frac{16}{\efsq}\right) + 2\tau^2\ln\frac{16}{\efsq}\right) \\ -8\tau  + 2A\tau^3  = 0,
%\end{multline}
%%%%%% Important!!!
%where $A = (A_{I3}\pi -41/6)\pi$. 
%%%%%
%\begin{multline}
%1 + \ln\frac{16}{\efsq}  -8w  \\ + \left(\efsq\right)^2\left[\frac{1}{576}\left(-2 + 3\ln\frac{16}{\efsq}\right) + \frac{w}{8}\left(1 - 2\ln\frac{16}{\efsq}\right) + 2w^2\ln\frac{16}{\efsq} + 2(A_{I3}\pi -41/6)\pi w^3\right]   = 0.
%\end{multline}
\begin{multline}
w = w_0(\efsq) \\
+ \frac{\left(\efsq\right)^2}{8}\left[\frac{1}{576}\left(-2 + 3\ln\frac{16}{\efsq}\right) + \frac{w}{8}\left(1 - 2\ln\frac{16}{\efsq}\right) + 2w^2\ln\frac{16}{\efsq} + 2(A_{I3}\pi -41/6)\pi w^3\right]   = 0.
\end{multline}
The leading-order solution of this equation yields the 
%expression~(\ref{eq:w0_def})  %%%% Reference to main text
expression~(\ref{eq:w0_def})  %%%% Reference to main text, {eq:w0_def}
of~the~main text. 
So we have up to subleading order
%\begin{multline}
%w(\efsq) \simeq w_0(\efsq) + \\+ \frac{\left(\efsq\right)^2}8\left[\frac{1}{576}\left(-2 + 3\ln\frac{16}{\efsq}\right) + \frac{w_0(\efsq)}{8}\left(1 - 2\ln\frac{16}{\efsq}\right) + 2w^2_0(\efsq)\ln\frac{16}{\efsq} + 2Aw^3_0(\efsq)\right].
%\end{multline}
%So we get semi-analytic solution
\begin{equation}
w(\efsq) \simeq w_0(\efsq) + \gamma\left(w_0\left(\efsq\right)\right)\left(\efsq\right)^2 ,
\end{equation}
where 
\begin{equation}\label{eq:appendix:gamma_def}
\gamma(w) = \frac18\left(Bw^3 - 4w^2 + 5w/12 - 5/576\right),
\end{equation}
and $B = 2A_{I3}\pi^2 -41\pi/3 + 16 = 7.11111$. 
%by~Eq.~(\ref{eq:gamma_def}).
%\begin{equation}
%\gamma(\efsq) = \frac{1}{2048}\left(
%\left(A + 8\right)\LL{3} + \left(3A + 8\right)\LL{2}  + \left(3A + 16/3\right)\ln\frac{16}{\efsq}
% + A + 28/9\right).
%\end{equation}
%Another form for $\gamma(\efsq)$ reads

Analogous program can be implemented for  Eq.~(\ref{eq:appendix:E_def})
\begin{equation}
\mathcal{E}_{\rm sq}(\tau, \efsq) = \mathcal{E}_{\rm bnd}(\tau) + \mathcal{E}_0(\tau) + \mathcal{E}_{\rm sing}(\tau, \efsq),
\end{equation}
where
\begin{equation}
	\mathcal{E}^{\rm sq}_{\rm bnd}(\tau) = \int\limits_{\varepsilon_1(\tau)}^{\varepsilon_1}\varepsilon\rho(\varepsilon, \tau)d\varepsilon,\\
	\mathcal{E}^{\rm sq}_0(\tau) =  \int\limits_{\varepsilon_1}^0\varepsilon\rho(\varepsilon,\tau)d\varepsilon,\\
	\mathcal{E}^{\rm sq}_{\rm sing}(\tau, \efsq) = \int\limits_0^{\efsq}\varepsilon\rho(\varepsilon, \tau)d\varepsilon.
\end{equation}
We get analogously to~the~derivation~(\ref{eq:I_bnd_preliminary_exp}) and (\ref{eq:I_bnd_exp})
\begin{equation}
\mathcal{E}^{\rm sq}_{\rm bnd}(\tau) = \frac{8 \tau }{\pi }+\frac{4 \tau ^2}{\pi }+\frac{14 \tau ^3}{\pi }+\frac{265 \tau ^4}{6 \pi }+O\left(\tau ^5\right)
\end{equation}
\begin{multline}
\mathcal{E}^{\rm sq}_{\rm sing}(\tau,\efsq) = \frac{\left(\efsq\right)^2}{2\pi^2}\left(\frac{1 + 2\ln\frac{16}{\efsq}}4 
\right.\\
\left.
+ \frac{\left(\efsq\right)^2}{1024}\left(-3 + 4\ln\frac{16}{\efsq}\right) + \frac{\tau\efsq}{18}\left(2 - 3\ln\frac{16}{\efsq}\right) + \frac{\tau^2}2\left(-1 + 2\ln\frac{16}{\efsq}\right)\right) + o(\text{4th order}).
\end{multline}
\begin{equation}
	\mathcal{E}^{\rm sq}_0(\tau) = \mathcal{E}_0(0) + A_{\mathcal{E}1}\tau + A_{\mathcal{E}2}\tau^2 + A_{\mathcal{E}3}\tau^3 + A_{\mathcal{E}4}\tau^4 + o(\tau^4),
\end{equation}
where $\mathcal{E}_0(0) = -8/\pi^2$, 
\begin{equation}\label{eq:appendix:AE_def}
A_{\mathcal{E}n} = \frac1{n!}\int\limits_{\varepsilon_1}^0\varepsilon\left.\frac{\partial^n\rho(\varepsilon,\tau)}{\partial\tau}\right|_{\tau=0}d\varepsilon,
\end{equation}
and numerical calculation yields
\begin{eqnarray}
A_{\mathcal{E}1} &=& 2 - 8/\pi,\\
A_{\mathcal{E}2} &=& -1.81362,\\
A_{\mathcal{E}3} &=& -14/\pi,\\
A_{\mathcal{E}4} &=& -14.455. 
\end{eqnarray}

Summing all contributions we have
\begin{multline}
\mathcal{E}_{\rm sq}(\tau,\efsq) = -8/\pi^2 + 2\tau + \left(A_{\mathcal{E}2} + \frac{4}{\pi}\right)\tau^2 +  \left(A_{\mathcal{E}4} + \frac{265}{6\pi}\right)\tau ^4 \\
+ \frac{\left(\efsq\right)^2}{2\pi^2}\left(\frac{1 + 2\ln\frac{16}{\efsq}}4 + \frac{\left(\efsq\right)^2}{1024}\left(-3 + 4\ln\frac{16}{\efsq}\right) + \frac{\tau\efsq}{18}\left(2 - 3\ln\frac{16}{\efsq}\right) + \frac{\tau^2}2\left(-1 + 2\ln\frac{16}{\efsq}\right)\right).
\end{multline}
%where $A_2 = A_{\mathcal{E}2} + {4}/{\pi}$, $A_4 = A_{\mathcal{E}4} + 265/(6\pi)$. 
We now substitute $\tau = w \efsq = \left(w_0 + \gamma\left(\efsq\right)^2\right)\efsq$ to obtain 
\begin{equation}
\mathcal{E}_{\rm sq}(\tau,\efsq) = -8/\pi^2 + 2\tau + \sum_{k = 2,4}\sum_{n = 0}^k a^{\rm sq}_{kn}w^n_0(\efsq)\left(\efsq\right)^k,
\end{equation}
where coefficients $a^{\rm sq}_{kn}$ read %are presented in the main text (\ref{eq:acoef1}-\ref{eq:acoef2}).
\begin{eqnarray}
\label{eq:acoef1}
a^{\rm sq}_{20} &=& -1/(8\pi^2) = -0.0126651,\\
a^{\rm sq}_{21} &=& 2/\pi^2 = 0.202642,\\
a^{\rm sq}_{22} &=& A_{\mathcal{E}2} + {4}/{\pi} = -0.54038,\\
a^{\rm sq}_{40} &=& -7/(2048\pi^2) = -0.000346313,\\
a^{\rm sq}_{41} &=& -5A_{\mathcal{E}2}/2304 + (89 - 5\pi)/(576\pi^2) = 0.0168282,\\
a^{\rm sq}_{42} &=&  5A_{\mathcal{E}2}/48 + (5\pi - 17)/(12\pi^2) = -0.199828,\\
a^{\rm sq}_{43} &=& 4(1 - \pi)/\pi^2 - A_{\mathcal{E}2} = 0.945664,\\
\label{eq:acoef2}
a^{\rm sq}_{44} &=& \frac14A_{\mathcal{E}2}B
	+ 2\pi A_{I3} +  A_{\mathcal{E}4} + \frac{361 - 82\pi}{6\pi} = -1.35695.
% old variant for a_{44} &=& (A_{\mathcal{E}2} + {4}/{\pi})(A_{I3}\pi^2 - 41\pi/6 + 8)/2 + A_{\mathcal{E}4} + 265/(6\pi),
\end{eqnarray}

\textbf{Sc lattice.} We expand the free energy per one spin projection 
\begin{equation}\label{eq:Appendix:energy_PM_phase}
    \mathcal{E}_{\rm sc}(\tau) = \frac1{N}\sumk t^{\rm sc}_\mathbf{k}\theta(\efsc - t^{\rm sc}_\mathbf{k}(\tau)), 
\end{equation}
for sc lattice with electron spectrum
\begin{equation}
    t^{\rm sc}_\mathbf{k}(\tau) = [-2(\cos k_x + \cos k_y + \cos k_z) + 4\tau(\cos k_x\cos k_y + \cos k_y\cos k_z + \cos k_x\cos k_z)]t. 
\end{equation}
where the Fermi energy $\efsc$ is determined from the equation
\begin{equation}\label{eq:Appendix:filling_PM_phase}
    I_{\rm sc}(\tau) =  \frac1{N}\sumk \theta(\efsc - t^{\rm sc}_\mathbf{k}(\tau)) = 1/2.
\end{equation}
Below we omit the argument of $t^{\rm sc}_\mathbf{k}$ for brevity. 
We directly differentiate Eq.~(\ref{eq:Appendix:filling_PM_phase}) with respect to $\tau$ to obtain
\begin{equation}
    \frac1{N}\sumk \partial_{\tau}(\efsc - t^{\rm sc}_\mathbf{k})\delta(\efsc - t^{\rm sc}_\mathbf{k}(\tau)) = 0, 
\end{equation}
so we obtain 
\begin{equation}\label{eq:deriv_of_Ef_sc}
    \partial_{\tau}\efsc = \rho^{\rm sc}_1(\efsc)/\rho^{\rm sc}_0(\efsc),
\end{equation}
where the partial DOS 
\begin{equation}
    \rho^{\rm sc}_n(E;\tau) = \frac1{N}\sumk [\partial_{\tau}t^{\rm sc}_\mathbf{k}(\tau)]^n\delta(E - t^{\rm sc}_\mathbf{k}(\tau))
\end{equation}
is introduced and $\rho^{\rm sc}_0(E;\tau) = \rho_{\rm sc}(E,\tau)$. 
We expand $\mathcal{E}_{\rm sc}(\tau)$ in powers of $\tau$ 
to calculate
\begin{equation}\label{eq:first_deriv}
	\frac{\partial \mathcal{E}_{\rm sc}}{\partial \tau} 
	= \frac1{N}\sumk [t^{\rm sc}_\mathbf{k}\partial_{\tau}(\efsc - t^{\rm sc}_\mathbf{k})\delta(\efsc - t^{\rm sc}_\mathbf{k}) + \partial_{\tau}t^{\rm sc}_\mathbf{k}\theta(\efsc - t^{\rm sc}_\mathbf{k})].
\end{equation}
Due to Eq.~(\ref{eq:Appendix:filling_PM_phase}) the first term vanishes. Using the bipartite character of the lattice, $E_{\rm F}(0) = 0$, $\partial_{\tau}t^{\rm sc}_{\mathbf{k} + \mathbf{Q}} = \partial_{\tau}t^{\rm sc}_{\mathbf{k}}$, $t^{\rm sc}_{\mathbf{k} + \mathbf{Q}}(\tau=0) = -t^{\rm sc}_{\mathbf{k}}(\tau=0)$,  we shift summation momentum $\mathbf{k}\rightarrow\mathbf{k} + \mathbf{Q}$ in the latter formula at $\tau = 0$ and obtain
\begin{equation}
    	\left.\frac{\partial \mathcal{E}_{\rm sc}}{\partial \tau}\right|_{\tau = 0} 
    	= \frac1{N}\sumk \partial_{\tau}t^{\rm sc}_\mathbf{k + \mathbf{Q}}\theta(\efsc - t^{\rm sc}_{\mathbf{k}+ \mathbf{Q}}(\tau = 0))] = 0.
\end{equation}
%Thus, in the first term (\ref{eq:first_deriv}) we perform the replacement $t^{\rm sc}_\mathbf{k}\rightarrow \efsc$ due to the delta-function and get vanishing of the first term. 
From the Eq.~(\ref{eq:first_deriv}) we get
\begin{equation}
	\frac{\partial^2 \mathcal{E}_{\rm sc}}{\partial \tau^2} = \sumk 
	\partial_{\tau}t^{\rm sc}_\mathbf{k}\cdot\partial_{\tau}(\efsc-t^{\rm sc}_\mathbf{k})\delta(\efsc - t^{\rm sc}_\mathbf{k}) = (\partial_{\tau}\efsc)\rho_1(\efsc; \tau) - \rho_2(\efsc; \tau).
\end{equation}
Using Eq.~(\ref{eq:deriv_of_Ef_sc}) we obtain
\begin{equation}\label{eq:second_deriv_Ef_sc}
	\frac{\partial^2 \mathcal{E}_{\rm sc}}{\partial \tau^2} = \frac{\left(\rho^{\rm sc}_1(E;\tau)\right)^2}{\rho_{\rm sc}(E;\tau)} - \rho^{\rm sc}_2(E;\tau).
\end{equation}
We calculate $\rho^{\rm sc}_n(\efsc = 0)$, that corresponds to the case of half-filling,
\begin{equation}
	\rho_n(\efsc = 0) = \frac{4^n}{2t}\int \frac{d\mathbf{k}}{(2\pi)^3} (\cos k_x\cos k_y + \cos k_y\cos k_z + \cos k_x\cos k_z)^n\delta(\cos k_x + \cos k_y + \cos k_z).
\end{equation}
Since for the integration domain $\cos k_z = -\cos k_x - \cos k_y$,
\begin{equation}
	\rho_n(\efsc = 0) = \frac{(-4)^n}{2t}\int \frac{d\mathbf{k}}{(2\pi)^3} (\cos^2 k_x + \cos^2 k_y + \cos k_x\cos k_y)^n\delta(\cos k_x + \cos k_y + \cos k_z).
\end{equation}
Introducing $r_a = \cos k_a$, $a = x,y,z$ we have
\begin{equation}
	\rho_n(\efsc = 0) = \frac{(-4)^n}{2t\pi^3}\int\limits_{-1}^{+1} \frac{dr_x}{\sqrt{1-r_x^2}}\int\limits_{-1}^{+1}\frac{dr_y}{\sqrt{1-r_y^2}}\int\limits_{-1}^{+1}\frac{dr_z}{\sqrt{1-r_z^2}}  (r_x^2 + r^2_y + r_xr_y)^n\delta(r_x + r_y + r_z).
\end{equation}
We take into account that the regions $r_x < 0, r_y>0$ and $r_x > 0, r_y<0$ yield the same contributions; the same takes place also for  $r_x, r_y< 0$ and $r_x, r_y> 0$ (we account this by adding the factor of 2). 
\begin{multline}
	\rho_n(\efsc = 0) = \frac{(-4)^n}{\pi^3t} \left[\int\limits_{0}^{+1} \frac{dr_x}{\sqrt{1-r_x^2}}\int\limits_{0}^{+1}\frac{dr_y}{\sqrt{1-r_y^2}}\frac{(r_x^2 + r^2_y + r_xr_y)^n}{\sqrt{1-(r_x+r_y)^2}}  \theta(1-r_x+r_y) + \right. \\
	\left. +
	\int\limits_{0}^{+1} \frac{dr_x}{\sqrt{1-r_x^2}}\int\limits_{0}^{+1}\frac{dr_y}{\sqrt{1-r_y^2}}\frac{(r_x^2 + r^2_y - r_xr_y)^n}{\sqrt{1-(r_x-r_y)^2}}  
	\right].
\end{multline}
Direct numerical calculation yields 
$\rho^{\rm sc}_0(0;0) = 0.142127$, $\rho^{\rm sc}_1(0;0) = -0.353038$,	$\rho^{\rm sc}_2(0;0) = 1.023248$ 
%$\rho^{\rm sc}_0(0;0) = 0.14212727597525068\cdot t^{-1}$, $\rho^{\rm sc}_1(0;0) = -0.3530378926770085\cdot t^{-1}$,	$\rho^{\rm sc}_2(0;0) = 1.0232475551639986\cdot t^{-1}$ 
then the formula (\ref{eq:second_deriv_Ef_sc})
yields at $\tau = 0$,
\begin{equation}
	a^{\rm sc}_{2,\rm PM} \equiv \frac{\left[\rho^{\rm sc}_1(0;0)\right]^2}{\rho_{\rm sc}(0;0)} - \rho^{\rm sc}_2(0;0) = -0.146317.
\end{equation}
Therefore we get to the leading order  the free energy per one spin projection
$$
\mathcal{E}_{\rm sc}(\tau) \approx \mathcal{E}_{\rm sc}(0) + {a^{\rm sc}_{2,\rm PM}}\tau^2/2.
$$
%Eq.~(\ref{eq:delta_F_HFA_PM_sc}).


\begin{thebibliography}{99}

%MIT review
\bibitem{1998:Imada}  M.Imada, A. Fujimori, Y. Tokura, Rev. Mod. Phys.~\textbf{70}, 1039 (1998)
%
\bibitem{book:Gebhard} F.~Gebhard. The Mott Metal-Insulator Transition: Models and Methods, -- Springer. Berlin--Heidelberg (1997).

\bibitem{book:Mott} N.F. Mott, Metal-Insulator Transitions, Taylor \& Francis Ltd. (1974).

%First-order metal-insulator transitions in the extended Hubbard model due to self-consistent screening of the effective interaction
\bibitem{2018:Schuler} M.~Schuler, E.G.C.P.~van~Loon, M.I.~Katsnelson, and T.O.~Wehling, Phys.~Rev.~B~\textbf{97}, 165135~(2018).

\bibitem{1951:Slater} J.C.Slater, Phys. Rev.~\textbf{51}, 538 (1951).

\bibitem{1949:Mott} N.F.~Mott, Proc. Phys. Soc. London, Ser. A~\textbf{62}, 416 (1949).

%%% Experiment:
% NaOsO3 Slater examples:
% Continuous metal-insulator transition of the antiferromagnetic perovskite NaOsO3
\bibitem{2009:Shi}  Y.G. Shi, Y.F. Guo, S. Yu, M. Arai, A.A. Belik, A. Sato, K. Yamaura, E. Takayama-Muromachi, H.F. Tian, H.X. Yang, J. Q. Li, T. 
Varga, J. F. Mitchell, and S. Okamoto, Phys.~Rev.~B~\textbf{80}, 161104(R) (2009).
% Magnetically driven metal-insulator transition in NaOsO3
\bibitem{2012:Calder} S. Calder, V.O. Garlea, D.F. McMorrow, M.D. Lumsden, M.B.~Stone, J.C.~Lang, J.W. Kim, J.A.~Schlueter, Y.G. Shi, K. Yamaura, Y.S. Sun, Y. Tsujimoto, and A.D. Christianson, Phys.~Rev.~Lett.~\textbf{108}, 257209 (2012).
% Structural and correlation effects in the itinerant insulating antiferromagnetic perovskite NaOsO3
\bibitem{2013:Jung} M.-C. Jung, Y.-J. Song, K.-W. Lee, and W. E. Pickett, 
Phys.~Rev.~B~\textbf{87}, 115119 (2013).

%Metal–Insulator Transitions in Pyrochlore Oxides Ln2Ir2O7, Ln = Nd, Sm, Eu, Gd, Tb, Dy, and Ho
\bibitem {2011:Matsuhira} K. Matsuhira, M. Wakeshima, Y. Hinatsu, and S. Takagi, J.~Phys.~Soc.~Jpn.~\textbf{80}, 094701 (2011).

% Continuous metal-insulator transition in the pyrochlore Cd2Os2O7
\bibitem{2001:Mandrus} D. Mandrus, J.R. Thompson, R. Gaal, L. Forro, J.C. 
Bryan, B.C. Chakoumakos, L.M. Woods, B.C. Sales, R. S. Fishman, and V. Keppens, Phys.~Rev.~B~\textbf{63}, 195104 (2001).


% Rare example of Slater sceraio of MIT in Pb2CaOsO6
\bibitem{2020:Pb2CaOsO6} H. Jacobsen, H.L. Feng, A.J. Princep, M.C. Rahn, 
Y. Guo, J. Chen, Y. Matsushita, Y. Tsujimoto, M. Nagao, D. Khalyavin, P. Manuel, C.A. Murray, Ch. Donnerer, J.G. Vale, M.M.~Sala, K.~Yamaura, A.T.~Boothroyd, Phys.~Rev.~B~\textbf{102}, 214409 (2020).

%"Magnetic field frustration of the metal-insulator transition in V2O3"
%Comment: The nature of MIT transition (Mott or Slater) is discussed, see detaily
\bibitem{2020:Trastoy} J.~Trastoy, A.~Camjayi, J.~del~Valle, Y.~Kalcheim, 
J.-P.~Crocombette, D.A.~Gilbert, J.A.~Borchers, J.E.~Villegas, D.~Ravelosona, M.J.~Rozenberg, and I.K.~Schuller, Phys.~Rev.~B~\textbf{101}, 245109 (2020).

%New phase boundary in highly correlated, barely metallic V2O3
\bibitem{1991:Carter:V2O3} S. A. Carter, T. F. Rosenbaum, J. M. Honig, and J. Spalek, Phys. Rev. Lett.~\textbf{67}, 3440~(1991).

%Metal-insulator transition and magnetic properties in the NiS2−xSex system
\bibitem{1992:Sudo:NiSSex} S.~Sudo, J. Magn. Mag. Mat.~\textbf{114}, 57 (1992).

% Optical Signature of a Crossover from Mott- to Slater-Type Gap in Sr2 Ir1 − xRhx O4
%Comment: Non-trivial transition from Mott to Slater scenario in Sr_2Ir_{1-x}Rh_xO_4
\bibitem{2020:Xu} B. Xu, P. Marsik, E. Sheveleva, F. Lyzwa, A. Louat, V. Brouet, D. Munzar, and C. Bernhard, Phys. Rev. Lett.~\textbf{124}, 027402 (2020).

%%%%%%%%%%%%%%%%%%%%%%%%%%%%%%%%%%%%%

\bibitem{Hubbard-I} J.~Hubbard, Proc. Roy. Soc. Series~A.~\textbf{276}, 238 (1963).

\bibitem{Hubbard-III} J.~Hubbard, Proc. Roy. Soc. Series~A.~\textbf{281}, 
401 (1964)

\bibitem{1984:Katsnelson} M.I.~Katsnelson and V.Yu.~Irkhin, J. Phys. C~\textbf{17}, 4291 (1984).

\bibitem{Spalek} P.~Korbel, W.~Wójcik,  A.~Klejnberg \textit{et al.}, Eur. Phys. J. B~\textbf{32}, 315 (2003).


\bibitem{fRG_review:Salmhofer} M. Salmhofer and C. Honerkamp, Prog. Theor. Phys.~\textbf{105}, 1 (2001).

%\bibitem{Parquet} I.T.Diatlov, V.V. Sadakov, K.A.~Ter-Martirosyan, Sov. Phys. JETP \textbf{5}, 631 (1957).

\bibitem{fRG_T_flow} C.~Honerkamp and M.~Salmhofer, Phys.~Rev.~B~\textbf{64}, 184516 (2001).

%\bibitem{2004:Zarubin} V.Yu. Irkhin and A.V. Zarubin,  Europ. Phys. J. B \textbf{38}, 563 (2004).

%\bibitem{2017:Igoshev} P.A.~Igoshev, E.E.~Kokorina, I.A.~Nekrasov, Physics of Metals and Metallography~\textbf{118}, 207~(2017).

\bibitem{1990:Schulz} H. J. Schulz, Phys. Rev. Lett.~\textbf{64}, 1445 (1990).

\bibitem{2015:Igoshev} P.A.~Igoshev, M.A.~Timirgazin, V.F.~Gilmutdinov, A.K.~Arzhnikov and V. Yu.~Irkhin, J. Phys.: Cond. Matt.~\textbf{27}, 446002 (2015).

\bibitem{2007:Igoshev} P.A.~Igoshev, A.A.~Katanin, V.Yu.~Irkhin, JETP~\textbf{105}, 1043 (2007).

\bibitem{2010:Igoshev} P.A.~Igoshev, M.A.~Timirgazin, A.A.~Katanin, A.K.~Arzhnikov and V.Yu.~Irkhin, Phys.~Rev.~B~\textbf{81},
094407 (2010).

\bibitem{2011:Igoshev} P.A.~Igoshev, V.Yu.~Irkhin, and A.A.~Katanin, Phys.~Rev.~B~\textbf{83}, 245118 (2011).

%"Intermediate coupling model of the cuprates, Advances in Physics"
%Comment GW theory applied to cuprate. I don't lile this paper. See it detaily!!!
\bibitem{2014:Markiewicz} T.~Das, R.S.~Markiewicz, and A.~Bansil, Adv. Phys.~\textbf{63}, 151 (2014).

%"Frustration of antiferromagnetism in the t-t' Hubbard model at weak coupling
%Comment: HFA treatment of AFM, metal phase, asymptotics. Very close to us!!
\bibitem{1998:Hofstetter} W. Hofstetter and D. Vollhardt, Ann. Phys.~\textbf{7}, 48 (1998).

\bibitem{2019:Igoshev_JETP_MIT} P.A.~Igoshev, V.Yu.~Irkhin, JETP~\textbf{128}, 909 (2019).

%"Quantum criticality around metal-insulator transitions of strongly correlated electron systems"
%Comment: MIT is considered as a quantum phase transition within HFA!!! Our expansions are really useful!
\bibitem{2007:Misawa} T.~Misawa and M.~Imada, Phys.~Rev.~B~\textbf{75}, 115121~(2007).

%%%%%%%%%%%%%%%%%%%%%%%%%%%%%%%%%%%%

%"Dynamical Mean Field Theory of the Antiferromagnetic Metal to Antiferromagnetic Insulator Transition"
%Comment: Consideration of MIT within DMFT on the Bethe lattice with t, t' and with direct exchange
\bibitem{1999:Chitra} R.~Chitra and G.~Kotliar, Phys.~Rev.~Lett.~\textbf{83}, 2386 (1999).

\bibitem{2016:Timirgazin} M. A. Timirgazin, P. A. Igoshev, A. K. Arzhnikov, V. Yu. Irkhin,  J.~Low.~Temp.~Phys. \textbf{185}, 651 (2016).

%%%%%%%%%%%%%%%%%%%%%%%%%%%%%%%%%%%%
%"On the Metal-Insulator Transition in a Two-Dimensional Hubbard model"
%Comment: The first considering of MIT problem within HFA for Hubbard model on the square lattice
\bibitem{1996:Kondo} H. Kondo and T. Moriya, J.~Phys.~Soc.~Jpn.~\textbf{65}, 2559 (1996).

%"Indications of a metallic antiferromagnetic phase in the two-dimensional U-t-t' model"
%Comment: Mean-field and Monte-Carlo
\bibitem{1997:Duffy} D. Duffy and A. Moreo, Phys. Rev.~B~\textbf{55}, R676 (1997).

%"Collinear antiferromagnetic state in a two-dimensional Hubbard model at half filling"
%Comment: Mean-field phase diagram and MIT at interval of large t'  (CAF phase with Q = (0,pi) occurs at large t')
\bibitem{2010:Yu} Z.-Q. Yu and L. Yin, Phys.~Rev.~B~\textbf{81}, 195122 (2010).

% Giant Van Hove Density of States Singularities and Anomalies of Electron and Magnetic Properties in Cubic Lattices
\bibitem{2019:Igoshev_FMM} P.A.~Igoshev, V.Yu.~Irkhin, Physics of Metals and Metallography~\textbf{120}, 1282 (2019).

%Electron Spectrum Topology and Giant Singularities of the Electron Density of States in Cubic Lattices
\bibitem{2019:Igoshev_JETP} P.A.~Igoshev, V.Yu.~Irkhin, JETP Letters~\textbf{110}, 727~(2019).


% The Occurrence of Singularities in the Elastic Frequency Distribution of a Crystal
\bibitem{1953:vanHove} L. van Hove, Phys.~Rev.~\textbf{89}, 1189 (1953).

%"Two-dimensional Hubbard model with nearest- and next-nearest-neighbor hopping"
%Comment: Mean-field and quantum Monte-Carlo ground state phase diagram. Early mean-field and Monte-Carlo:
\bibitem{1987:Hirsch} H. Q. Lin and J. E. Hirsch, Phys.~Rev.~B~\textbf{35}, 3359 (1987).


%"Magnetic properties and Mott transition in the square-lattice Hubbard model with frustration"
%Comment: Variational Claster Approximation, details ???
\bibitem{2013:Yamada} A. Yamada, K.~Seki, R.~Eder, and Y.~Ohta Phys.~Rev.~B~\textbf{88}, 075114 (2013).

%"Magnetism and d-wave superconductivity on the half-filled square lattice with frustration"
%Comment: again variational cluster approximation (VCA), SC is also considered. It is strange that Uc for AFM phase within VCA is smaller than within HFA. MIT is also considered and strong finite-size effects at small t'(<0.2) are found. See it detaily!!!
\bibitem{2008:Nevidomskyy} A.H.~Nevidomskyy, C.~Scheiber, D.~Senechal, and A. M. S. Tremblay, Phys.~Rev.~B~\textbf{77}, 064427 (2008).

%Mott Transitions and d-Wave Superconductivity in Half-Filled-Band Hubbard Model on Square Lattice with Geometric Frustration
%Comment: the consideration within the variational Monte-Carlo approach. Non-direct construction of the ground state phase diagram. At t' > 0.25t AFM insulator state is destroyed by SC or non-magnetic insulator state occurs. No metalicity discission
\bibitem {2006:Yokoyama} H. Yokoyama, M. Ogata and Y. Tanaka, J.~Phys.~Soc.~Jpn.~\textbf{75}, 114706 (2006).
%"Metal-insulator transition and strong-coupling spin liquid in the t−t' Hubbard model"
%Comment: This is improved variational Monte-Carlo approximation which yields some new results only at large t'> 0.5t (including spin liquid) 
\bibitem{2009:Becca} F.~Becca, L.F.~Tocchio, and S.~Sorella, Journal of Physics: Conference Series \textbf{145}, 012016 (2009).

%"Role of backflow correlations for the nonmagnetic phase of the t-t' Hubbard model"
%Comment: The same as previous paper
\bibitem{2008:Tocchio} L.F.~Tocchio, F.~Becca, A.~Parola, and S.~Sorella Phys.~Rev.~B~\textbf{78}, 041101(R)~(2008).

% Something about PATH-INTEGRAL RENORMALIZATION GROUP METHOD (PRG), see it!!!
%"Path-Integral Renormalization Group Method for Numerical Study on Ground States of Strongly Correlated Electronic Systems"
\bibitem{2001:Kashima} T. Kashima and M. Imada, J. Phys.~Soc.~Jpn. \textbf{70}, 3052 (2001).
%"Nonmagnetic Insulating States near the Mott Transitions on Lattices with Geometrical Frustration and Implications for к-(ET)2 Cu2(CN)3"
\bibitem{2001:Morita} H. Morita, S. Watanabe, and M. Imada, J.~Phys.~Soc.~Jpn.~\textbf{71}, 2109 (2002).

%"Gapless quantum spin liquid, stripe, and antiferromagnetic phases in frustrated Hubbard models"
%Comment: also PRG. It is found that at t'>~0.25t on the square lattice non-magnetic MIT phase occurs. At t'<~0.25t PM-AFM phase transition occurs. It is found that Uc is close to SBA and Monte-Carlo result. See details!!!
\bibitem{2006:Mizusaki} T.~Mizusaki, and M.~Imada, Phys.~Rev.~B~\textbf{74}, 014421 (2006).

\bibitem{1986:Kotliar} G.~Kotliar and A.E.~Ruckenstein, Phys.~Rev.~Lett.~\textbf{57}, 1362 (1986).
%"Impact of magnetic frustration on the Mott transition within a slave boson mean-field theory"
%Comment: Very close to our consideration
\bibitem{2000:Yang}
I. Yang, E. Lange, and G. Kotliar, Phys.~Rev.~B~\textbf{61}, 2521 (2000).

%From Slater to Mott–Heisenberg physics: the antiferromagnetic phase of the Hubbard model
%Comment: The qualitetive conclusion of validity of Slater picture of anitiferromagnetism at small U for hypercubic lattice is stated. Something strange... 
\bibitem{2003:Pruschke}  Th.~Pruschke and R.~Zitzler, J. Phys.: Condens. Matter~\textbf{15}, 7867 (2003).

%Phase Diagram of the Frustrated Hubbard Model
\bibitem{2004:Zitzler} R. Zitzler, N.-H. Tong, Th. Pruschke, and R. Bulla, Phys. Rev. Lett.~\textbf{93}, 016406 (2004).

%Half-filled Hubbard model on a Bethe lattice with next-nearest-neighbor hopping
\bibitem{2009:Peters} R. Peters and Th. Pruschke, Phys.~Rev.~B~\textbf{79}, 045108 (2009).

%\bibitem{Supplemental} Supplemental Material for: Metal-insulator transition and antiferromagnetism  in the generalized Hubbard model: 
%a treatment of correlation effects. In section 1 a~general derivation of useful expansion of~the~lattice sum $G(\Delta)$ with respect to $\Delta$ and the connection of singularity of $G(\Delta)$ at $\Delta = 0$ and the singularity of DOS $\rho(\epsilon) $ at $\epsilon = 0$. 
%The asymptotics for the density of states for the square, simple cubic and body-centered cubic lattices is considered and used  for analytical investigation of $G(\Delta)$ and free energy of AFM insulator phase for these lattices. 
%In section 2 an analytical expansion of free energy of the paramagnetic phase for the square and simple cubic lattice with respect to $t'$ with accurate treating the strong van Hove singularity in the former case is presented. 

%"Effects of interaction strength, doping, and frustration on the antiferromagnetic phase of the two-dimensional Hubbard model"
%Comment: cellular 2x2 DMFT investigation of AFM phase at finite T and doping. See it detaily!!!
\bibitem{2017:Frantino} L. Fratino, M. Charlebois, P. Semon, G. Sordi, and A. M. S. Tremblay, Phys.~Rev.~B~\textbf{96}, 241109(R) (2017).

%"Conditions for magnetically induced singlet d-wave superconductivity on the square lattice"
%Comment: the phase diagram of AFM and SC phases is constructed at finite temperature within two-particle self-consistent approximation. The agreement with our and other investigation is not seen. See details!!!
%\bibitem{2008:Hassan} S.R.~Hassan, B.~Davoudi, B.~Kyung, and A.-M.~S.~Tremblay, Phys.~Rev.~B~\textbf{77}, 094501 (2008).

% Phase separation in the particle-hole asymmetric Hubbard model
\bibitem{2007:Eckstein} M. Eckstein, M. Kollar, M. Potthoff, and D. Vollhardt, Phys.~Rev.~B~\textbf{75}, 125103~(2007).
 

%"Mott Transition, Antiferromagnetism, and d-Wave Superconductivity in Two-Dimensional Organic Conductors"
%Comment: Method is cellular DMFT, the lattice is triangular, with spectrum e(k) = -2(cos kx + cos ky) - 4t'cos(kx + ky). We miss it since it is not square lattice
%\bibitem{2006:Kyung} B. Kyung and A.-M.~S.~Tremblay, Phys.~Rev.~Lett. \textbf{97}, 046402 (2006).

\bibitem{1992:Fresard} R.~Fresard and P.~W\"{o}lfle, J. Phys.: Cond. Matt.~\textbf{4}, 3625 (1992).



\bibitem{2018:Igoshev} P.A. Igoshev, M.A. Timirgazin, A.K. Arzhnikov, V.Yu. Irkhin, J. of Magn. and Magn. Materials \textbf{459}, 311~(2018).


%"Magnetic phase diagram of an extended Hubbard model at half filling: Possible application for strongly correlated iron pnictides"
%Comment: Chinese work on the t-t'-t'' Hubbard model within SBA approximation at half-filling. Very crude phase diagram with small points. It is not valuable for us.
%\bibitem{2012:Li} T.-J. Li, Y.-M. Quan, D.-Y. Liu, L.-J. Zou, J. of Magn. and Magn. Materials \textbf{324},  1046 (2012).


\bibitem{2016:FCC-Timirgazin}  M. A. Timirgazin, P. A. Igoshev, A. K. Arzhnikov, V. Yu. Irkhin, J. Phys.: Cond. Matt.~\textbf{28}, 505601 (2016).

%\bibitem{eq:delta_F_HFA_sq_expansion} Eq.~(43) of Supplemental material for: Metal-insulator transition and antiferromagnetism  in the generalized Hubbard model: a treatment of correlation effects.
%\bibitem{eq:delta_F_AFM_HFA_sc_expansion} Eq.~(69) of Supplemental material for: Metal-insulator transition and antiferromagnetism  in the generalized Hubbard model: a treatment of correlation effects.
%\bibitem{eq:delta_F_AFM_HFA_bcc_expansion} Eq.~(85) of Supplemental material for: Metal-insulator transition and antiferromagnetism  in the generalized Hubbard model: a treatment of correlation effects.
\end{thebibliography}

\begin{thebibliography}{5}
\bibitem{1969:Jelitto} R.J.~Jelitto, J. Phys. Chem. Solids~\textbf{30}, 609 (1969).
\bibitem{1971:Katsura} S. Katsura and T. Horiguchi, J.~Math.~Phys.~\textbf{12}, 230 (1971).

\end{thebibliography}
\end{document}